\documentclass[aps,prb,showpacs,twocolumn]{revtex4-2}
\usepackage{amsmath}
\usepackage{amsfonts}
\usepackage{amssymb}
\usepackage{physics} 
\usepackage{bm} 

\usepackage{upgreek} 
\usepackage{graphicx}

\usepackage{hyperref} 
\usepackage{color} 
\usepackage{xspace} 
\usepackage{braket} 
\usepackage{todonotes}
\usepackage{graphicx}  
\usepackage{import}
\usepackage{changes} 

\usepackage{overpic} 

\definecolor{BlueO}{RGB}{126,155,209}
\definecolor{DarkBlueO}{RGB}{0,38,48}

\hypersetup{
	colorlinks=true,
	linkcolor=DarkBlueO,    
	urlcolor=BlueO,
	citecolor=BlueO,
}

\newcommand{\eexp}{\mathrm{e}} 
\newcommand{\imag}{\mathrm{i}}
\newcommand{\Heff}{\mathcal H_\mathrm{eff}}
\newcommand{\Hn}[1]{\mathcal H_{#1}}
\newcommand{\Oeff}{\mathcal O_\mathrm{eff}}
\newcommand{\On}[1]{\mathcal O_{#1}}
\newcommand{\Itot}{I_\mathrm{tot}}
\newcommand{\Urot}{U_\mathrm{rot}}
\newcommand{\gs}{\mathrm{ref}} 

\renewcommand{\vec}[1]{\boldsymbol{\mathbf{#1}}}

\definechangesauthor[name=Kai, color=blue]{KPS}
\definechangesauthor[name=Max, color=green]{MH}
\definechangesauthor[name=Andi, color=red]{AS}

\newcounter{subfigure}[figure]
\newcommand{\labelsubfig}[2]{\setcounter{subfigure}{#1}\refstepcounter{subfigure}\label{#2}}

\begin{document}
	
\title{Dynamic structure factor of the antiferromagnetic Kitaev model in large magnetic fields}
\author{Andreas Schellenberger}
\email{andreas.schellenberger@fau.de}
\author{Max H\"ormann}
\email{max.hoermann@fau.de}
\author{Kai Phillip Schmidt}
\email{kai.phillip.schmidt@fau.de}

\affiliation{Friedrich-Alexander-Universit\"at Erlangen-N\"urnberg, Department of Physics, Staudtstraße 7, 91058 Erlangen, Germany}

\begin{abstract}
We investigate the dynamic structure factor of the antiferromagnetic Kitaev honeycomb model in a magnetic field by applying perturbative continuous unitary transformations about the high-field limit. One- and two-quasiparticle properties of the dressed elementary spin-flip excitations of the high-field polarized phase are calculated which account for most of the spectral weight in the dynamic structure factor. We discuss the evolution of spectral features in these quasiparticle sectors in terms of one-quasiparticle dispersions, two-quasiparticle continua, the formation of antibound states, and quasiparticle decay. In particular, a comparably strong spectral feature above the upper edge of the upmost two-quasiparticle continuum represents three antibound states which form due to nearest-neighbor density-density interactions.
\end{abstract}

\maketitle

\section{Introduction} 

In the field of quantum many-body systems quantum spin liquids (QSLs) offer very different physical behavior than known from conventional systems, which includes long-range entangled topological order and fractionalized excitations with anyonic particle statistics \cite{Savary2016,Balents2010}. The driving mechanism for the formation of QSLs is typically frustration, referring to the balanced contribution of competing interactions \cite{Balents2010}. 
Apart from the fascinating fundamental attributes of QSLs, potential applications exploit topological properties like (non-Abelian) statistics of anyons to construct quantum memories and quantum computers \cite{Kitaev2003, Kitaev_2006, Freedman2002}. Here the topological nature is beneficial due to a strong protection from decoherence \cite{Kitaev2003,Kitaev_2006}, which is a major challenge for other quantum computing platforms \cite{Preskill1998, Castelvecchi2017,Arute2019,Friis2018}.
The experimental search for QSLs is therefore of high interest although the identification of those is very challenging \cite{Savary2016, Liu2011,Choi2012, Gradner2010, Ross2009, Ross2011, Tokiwa2016, Fak2012}. The lack of directly measurable characteristic properties brings spectroscopic experiments with inelastic neutron scattering, having a high energy and momentum resolution, in a prominent role, since rich dynamical spin-spin correlation functions can be accessed in terms of the dynamic structure factor (DSF) \cite{Savary2016, Fak2012, Qi2009, Han2012, Dodds2013,Punk2014, Morampudi2017}.

A prominent example of a microscopic model featuring different QSLs is the Kitaev honeycomb model \cite{Kitaev_2006}, which was invented by Kitaev in 2006 and can be solved exactly in terms of Majorana fermions. The ground state phase diagram contains topologically ordered gapped QSLs with Abelian anyons connected to the toric code \cite{Kitaev2003} in perturbative limits \cite{Kitaev_2006, Schmidt_2003} and a gapless QSL with algebraically decaying correlations around the case of uniform Kitaev couplings. The experimental realization of the Kitaev model is under active research in recent years with a variety of different material candidates \cite{Jackeli2009,Chaloupka_2010, Chaloupka2013, Plumb2014, Sears2015, Choi2012,Liu2011,Winter2017, Kasahara2018, Sanders2021, Wolter_2017,Trebst2022}. 
While prominent Kitaev couplings originating from strong spin-orbit coupling have already been identified, the simultaneous presence of Heisenberg interactions yields conventional long-range spin order at low temperatures \cite{Singh_2012,Liu2011,Choi2012,Ye_2012,Chaloupka2013}. Interestingly, this unwanted spin order can be melted by an external magnetic field and indications for the realization of a QSL at intermediate field values have been found \cite{Yadav2016, Janssen2016, Zheng2017,Baek_2017,Sears_2017,Leahy_2017, Yokoi2021, Sanders2021,Lin_2021,Hentrich_2018,Yu_2018, Modic_2020}. 

On the theoretical side, the Kitaev honeycomb model in a field has been investigated in several works. Along the [111]-field direction, the quantum phase diagram is qualitatively different for ferro- and antiferromagnetic Kitaev interactions \cite{Gohlke2018,Zhu2018,TrebstPhases2019,Jiang2011,Nasu2018}: Numerical investigations for ferromagnetic isotropic Kitaev interactions consistently show a direct transition between the gapped Kitaev QSL with non-Abelian topological order at finite fields and the high-field polarized phase. In contrast, for antiferromagnetic Kitaev interactions, several numerical approaches show the presence of an intermediate phase \cite{Gohlke2018,Zhu2018, Liang2018,TrebstPhases2019,Jahromi2021,Jiang_2018,Jiang2020} whose concrete nature is a field of current research. Interestingly, these considerations can be extended to varying the structure of the lattice \cite{Yao_2007,Yang_2007,Becker_2015, Hickey_2021} or to investigating spin values larger than 1/2 \cite{Jahromi2021,Zhu_2020,Trebst_2020_S1,Lee2020}. 

It is therefore an interesting question how the DSF evolves in the different quantum phases when changing the ratio between Kitaev coupling and magnetic field. Indeed, on the one hand, a quantitative understanding of the DSF is crucial for identifying the different phases and, on the other hand, the properties of excitations. This includes the formation of quasiparticles, (anti-)bound states, as well as decay processes that are important for a physical understanding of the underlying processes. In this paper we investigate the DSF of the antiferromagnetic Kitaev honeycomb model in a [111]-field within the high-field polarized phase. High-order series expansions using the pCUT method \cite{Knetter2000,Knetter2003} are used to calculate the one- and two-quasiparticle (QP) contributions to the DSF in terms of dressed spin-flip excitations. This extends recent calculations for the 1QP gap and static structure factor \cite{Jahromi2021} applying the same approach.

Apart from prominent spectral signatures of the two 1QP bands and the presence of three 2QP continuum contributions, we identify the appearance of three antibound states in the DSF above the 2QP continua, confirming a recent conjecture by \cite{Gohlke2018} based on density matrix renormalization group (DMRG) calculations. The three antibound states originate from the repulsive nearest-neighbor interactions and the reduced bandwidth for increasing Kitaev couplings. In addition, we discuss the occurrence of quasiparticle decay, which is observed most prominently in the upper 1QP band, and its connection to the convergence of the calculated high-order series expansions.  

The paper is structured as follows: In Sec.~\ref{Sec:Model} we introduce the model and its known zero-temperature properties. In Sec.~\ref{Sec:SeriesExpansion} we explain the technical aspects of the pCUT method with respect to the calculation of the 1QP and 2QP sectors and we discuss the convergence and extrapolation of the obtained series expansions. The physical results of the 1QP and 2QP contributions to the DSF are presented in Sec.~\ref{Sec:Results}. Last, we draw conclusions in Sec.~\ref{Sec:Conclusion}. 

\section{Model} 
\label{Sec:Model}

\begin{figure}[b]
    \includegraphics[width=0.35\textwidth]{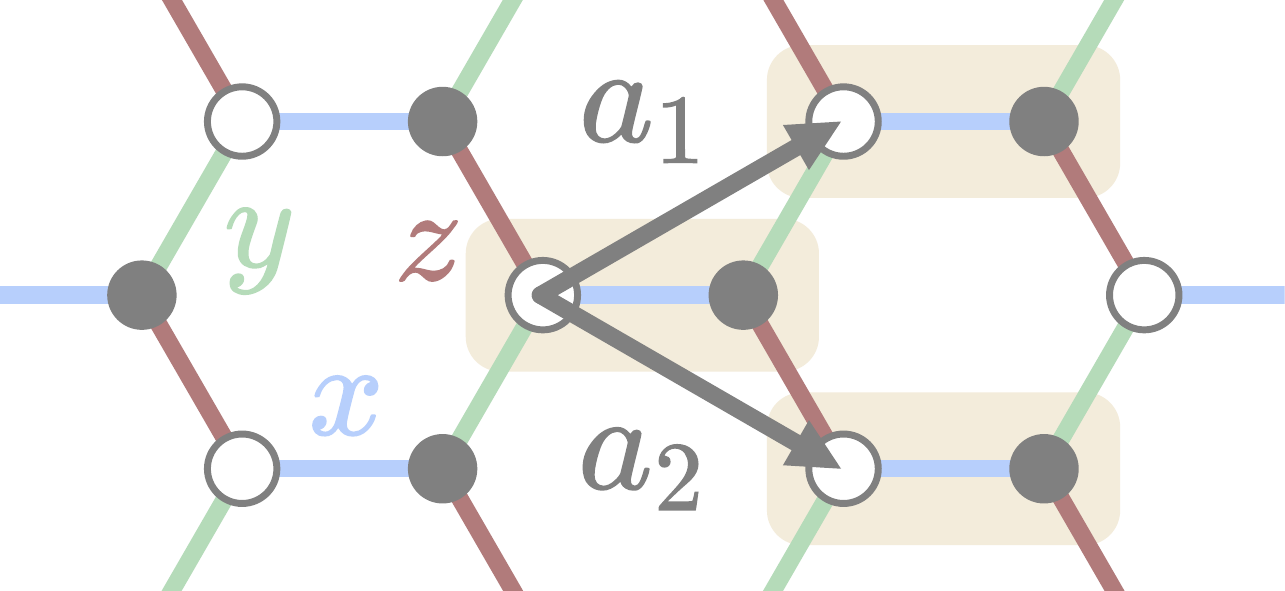}
    \caption{The honeycomb lattice with sites as gray and white circles illustrating spins 1/2 and the three bond types of the Kitaev model colored and labeled by $x,y,z$. The primitive cells are colored in light orange, with the corresponding lattice vectors $a_1,a_2$ spanning the infinite lattice.}
    \label{Fig:GridXYZ}
\end{figure}

The Hamiltonian describing the Kitaev honeycomb model in a uniform magnetic field can be written as
\begin{equation}
    H = - h\sum_{i, \alpha} \sigma^\alpha_i  + \sum_{\braket{i,j}\in\alpha} J_\alpha \sigma^\alpha_i \sigma^\alpha_j\,,
    \label{Eq:Hamiltonian}
\end{equation}
where the first term denotes the coupling of the spins to a magnetic field $h>0$ in the [111] direction due to the intrinsic magnetic moment. 
For $J_\alpha<0$ ($J_\alpha>0$) the second term describes the (anti-)ferromagnetic Kitaev coupling between neighboring spins \cite{Kitaev_2006, TrebstPhases2019,Gohlke2018}. Here, we will stick to uniform antiferromagnetic couplings $J_\alpha =: J >0$ and large magnetic fields ${J< h = 1}$. The Kitaev interaction of two neighboring spins $i,j$ depends on the orientation $\alpha \in \{x,y,z\}$ of the bond as visualized in \mbox{Fig.~\ref{Fig:GridXYZ}}. The spins are distributed on an infinite 2D honeycomb lattice, exhibiting three different bond directions denoted as $x,y,z$.

For a vanishing magnetic field $h=0$, the bare Kitaev model is exactly solvable. By transforming the spins into Majorana fermions, the elementary excitation can be modeled as a single free Majorana fermion coupled to a static $\mathbb{Z}_2$ gauge field forming a QSL \cite{Kitaev_2006}. The behavior of the phase depends on the choice of the $J_\alpha$ parameters and can be either gapped (A phase) or gapless (B phase). We force the bare model into the B phase if and only if the triangle inequalities 
\begin{align*}
    |J_x|\leq |J_y|+|J_z|\,,\; |J_y| \leq |J_x|+|J_z|\,, \; |J_z|\leq |J_x| + |J_y|
\end{align*}
are fulfilled. By having isotropic Kitaev couplings, the bare Kitaev model is therefore located in the center of the B phase \cite{Kitaev_2006}.

Adding the magnetic field breaks time-reversal symmetry and opens a gap in the B phase. This gapped phase gives rise to non-Abelian anyons offering the needed complexity of braiding rules to design a topological quantum computer \cite{Kitaev_2006}. When driving the magnetic field to large values, we end up in the polarized phase, where all spins are aligned along the field in the limit $J=0$. For moderate magnetic-field strengths the behavior of the system depends on the sign of $J$. While for ferromagnetic couplings the system displays a direct first-order phase transition between the two described phases, for the antiferromagnetic case several works found evidence for the existence of an intermediate phase. Although most of the studies propose the phase to be a gapless quantum spin liquid \cite{Gohlke2018,Zhu2018, Liang2018,TrebstPhases2019,Jahromi2021,Jiang_2018}, current research found indication against this suggested nature of the phase \cite{Jin2021} and towards a gapped spin liquid \cite{Jiang2020, Zhang2022}.

Here we focus on the polarized phase at large magnetic fields. For vanishing Kitaev interactions we obtain the ground state trivially with all spins pointing in the direction of the magnetic field. The polarized phase extends to parameter values $h=1, J < J_\mathrm{crit}$. 
The specific value of $J_\mathrm{crit}$ depends on the used method, e.g., $J\approx 1.39$ for DMRG \cite{Gohlke2018,Zhu2018}, $J\approx 1.22$ for tensor networks \cite{Jahromi2021}, and $J\approx 1.15$ for pCUT \cite{Jahromi2021}. The discrepancy between the different values is likely due to the (almost) gapless nature of the intermediate phase and the extremely flat closing of the gap (with critical exponent larger than 1) in the high-field phase. 
In the following, to simplify notation and calculations, we apply a rotation $\Urot$ on $H$ to align the magnetic field along the \mbox{$z$-axis}. In this rotated basis, we can choose the $z$-axis as our quantization axis, denoting the ground state for $J=0$ as $\ket{\mathrm{ref}}\equiv\ket{\uparrow\uparrow\cdots}$. Elementary excitations above the ground state are local spin flips at site $i$ denoted as $\ket{\downarrow_i}$. By rescaling the rotated Hamiltonian with $1/2h$, a spin flip costs an energy quanta of 1. We rewrite $H$ exactly in terms of hard-core boson annihilation (creation) operators $b_i^{(\dagger)}$ by applying the Matsubara-Matsuda transformation \cite{Matsubara1956,Guo2012}
\begin{align*}
    \sigma^x_i = b^\dagger_i + b_i\,, \quad \sigma^y_i = i (b_i-b_i^\dagger)\,, \quad \sigma^z_i = 2n_i - 1
\end{align*}
and obtain the normalized and rotated \mbox{Hamiltonian $\mathcal H$} as
\begin{align}
    \mathcal H := \frac{\Urot^\dagger H \Urot}{2h}  = E_0 + \sum_i n_i + J\sum_{\substack{\braket{i,j}\in \alpha\\m\in\mathcal M}} t_m^\alpha(i,j)\,,
    \label{Eq:RotatedHamiltonian}
\end{align}
where $E_0 = -N/2$ is the bare ground-state energy, $N$ the number of spins, and $n_i=b_i^\dagger b_i^{\phantom{\dagger}}$ the local number operator at site $i$. The last sum combines the transformed Kitaev terms. The expression $t^\alpha_m(i,j)$ contains all terms that add or remove $m$ hard-core bosons from the sites $i,j$ being connected by an $\alpha$-bond. For $\mathcal{H}$ the hard-core boson number changes by a maximum of two, written as ${\mathcal M = \{0,\pm 1,\pm 2\}}$. The concrete form of the $t^\alpha_m(i,j)$ is given by
\begin{align*}
    t_0^\alpha(i,j) &= C^z \left[-(2n_i-1)(2n_j-1) -  2(b^\dagger_i b_j + b_i b_j^\dagger)\right]\,,\\
    t_{- 1}^\alpha(i,j) &=  \sqrt{2} C^\alpha (-b_i-b_j+2b_i n_j +2b_j n_i)\,,\\
    t_{- 2}^\alpha(i,j) &=  2 \overline{C^\alpha} b_i b_j\,,
\end{align*}
with ${C^x = \overline{C^y} =  e^{i \pi/3}/6\sqrt{3}}$, ${C^z = -C^x - C^y}$, and the relation ${t^\alpha_m(i,j)=\left[t^\alpha_{-m}(i,j)\right]^\dagger}$. 
We further denote ${\ket{i_1,\dots,i_n}=b^\dagger_{i_1}\cdots b^\dagger_{i_n}\ket{\gs}}$ as an $n$QP state with hard-core bosons at the pairwise distinct sites $i_1, \dots, i_n$ and $\ket{\gs}$ as the QP-vacuum. Later on we specify the position of the QPs as ${\ket{\vec R_1,\dots,\vec R_n;\vec r_1,\dots,\vec r_n} \equiv \ket{i_1,\dots,i_n}}$, where $\vec R_j$ is the position of the primitive cell (shown as the orange region in Fig.~\ref{Fig:GridXYZ}) and $\vec r_j$ the position within the primitive cell (displayed as white/gray circles in Fig.~\ref{Fig:GridXYZ}) of site $i_j$.

\section{Series expansion}
\label{Sec:SeriesExpansion}
In order to obtain physical quantities like the dispersion or correlation functions in the polarized phase, we perform high-order series expansions about the high-field limit $J=0$. For that purpose we apply the pCUT method \cite{Knetter2000}, which yields a QP-conserving effective Hamiltonian. As a consequence, we can discuss different QP channels, corresponding to effective few-body problems, separately from each other. To enlarge the range of convergence of the obtained series expansions, we further use extrapolation techniques if applicable \cite{PadeApproximants1996,NumericalRecipies2007}. For a detailed introduction to pCUT see \cite{Knetter2000, Knetter2003_JoPA, PhdKnetter2003}.

\subsection{Transformation of the Hamiltonian and observables}
\label{Sec:pCUT}
We define the magnetic field term as the unperturbed part and the Kitaev interactions as the perturbation. 
Using pCUT, we transform $\mathcal H$ into a block-diagonal form $\Heff$, where only particle-conserving processes contribute. Therefore, the effective Hamiltonian can be written down as
\begin{align}
    \Heff = \sum_{n=0}^\infty \Hn{n}\,,
    \label{Eq:HeffAsHn}
\end{align} 
where the $\Hn{n}$ are defined as a sum of normal ordered particle-conserving $n$QP operators, meaning that $\Hn{n}$ only contributes to $m$QP states with $m\geq n$. The coefficients in $\Hn{n}$ are expressed as series expansions in $J$ \cite{Knetter2003_JoPA}.

To calculate the DSF of the system, we have to apply the same transformation onto the initial observables $\sigma^\alpha_{ r}$, acting on the $\alpha$-component of a single spin $r$ at position $\vec r$ in the lattice. 
First, we calculate $\mathcal O^\alpha(\vec r) = \Urot^\dagger \sigma^\alpha_{r} \Urot ^{\phantom{\dagger}}$, as done for $H$ in Eq.~\eqref{Eq:RotatedHamiltonian}.
Restricting ourselves to zero temperature, the effective observable $\Oeff^\alpha$ of $\mathcal O^\alpha$ only acts onto the bare vacuum $\ket{\gs}$ given as
\begin{align*}
    \Oeff^\alpha(\vec r) \ket{\gs} =&  \sum_{n=0}^\infty \On{n}^\alpha(\vec r) \ket{\gs} \,,
\end{align*}
with $\On{n}^\alpha(\vec r)$ creating $n$ quasiparticles out of the particle vacuum \cite{Knetter2003, Knetter2001}. Note that, in contrast to $\Heff$, the effective observable is not particle conserving.

\subsection{Diagonalizing quasiparticle sectors}
\label{Sec:DiagonalizingSingleQP}

To be able to determine the amplitudes of operators in $\Hn{n}, \On{n}^\alpha(\vec r)$ in the thermodynamic limit by only considering finite clusters, we exploit the linked cluster theorem \cite{Gelfand1990}. By using the cluster additivity of $\Heff, \Oeff^\alpha(\vec r)$, we can restrict ourselves to finite clusters with the size being proportional to the calculated order in perturbation, as the maximum hopping distance is limited by it \cite{Knetter2003_JoPA,Knetter2003}. For general information on the linked cluster theorem we refer to Refs.\ \cite{Knetter2003_JoPA,PhdKnetter2003,PhdCoester2015} and for the concrete application studied in this work we refer to the Appendix.

\paragraph*{One-quasiparticle sector}
To use the translational invariance of $\Hn{1}$, we Fourier transform the real-space states into
\begin{align}
    \ket {\vec k;\vec r} = \frac{1}{\sqrt N} \sum_{\vec R} \eexp^{\imag \vec k \vec R} \ket{\vec R;\vec r}\,,
    \label{Eq:State1QPKspace}
\end{align}
where $\vec k$ denotes the momentum and $\vec r$ the position of the quasiparticle in the primitive cell. The sum goes over all primitive cells in the lattice.
As the momentum is conserved by $\Heff$, we only have to consider terms of the form $\braket{\vec k, \tilde{\vec r} | \Hn{1} | \vec k,\vec r}$ varying $\tilde{\vec r},\vec r$. As the primitive cell inhibits two sites, we can separate $\Hn{1}$ into $2 \times 2$ blocks of fixed momentum $k$ and obtain the eigenenergies $\omega_1(\vec k)$ and $\omega_2(\vec k)$ as a series expansion, by diagonalizing the \mbox{$2\times 2$ matrices} order by order in $J$.

\begin{figure}
    \includegraphics[width=0.35\textwidth]{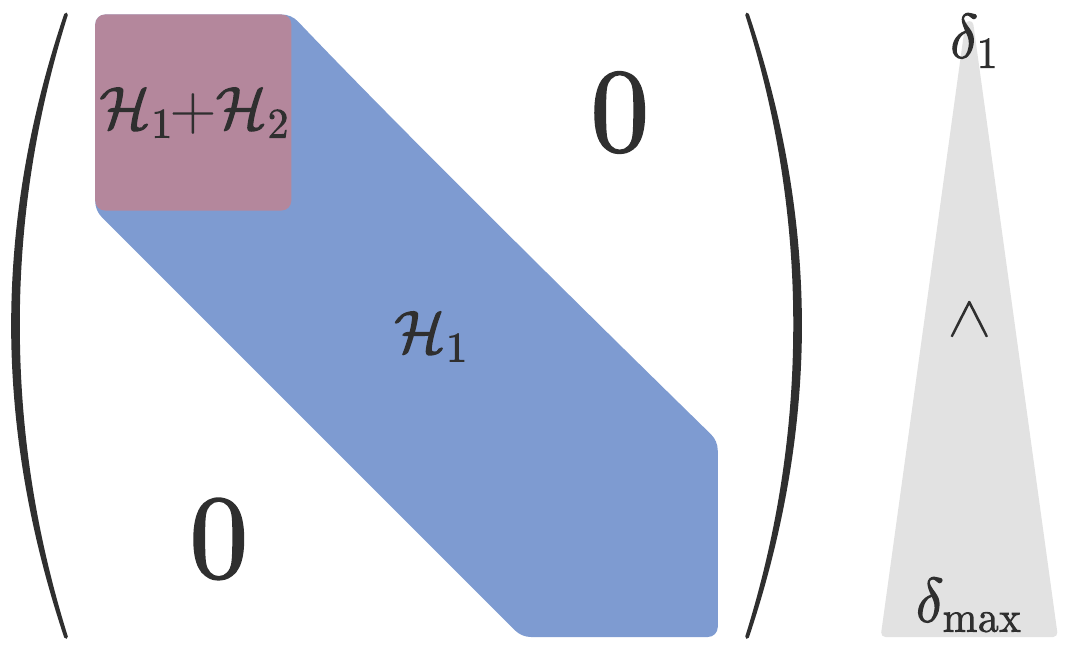}
    \caption{Visualization of the 2QP Hamiltonian matrix for the calculation of the 2QP energies for fixed total center-of-mass momentum $\vec k$. Matrix elements of $\Hn{1}$ form a band along the diagonal, while the matrix elements of the 2QP interactions $\Hn{2}$ are restricted to small distances. The basis is chosen in momentum space according to Eq.~\eqref{Eq:State2QPKspace} with increasing distances when `going down' the matrix, as indicated on the right side.}
    \label{Fig:H2H1DistanceMatrix}
\end{figure}

\paragraph*{Two quasiparticle sector}
For two QPs, we have to determine the 2QP interactions of $\Hn{2}$ and then construct the effective 2QP Hamiltonian matrix by evaluating matrix elements of $\Hn{1}+\Hn{2}$. As for the 1QP case in Eq.~\eqref{Eq:State1QPKspace}, we exploit the translational invariance by introducing the conserved center-of-mass momentum $\vec k$ and using the 2QP basis 
\begin{align}
    \ket{\vec k,\vec \delta;\vec r_1,\vec r_2} = \frac{1}{\sqrt N} \sum_{\vec R} \eexp^{\imag \vec k(\vec R+\vec \delta/2)}\ket{\vec R,\vec R+\vec \delta;\vec r_1,\vec r_2}
    \label{Eq:State2QPKspace}
\end{align}
with $\vec \delta$ being the distance between the two quasiparticles in real space \cite{PhdKnetter2003}. One is therefore left with separated subspaces of fixed $\vec k$ and varying $\vec \delta,\vec r_1,\vec r_2$.
As the distance between the QPs can get arbitrarily large, we restrict ourselves to a finite maximum absolute distance $\delta_\mathrm{max}$ to obtain matrices of finite size. The resulting 2QP Hamiltonian matrix is sketched in Fig.~\ref{Fig:H2H1DistanceMatrix} for finite order sorting the 2QP basis by growing absolute distances $|\vec \delta|\leq \delta_\mathrm{max}$ when going from top to bottom. The 1QP hopping amplitudes of $\Hn{1}$ give rise to a `band' of matrix elements along the diagonal of the matrix with a `width' depending on the perturbative order, as only one of the two QPs is affected by $\Hn{1}$. The matrix elements of $\Hn{2}$ form a block for small distances, as the 2QP interactions do depend on $\vec \delta$ and larger distances only start to contribute at higher perturbative orders. We obtain the energies of the 2QP sector for a given momentum $\vec k$ by tridiagonalizing the 2QP Hamiltonian matrix using the Lanczos algorithm \cite{SchmidtPhD,Knetter2003}.

Additionally, we can calculate the edges of the 2QP continua in the thermodynamic limit by using the free particle approximation (FPA) \cite{Feynman_lectures}. Therefore, we ignore the 2QP interaction in $\Hn{2}$ and add up the energies $\omega_i(\vec k)$ of two independent QPs as
\begin{align}
    \omega_{i_1,i_2}(\vec k_\mathrm{tot}) \in \{\omega_{i_1}(\vec k_1) + \omega_{i_2}(\vec k_2) \mid \vec k_1+\vec k_2=\vec k_\mathrm{tot}\}\,,
    \label{Eq:FPA2QPEnergies}
\end{align}
where $\omega_{i_1,i_2}(\vec k_\mathrm{tot})$ is the total energy of the 2QP state with total momentum $\vec k_\mathrm{tot}$ \cite{Adelhardt2020}. By maximizing (minimizing) Eq.~\eqref{Eq:FPA2QPEnergies} for the three combinations of ${i_1,i_2 \in \{1,2\}}$, we can calculate the maxima (minima) of the three continua for fixed $\vec k_\mathrm{tot}$.

\paragraph*{Observable}
Analogously to the Fourier transform of the states, we define the global observable in momentum space representation as
\begin{align}
    \Oeff^\alpha(\vec k)\ket{\gs} = \frac{1}{\sqrt N} \sum_{\vec r} e^{i\vec k \vec r} \Oeff^\alpha(\vec r)\ket{\gs}\,.
    \label{Eq:ObservableEffectiveT0Momentum}
\end{align}
Note that, in contrast to Eqs.~\eqref{Eq:State1QPKspace} and \eqref{Eq:State2QPKspace}, the summation runs over all sites $\vec r$ on the lattice.

We can calculate the spectral weight $\Itot^\alpha$ of $\sigma^\alpha$ reading
\begin{align}
        \Itot^\alpha &= \sum_{n=0}^\infty I_n^\alpha = \sum_{n=0}^\infty \braket{\gs| (\On{n}^\alpha)^\dagger \On{n}^\alpha | \gs} \,,
        \label{Eq:Itot}
\end{align}
where $I_n^\alpha$ is the weight of the $\alpha$-component of the $n$QP channel \cite{Knetter2003}. We can further subdivide $I_n^\alpha$ into the momentum-resolved spin-components $S^\alpha(k,\omega)$ of the DSF, depending on the momentum $\vec k$ and the energy $\omega$, as
\begin{align}
    \label{Eq:SpectralDensityEff}
    S^\alpha&(\vec k,\omega ) = \sum_{n=0}^\infty S_n^\alpha(\vec k,\omega) \\
    =& \sum_{n=0}^\infty -\frac{1}{\pi} \mathrm{Im} \braket{ \gs |  (\mathcal O_{n}^\alpha)^\dagger \frac{1}{\omega - [ \Heff-E_0]+i0^+} \mathcal O_{n}^\alpha| \gs }
    \nonumber
\end{align}
again splitting $S^\alpha(\vec k,\omega)$ into the different QP channels $S_n^\alpha(\vec k,\omega)$ \cite{SchmidtPhD}. We obtain the full DSF ${S(\vec k, \omega) = \sum_{n=0}^\infty S_n (k,\omega)}$ by adding up the three components as $S_n (k,\omega) = \sum_\alpha S^\alpha_n(\vec k,\omega)$ \cite{Gohlke2018}. For $n=1$, we use Dirac's identity to obtain delta peaks at the energy of the eigenstates in the 1QP subspace. In the 2QP sector, we evaluate $S_2(\vec k, \omega)$ by a continued fraction approximation using the Lanczos algorithm \cite{SchmidtPhD,Knetter2003}. With this approach we reach large iterations of around 200 for a maximum distance $\delta_\mathrm{max}$ of 50 sites between the two QPs.
To mimic the continuous behavior in the thermodynamic limit, we add a small Lorentz broadening to $S_2(\vec k, \omega)$, by replacing $\omega \to \omega +i\delta_\omega$ with $\delta_\omega = 0.018$ to compare with \cite{Gohlke2018,Knetter2003}. To match the discrete spectrum of $S_1$ with the continua in $S_2$, we perform the same line broadening for the delta peaks in $S_1$ for Figs.~\ref{Fig:QualityDSF} and \ref{Fig:DSF2QP}.

\subsection{Convergence and extrapolation}
\label{Sec:ConvergenceExtrapolation}

\begin{figure*}
    \includegraphics[width=\textwidth]{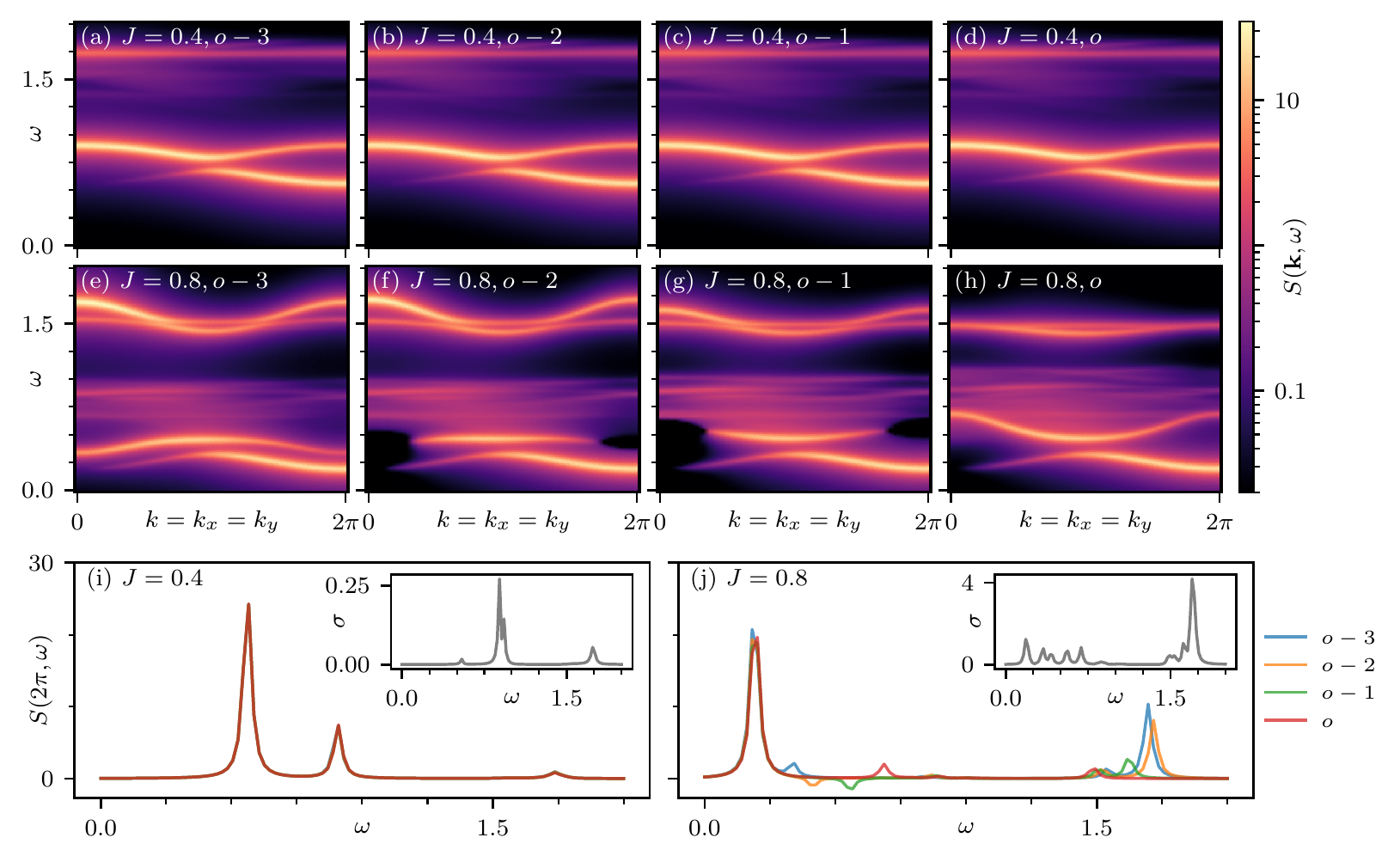}
    \labelsubfig{0}{Fig:QDSF_a}
    \labelsubfig{3}{Fig:QDSF_d}
    \labelsubfig{4}{Fig:QDSF_e}
    \labelsubfig{5}{Fig:QDSF_f}
    \labelsubfig{6}{Fig:QDSF_g}
    \labelsubfig{7}{Fig:QDSF_h}
    \labelsubfig{8}{Fig:QDSF_i}
    \labelsubfig{9}{Fig:QDSF_j}
    \caption{Dynamical structure factor $S(\vec k,\omega)$ (adding up the 1QP (2QP) contributions $S_1$ ($S_2$)) for varying couplings and bare perturbative orders. The $\delta$-peaks in 1QP and 2QP are broadened with a Lorentzian, as given in Sec.~\ref{Sec:DiagonalizingSingleQP}. The maximum orders $o$ for the dispersion and the observable are 8 (7) and 7 (6) for 1QP (2QP). We denote the reduction of all maximum orders by $d$ as $o-d$. (a)-(d), (e)-(h): Varying the maximum order $o$ for fixed $J=0.4,0.8$ along the path $k=k_x=k_y$. (i), (j): $S(\vec k,\omega)$ for $k=2\pi$ and fixed $J$-value with varying maximum order. The inset shows the standard deviation $\sigma$ for all investigated orders of the respective plot. } 
    \label{Fig:QualityDSF}
\end{figure*}

As we treat the Kitaev couplings perturbatively, we have to ensure the convergence of our series expansions within the used parameter range and maximum order. Our calculations were performed on the computer up to order 8 (7) for the 1QP (2QP) Hamiltonian and up to order 7 (6) for the 1QP (2QP) observables. For each process we used an individual minimal cluster as discussed in the Appendix for minimizing computation time and memory usage.

We investigate the convergence regarding the two couplings $J=0.4$ and $J=0.8$, by varying the maximum order. In Fig.~\ref{Fig:QualityDSF} we plot the DSF for the 1QP and 2QP subspaces with varying orders using the bare series, including two bands in 1QP around $\omega\in [0.2,1]$ and up to three antibound states in 2QP around $\omega \in [1.5, 1.75]$. We denote $o-d$ as a reduction of the respective maximum order by $d$, as introduced in the last paragraph. We leave the physical interpretation of the obtained results to the later discussion and concentrate on the changes between the different orders. 

For $J=0.4$ we obtain a very good convergence, as we can not identify quantitative changes in Figs.~\ref{Fig:QualityDSF}\ref{Fig:QDSF_a}-\ref{Fig:QDSF_d}. This is supported by Fig.~\ref{Fig:QualityDSF}\ref{Fig:QDSF_i}, as all orders overlap almost perfectly. The increased standard deviation $\sigma$ in the inline plot can be traced back to a slight deviation of the upper band's energy in 1QP.

For $J=0.8$ the quality of the convergence decreases, as can be seen in Figs.~\ref{Fig:QualityDSF}\ref{Fig:QDSF_e}-\ref{Fig:QDSF_h}. While the lower 1QP band converges well, the upper 1QP band changes its dispersion and spectral density quite drastically, as shown in Fig.~\ref{Fig:QualityDSF}\ref{Fig:QDSF_j}. For lower orders we obtain nonphysical negative spectral weights, visualized with black areas in Figs.~\ref{Fig:QualityDSF}\ref{Fig:QDSF_f} and \ref{Fig:QualityDSF}\ref{Fig:QDSF_g}. The upper antibound state moves down in energy for larger maximum orders, while the overall spectral weight decreases. In contrast, the lower antibound states change less, only moving slightly down in energy. The overall $\sigma$ lies an order higher than for $J=0.4$.

In case physical quantities are given as single series like 1QP dispersions or $n$QP spectral weights, one can improve the convergence by extrapolating the bare series results, as done in several works using pCUT \cite{lenke2021highorder,Adelhardt2020,SchmidtPhD,Jahromi2021, Hafez2010}.
Having the series expansion of a function $f$, we define a dlog-Pad\'e extrapolant as
\begin{align}
    P_\mathrm{dlog}[n,m]_f(x) := P[n,m]_{(\mathrm d/\mathrm d x)\ln f(x)}\,,
\end{align}
where 
\begin{align}
    P[n,m]_g(x):= \frac{P_n(x)}{Q_m(x)} = \frac{\sum_{k=0}^n p_kx^k}{\sum_{k=0}^m q_kx^k}
    \label{Eq:Pade}
\end{align}
is the Pad\'e extrapolant of an analytic function $g$ \cite{PadeApproximants1996,NumericalRecipies2007}. The coefficients $p_k,q_k$ are fixed by demanding the series expansion of $P[n,m]_g$ and $g$ to be the same up to order $n+m$. We can approximate $f$ by integrating and exponentiating $P_\mathrm{dlog}[n,m]_f$ \cite{lenke2021highorder}. This was done for the 1QP states in Fig.~\ref{Fig:DSF2QP}. Comparing Fig.~\ref{Fig:QualityDSF}\ref{Fig:QDSF_h} with Fig.~\ref{Fig:DSF2QP}\ref{Fig:DSF2QP_d}, we obtain a better convergence for the upper band, as it behaves as expected with respect to smaller perturbations.

Apart from calculating series only to finite perturbative order, quasiparticle decay can affect the convergence and extrapolation of the series. Indeed, within the pCUT method we assume that the QP sectors can be fully separated, i.e., states with different particle number are separated in energy. This is, however, in general only true in the perturbative limit but does not have to be the case for finite $J$. Here, as the upper 1QP band starts to overlap with the lowest 2QP continuum for $J\geq J_\mathrm{decay}\approx 0.57$ (as can be calculated using the FPA), the quasiparticles of the upper band decay as the lifetime becomes finite \cite{Verresen_2019}. We note that generalizations of generators for nonperturbative CUTs, which enable one to treat quasiparticle decay, are discussed in \cite{Fischer2010}.

We find further evidence of quasiparticle decay by the occurrence of poles around $J_\mathrm{decay}$ when calculating dlog-Pad\'e extrapolants of physical quantities. These poles can be extracted by investigating the roots of $Q_m$ \footnote{We sort out defective poles using the approach of Adelhardt et al.~\cite{Adelhardt2020}. Therefore we vary $n,m$ for $n+m\leq k-1$ and group the resulting extrapolants in families of constant $n-m$. If a family inhibits less than two members after removing the extrapolants with nonphysical poles, we discard the whole family. For approximating $f$, we take the mean of the highest order extrapolant of all families \cite{Adelhardt2020} }. 
Exemplarily, we plot the real part of the present poles $J_\mathrm{pole}$ for the spectral weights in Fig.~\ref{Fig:SpectralWeight} gathering around $J_\mathrm{decay}$. We explain the lacking occurrence of poles in $I_2^z$ with the little number of nondefective extrapolants for the calculated order. The only pole at $J\approx 0.33$ matches quite well with the starting overlap of 2QP and 3QP energies. The same analysis was done for the energy series of the 1QP bands, featuring a number of poles around $J_\mathrm{decay}$ in the upper band (not shown). The lower band shows no poles as it does not intersect with the 2QP states for the investigated $J$-values. In both cases, the standard deviation of the dlog-Pad\'es rises noticeably in the area of overlapping QP channels. 

We conclude that the calculated series show very good convergence for $J=0.4$, while our results for $J=0.8$ can only be trusted on a qualitative level, being explained by the use of bare perturbative series and the presence of quasiparticle decay. To increase quality, we use dlog-Pad\'e extrapolants for the 1QP sector, as can be seen in the difference between Fig.~\ref{Fig:QualityDSF}\ref{Fig:QDSF_h} and Fig.~\ref{Fig:DSF2QP}\ref{Fig:DSF2QP_d}. As we are not able to extract series expansions out of the 2QP Hamiltonian matrix in 2QP calculations, we stick to the bare series results in this sector.

\section{Results}
\label{Sec:Results}
In this section we present our results for the dispersion and the spectral quantities of the one- and two-quasiparticle sectors.
We start by analyzing the spectral weight of the observables $\sigma^\alpha$ to ensure the validity of our approach to consider solely the 1QP and 2QP dynamics. After that we discuss the dispersion and the formation and nature of antibound states. We conclude with a discussion of the DSF, comparing our results to DMRG data \cite{Gohlke2018}.

For the plots, we parametrize the momentum $\vec k$, which is used for the 1QP and 2QP states, with respect to the dual basis $\{\vec b_1,\vec b_2\}$ of the real-space basis $\{\vec a_1,\vec a_2\}$ as plotted in Fig.~\ref{Fig:GridXYZ}:
\begin{align*}
    \vec k = k_1 \vec b_1 + k_2 \vec b_2 \quad \text{with } \vec b_i \vec a_j = \delta_{ij} \,.
\end{align*}
Therefore, we can scan through all possible phases between the primitive cells by varying $k_1,k_2$ between $0$ and $2\pi$. 

\subsection{Spectral weights}
\label{Sec:SpectralWeight}

\begin{figure}
    \includegraphics[width=\columnwidth]{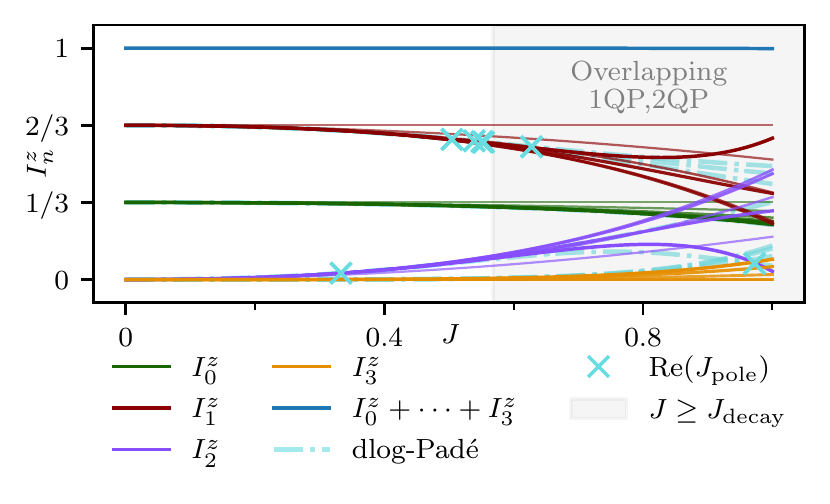}
    \caption{Spectral weights $I_n^z$ for varying $J>0$ up to order 7 (6) for $n\leq 2$ ($n=3$). The weights are decomposed into the $n$QP channels $I_n^z$ and plotted in different colors. The sum of all calculated weights is plotted in blue and is close to $I_\mathrm{tot}^z$, as discussed in the main text. The dlog-Pad\'e extrapolants are plotted in dotted lines as an indicator of quality. The real part of the poles $J_\mathrm{pole}$ of the plotted extrapolants are shown as crosses on the corresponding extrapolant. The coupling regime of intersecting 1QP and 2QP states starting at $J_\mathrm{decay}$ is colored in light gray.} 
    \label{Fig:SpectralWeight}
\end{figure}

First, we take a look at the spectral weights of $\sigma^\alpha$, as introduced in Eq.~\eqref{Eq:Itot}. For the chosen uniform \mbox{coupling $J$}, we obtain the same weights for all $\alpha$. Therefore, we stick to $\sigma^z$ and plot the weights $I^z_n$ of the lower $n$QP channels up to order 7 (order 6) for $n\leq 2$ ($n=3$) in Fig.~\ref{Fig:SpectralWeight}. The different $I_n^z$'s are plotted in various colors, also including lower orders and dlog-Pad\'e extrapolants to check validity. Using Eq.~\eqref{Eq:Itot} we expect the sum of all calculated $I_n^z$ being close to $I_\mathrm{tot}^z = 1$, as the $n$QP channel starts contributing only at $2(n-1)^{\mathrm{th}}$ order. When calculating the spectral weight of the higher QP channels for $J=0.4,0.8$ we obtain
\begin{align*}
    \sum_{i=3}^\infty |I_i^z(J=0.4)| \approx 0.001 \,,\quad  \sum_{i=3}^\infty |I_i^z(J=0.8)| \approx 0.03\,.
\end{align*}
This finding ensures that most of the intensity lies in the 1QP and 2QP channels for the investigated values of $J$ up to order 7, making our approach neglecting higher QP contributions feasible.
Under increasing Kitaev couplings, we obtain a shift of intensity mainly from the 1QP to the 2QP sector, while $I_0^z$ only shows a minor downwards-trend of similar slope as the upwards trend for $I_3^z$. We find a notable $I_2^z$ contribution in the investigated range, making a discussion of the 2QP subspace relevant. 

The quality of the shown dlog-Pad\'e extrapolants decreases in the region of spectral overlap of the 1QP and 2QP sectors starting at $J_\mathrm{decay}\approx 0.57$, due to the presence of quasiparticle decay, as discussed in Sec.~\ref{Sec:ConvergenceExtrapolation}. For the extrapolants of $I_1^z$, we indeed obtain a number of poles right before the intersection. 
Therefore, the weights $\sum_{i=3}^\infty I_i^z$ may differ noticeably within the region of overlapping sectors, as indicated by the large standard deviation of the extrapolants (not shown). 

\subsection{Excitation energies}
In the following we present the 1QP and 2QP energies, separately. While we can calculate the two 1QP bands in the thermodynamic limit and extract series expansions, we have to stick to finite systems for the 2QP discussion and insert values for $J$ in the bare series. We further calculate the edges of the 2QP continua in the thermodynamic limit using the FPA. 

\subsubsection{One-quasiparticle dispersion}
\begin{figure}
    \includegraphics[width=\columnwidth]{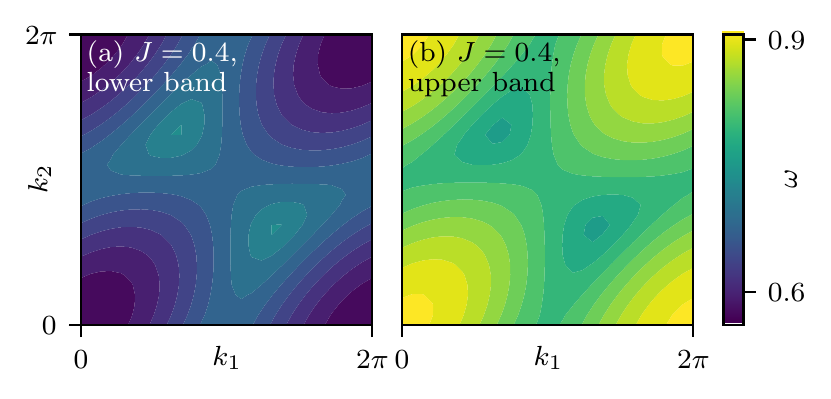}
    \caption{(a), (b): Dispersions $\omega_1(\vec k), \omega_2(\vec k)$ of the lower and upper 1QP band for $h=1, J=0.4$. The global minimum and maximum lies at $\vec k=0$. All energies are lowered by the Kitaev interaction due to the antiferromagnetic coupling.} 
    \label{Fig:1QPDispersionGap}
\end{figure}

Fig.~\ref{Fig:1QPDispersionGap} shows the dispersion of the 1QP sector for $J=0.4$. Both bands are symmetric to each other with a maximum band splitting for vanishing momenta. We observe a sixfold rotation symmetry, originating from the structure of the lattice and the choice of the parameters. The symmetry breaks when choosing nonuniform fields or couplings (not shown). We observe a strict trend towards smaller energies for both bands for increasing perturbation, as we start from an excitation energy of $1$ for $J=0$. By subclassifying the contributing processes in $\Hn{1}$, we can show that the negative contribution to the local hopping term $\braket{i|\Hn{1}|i}$, resulting in the strict downwards trend, comes from the antiferromagnetic coupling between neighboring spins in Eq.~\eqref{Eq:Hamiltonian}. We can reuse this finding for the discussion of the antibound states in the 2QP sector later on.

\subsubsection{Two-quasiparticle energies}
\begin{figure}
    \includegraphics[width=\columnwidth]{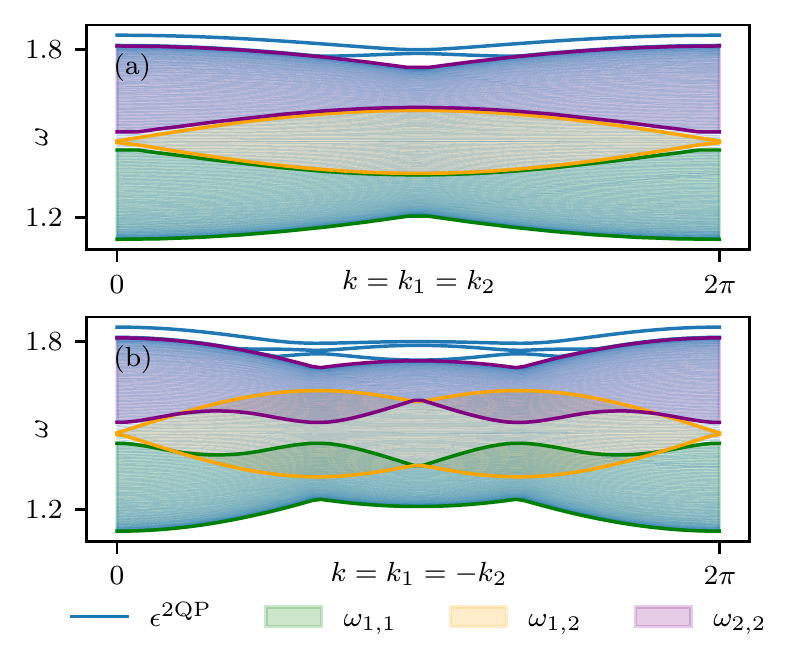}
    \labelsubfig{0}{Fig:Disp2QP_a}
    \labelsubfig{1}{Fig:Disp2QP_b}
    \caption{2QP spectrum for $J=0.4$. The eigenvalues $\epsilon^\mathrm{2QP}$ of the tridiagonalized 2QP Hamiltonian matrix (see Fig.~\ref{Fig:H2H1DistanceMatrix}) are plotted in blue. Using the FPA in Eq.~\eqref{Eq:FPA2QPEnergies}, the three continua are plotted in different colors. For (a) the 2QP spectrum is plotted along $k_1=k_2$, and for (b) along $k_1=-k_2$.} 
    \label{Fig:Dispersion2QP}
\end{figure}

Turning to the 2QP energies, the spectrum for ${J=0.4}$ is plotted in Fig.~\ref{Fig:Dispersion2QP} along two different paths in \mbox{$k$-space}. We obtain three continua, which form out of the two 1QP bands. While the three continua overlap in Fig.~\ref{Fig:Dispersion2QP}\ref{Fig:Disp2QP_b}, the dispersion in Fig.~\ref{Fig:Dispersion2QP}\ref{Fig:Disp2QP_a} features a small gap between the continua that increases for larger couplings (not shown). This originates from the structure of the 1QP bands, as for the path in Fig.~\ref{Fig:Dispersion2QP}\ref{Fig:Disp2QP_b} the two 1QP bands are closer to each other (see Fig.~\ref{Fig:1QPDispersionGap}). We find a good agreement of the energies obtained by the FPA and the 2QP matrix. When extrapolating $\delta_\mathrm{max} \to \infty$, the band edges converge to those of the FPA (not shown).

Above the highest 2QP continuum we obtain up to three antibound states. These start to build around the edges of the first Brillouin zone ($k\approx \pi$ in Fig.~\ref{Fig:Dispersion2QP}) and emerge fully out of the continuum for rising couplings.

\begin{figure*}
    \includegraphics[width=\textwidth]{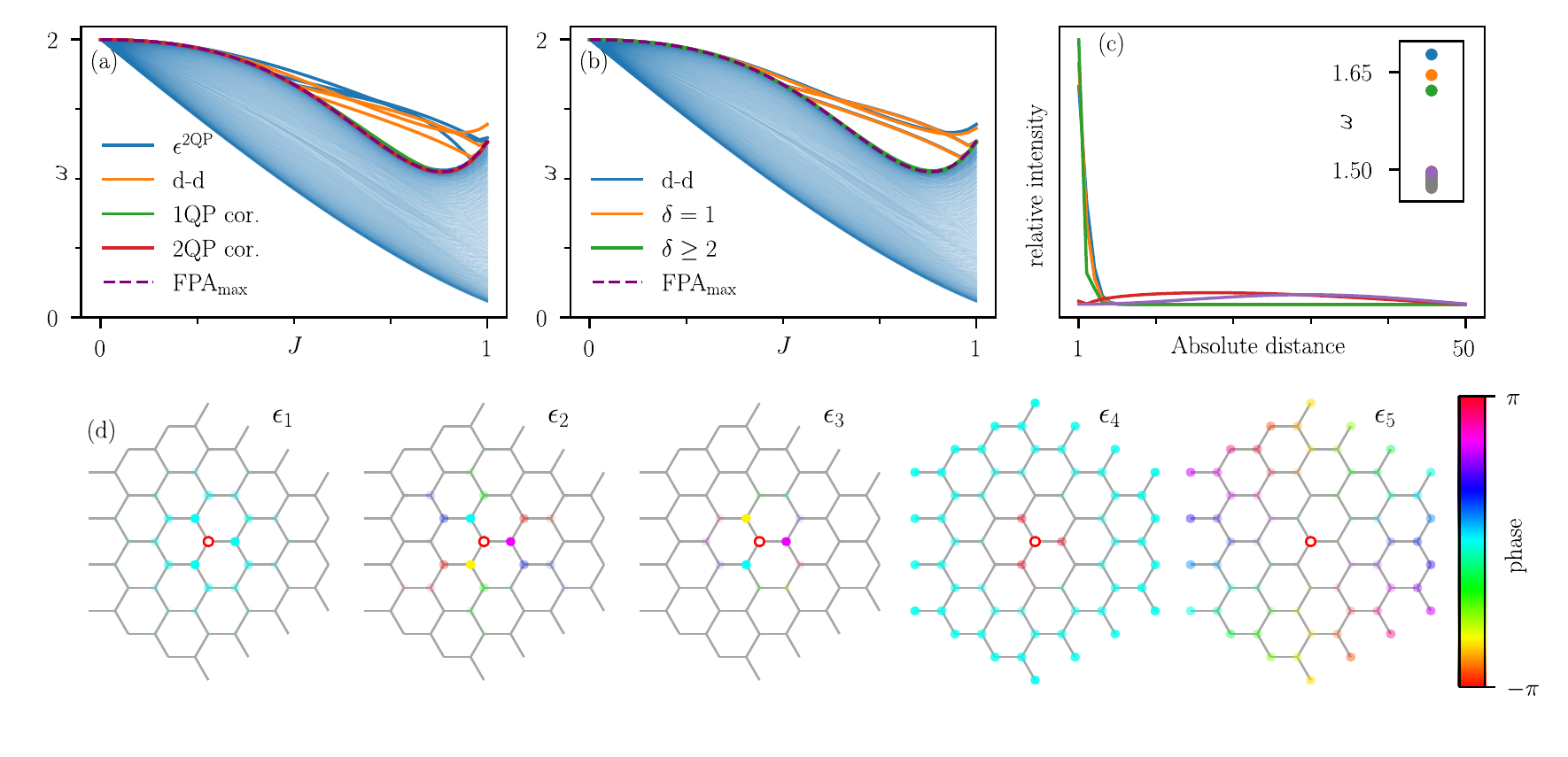}
    \labelsubfig{0}{Fig:H2Inter_a}
    \labelsubfig{1}{Fig:H2Inter_b}
    \labelsubfig{2}{Fig:H2Inter_c}
    \labelsubfig{3}{Fig:H2Inter_d}
    \caption{Classification of the antibound states for $J>0$ at $k=0$. (a): Comparison of the different types of correlated processes, as defined in the main text with `d-d' being the shortcut for density-density. Additionally, the energies $\epsilon^\mathrm{2QP}$ of the full model and the upper edge of the topmost continuum using the FPA are plotted. (b): Subgroups of the density-density interaction, only considering states with neighbored QPs ($\delta =1$) or non-neighbored QPs ($\delta \geq 2$). (c), (d): Eigenvectors of the five highest eigenenergies of the full Hamiltonian for $J=0.6$. In (c) the vectors are decomposed by the absolute distance of the two QPs in the respective state. The energies of the states are plotted in the inset. In (d) the distance between the QPs is defined between the red-circled site and the corresponding colored site. The intensity is given by the opacity and the relative phase by the color code. } 
    \label{Fig:H2Interaction}
\end{figure*}

To find the driving mechanism that leads to the formation of the antibound states, we divide the $\Hn{2}$ processes in three subgroups. Having the general process $\braket{\tilde i, \tilde j | \Hn{2} | i,j}$, we define `density-density interactions' as the diagonal process $\tilde i = i, \tilde j = j$, and `$n$QP correlated hoppings' as processes where $n$ QPs change their position. In Fig.~\ref{Fig:H2Interaction}\ref{Fig:H2Inter_a} we compare the different subgroups for fixed momentum $k=0$ and varying $J$. Therefore, we plot the three highest eigenvalues when only considering the effect of one of the subgroups. For comparison, the antibound states and the upper edge of the upmost continuum for the full Hamiltonian are plotted, using the FPA for the latter one.

Clearly, the density-density interaction is the driving force for the formation of the antibound states, as we obtain qualitatively the same antibound states when neglecting correlated hopping processes. The rising of the continuum's upper edge at $J\approx 1$ is a sign of divergence, as it is strongly dependent on the maximum order (not shown). 
To further classify the density-density interaction, we subdivide it into processes involving states of neighboring particles $(\delta=1)$ and non-neighboring ones ($\delta\geq 2$) in Fig.~\ref{Fig:H2Interaction}\ref{Fig:H2Inter_b}. As only neighboring particles seem to form the antibound state, we plot the amplitudes of the eigenvectors of the antibound states, decomposed into the (absolute) distance $\delta$ between the two particles (being the minimum number of sites between the QPs) in Figs.~\ref{Fig:H2Interaction}\ref{Fig:H2Inter_c} and \ref{Fig:H2Interaction}\ref{Fig:H2Inter_d} for $J=0.6$. In both plots we obtain a clear difference between the states within the continuum, spreading over all distances, and the three antibound states, which form different superpositions of the three nearest-neighbor states, only differing by the relative phase. The intensities peak at $\delta = 1$ and decay exponentially for larger distances. Therefore, we conclude that the number of three antibound states can be traced back to the number of three nearest-neighbor states in the chosen basis in Eq.~\eqref{Eq:State2QPKspace}. Note that the finite intensity for the states in the continua, as shown in Figs.~\ref{Fig:H2Interaction}\ref{Fig:H2Inter_c}--\ref{Fig:H2Inter_d}, is a finite-size effect and vanishes for $\delta_\mathrm{max}\to \infty$.

The strong anti-binding effect can be explained by the antiferromagnetic interaction of neighboring spins, as discussed in the 1QP section. The flipped spin of the quasiparticle interacts attractively with all neighboring spins being aligned antiparallel, whereas another neighboring quasiparticle with parallel spin is nonfavorable in energy. In contrast to six attractive neighbors for two non-neighboring particles, the number reduces to four when having two particles next to each other. As this effect does not depend on the direction for uniform parameters, we can explain the appearance of at most three antibound states.

All states in Fig.~\ref{Fig:H2Interaction}\ref{Fig:H2Inter_d} feature a uniform contribution of the distances as well as a uniformly changing relative phase, when moving around the red-circled origin. This is due to the high-symmetry point $\vec k=0$ and is broken when choosing $\vec k\neq 0$ (not shown).
By decomposing $\Hn{1},\Hn{2}$ into the different perturbative orders, we find the subleading orders being crucial for the formation of the antibound states. Restricting the Hamiltonian to order one, we do not observe any antibound state at ${\vec k=0}$, even for very high couplings $J\approx 1$, as the maximum energy converges to the edge of the upper continuum for $\delta_\mathrm{max} \to \infty$. This is due to the large leading order nearest-neighbor QP-hoppings which suppress the formation of antibound states. For higher orders, the density-density interaction increases monotonously while the nearest-neighbor hopping amplitude decreases for intermediate $J$, resulting in the emergence of antibound states for intermediate couplings. 

\subsection{Dynamic structure factor}

Next, we investigate the DSF of the 1QP and 2QP sectors, which are directly relevant for inelastic neutron-scattering experiments. The high relative spectral weight found in Sec.~\ref{Sec:SpectralWeight} ensures that we fetch the actual physical behavior quite well. We validate our findings further by comparing our results with the DSF calculated by Gohlke et al.\ using DMRG \cite{Gohlke2018}.

\subsubsection{One-quasiparticle dynamic structure factor}
\begin{figure}
    \includegraphics[width=\columnwidth]{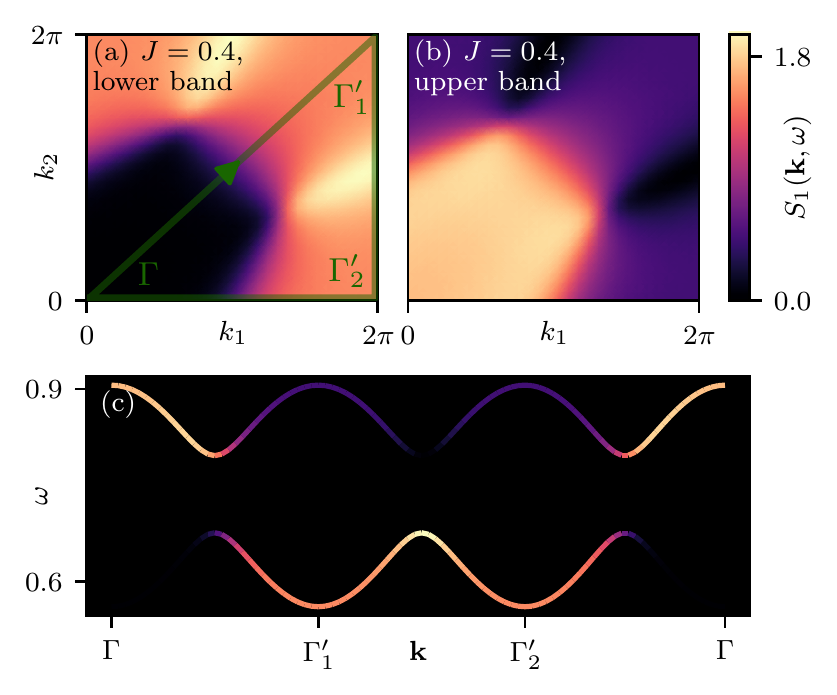}
    \labelsubfig{2}{Fig:DSF1QP_c}
    \caption{1QP DSF $S_1(\vec k,\omega)$ for $J=0.4$. (a), (b): Intensities $S_1(\vec k,\omega=\omega_{i}(\vec k))$ for the lower and upper band for varying momenta (the same as in Fig.~\ref{Fig:1QPDispersionGap}). (c): $S_1(\vec k,\omega)$ along the $k$-path illustrated in green in (a) for the same set of parameters.} 
    \label{Fig:DSF1QP}
\end{figure}

Starting with the 1QP sector, we investigate the DSF $S_1(\vec k,\omega)$ in Eq.~\eqref{Eq:SpectralDensityEff}, summing up the three spin components $S_1^\alpha$. Fig.~\ref{Fig:DSF1QP} shows the DSF $S_1$ for $J=0.4$. The two bands share a rather uneven distribution of intensity. For $k=0$, the lower band's intensity vanishes exactly. This originates from the different parity of the observable and the lower eigenstate at $k=0$, resulting in a vanishing scalar product. Following the path in Fig.~\ref{Fig:DSF1QP}\ref{Fig:DSF1QP_c}, we obtain an intensity shift from the upper to the lower band. These findings match qualitatively with those of Gohlke et al. \cite{Gohlke2018}. The sum of the two bands is close to a constant function for small \mbox{$J$-values}, while for higher couplings we observe a stronger $k$-dependency, coming from the induced hoppings. 

\subsubsection{Full dynamic structure factor}
\begin{figure*}
    \includegraphics[width=\textwidth]{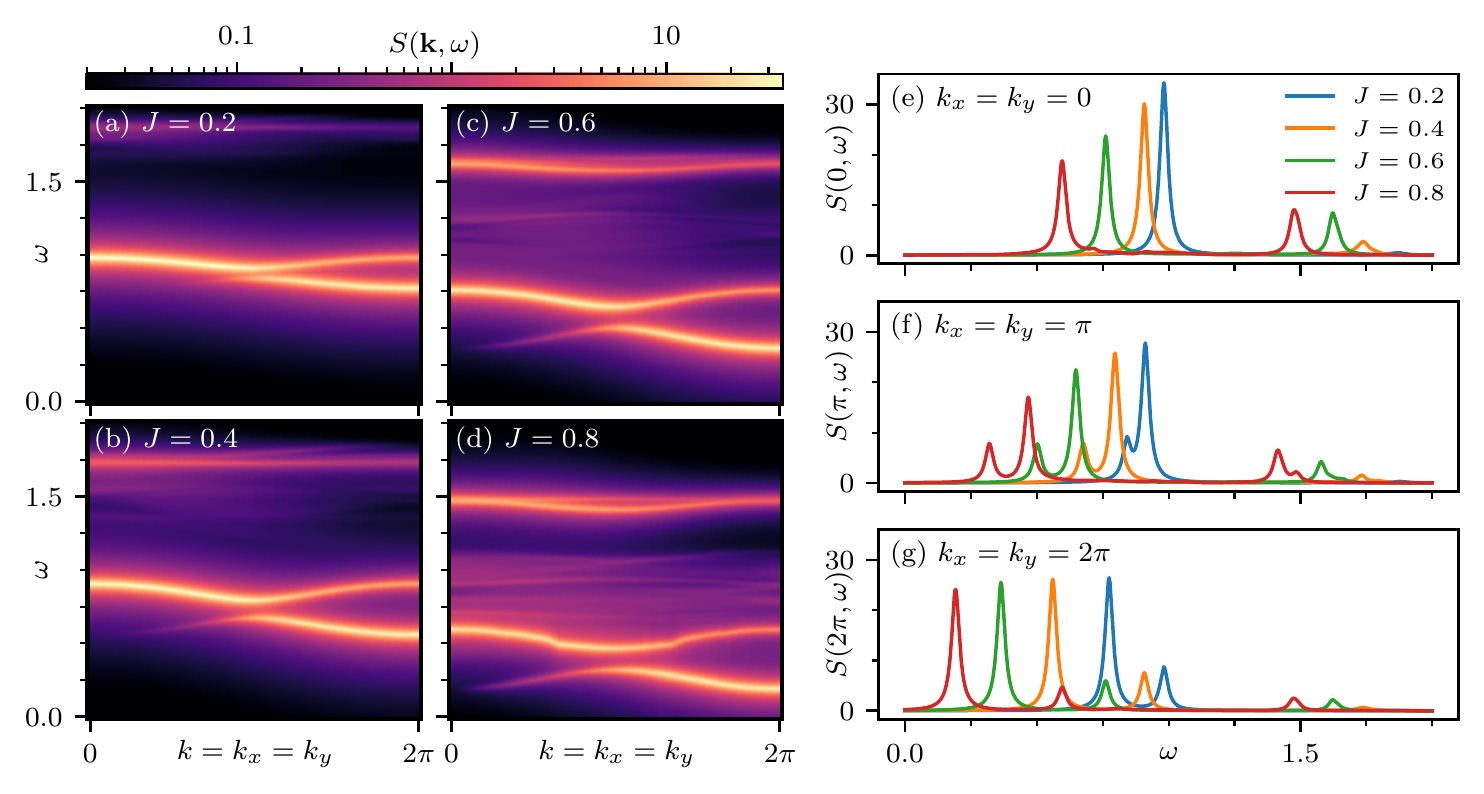}
    \labelsubfig{0}{Fig:DSF2QP_a}
    \labelsubfig{1}{Fig:DSF2QP_b}
    \labelsubfig{2}{Fig:DSF2QP_c}
    \labelsubfig{3}{Fig:DSF2QP_d}
    \labelsubfig{4}{Fig:DSF2QP_e}
    \labelsubfig{5}{Fig:DSF2QP_f}
    \labelsubfig{6}{Fig:DSF2QP_g}
    \caption{DSF $S(\vec k,\omega)$ (adding up $S_1$ and $S_2$) for varying $J$ values. The $\delta$-peaks in 1QP and 2QP are broadened with a Lorentzian, as given in in Sec.~\ref{Sec:DiagonalizingSingleQP}. (a)--(d): $S(\vec k,\omega)$ along the path $k_x=k_y$ for varying Kitaev couplings. (e)--(g): $S(\vec k,\omega)$ for fixed $\vec k$-value and varying Kitaev couplings.} 
    \label{Fig:DSF2QP}
\end{figure*}

In Fig.~\ref{Fig:DSF2QP} we added up the 1QP and 2QP contributions $S_1(\vec k,\omega)$ and $S_2(\vec k,\omega)$ to discuss the full DSF. While $S_1(\vec k,\omega)$ has been extrapolated using dlog-Pad\'e extrapolation, the bare series are used for $S_2(\vec k,\omega)$.
 
Starting with small interactions in Fig.~\ref{Fig:DSF2QP}\ref{Fig:DSF2QP_a}, we observe two bands from the 1QP subspace with large intensity and a small feature at the upper end of the upmost 2QP continuum. This matches with our discussion around Fig.~\ref{Fig:SpectralWeight}, as $I_2$ is small for small Kitaev couplings. When moving to larger $J$-values in Figs.~\ref{Fig:DSF2QP}\ref{Fig:DSF2QP_b}--\ref{Fig:DSF2QP_d}, the two 1QP bands move down in energy, as well as the feature at the upper end of the 2QP continuum. Meanwhile, the gap between the two modes in 1QP grows slightly, coming from the increased kinetic energy of the QPs due to the Kitaev interactions. When increasing $J$ further, the two modes again move closer to each other when approaching the quantum phase transition (see \cite{Gohlke2018}). Because of that also the 2QP continua grow only slightly in energy width, as can be explained by the FPA in Eq.~\eqref{Eq:FPA2QPEnergies}.

While the overall intensity of the 2QP channel grows, most of the weight moves to the emerging antibound states. For $J \gtrapprox 0.6$, we obtain a prominent peak in intensity for the antibound states, as can be seen more clearly in Figs.~\ref{Fig:DSF2QP}\ref{Fig:DSF2QP_e}--\ref{Fig:DSF2QP_g} with a linear scaling. With the chosen line broadening $\delta_\omega$ we can resolve up to two peaks at maximum, as can be seen in Fig.~\ref{Fig:DSF2QP}\ref{Fig:DSF2QP_f}. The dispersion of the three antibound states is almost flat in momentum space, only showing little dispersive behavior around $k_1=k_2=\pi$. The antibound states are only fully emerged for the larger couplings in Figs.~\ref{Fig:DSF2QP}\ref{Fig:DSF2QP_c} and \ref{Fig:DSF2QP}\ref{Fig:DSF2QP_d}, as can be seen in Fig.~\ref{Fig:H2Interaction}\ref{Fig:DSF2QP_a}, too. The spectral intensity in the continua rises for increasing $J$-values but stays an order of magnitude lower than the intensity of the antibound states.

The quality of the data decreases when investigating larger $J$-values. Starting at the spectral intersection of the upper 1QP band with the lowest 2QP continuum, we observe a rise of the standard deviation of the band when using dlog-Pad\'e extrapolation, as can be seen by the artificial kinks in the upper 1QP band in Fig.~\ref{Fig:DSF2QP}\ref{Fig:DSF2QP_d}. Nonetheless, we are able to qualitatively match with the DMRG results of Gohlke et al.\ in Fig.~7(a) \cite{Gohlke2018} corresponding to a coupling of $J\approx 0.86$ in our units \footnote{In discussion with Gohlke an error of $1/(4\pi)$ in the normalization of Fig.~7(a) in \cite{Gohlke2018} was found. Adding this factor to our data, we obtain reasonable matching results.}. We obtain the 1QP bands at the same energies and intensities, especially for the lower band with good precision. Due to the matching of energy and intensity, we can identify the sharp spectral feature in the DSF discussed in Ref.~\cite{Gohlke2018} as 2QP antibound states. We can furthermore confirm that the driving force for the antibound states is the repulsive nearest-neighbor density-density interaction. Looking at DMRG data for larger couplings, we observe the antibound states staying a noticeable feature in the DSF before smearing out in energy in the vicinity of the quantum phase transition. 

\section{Conclusion}
\label{Sec:Conclusion}

We applied the pCUT method to the high-field polarized phase of the antiferromagnetic Kitaev honeycomb model in a [111]-field and calculated the 1QP and 2QP contributions to the DSF. The 2QP sector contains three continua, originating from different combinations of the two 1QP bands and, for intermediate antiferromagnetic Kitaev couplings, we found three 2QP antibound states. By investigating different processes of the 2QP interaction and the structure of the corresponding eigenvectors, we found the driving effect for the formation of the antibound states being the nearest-neighbor density-density interactions, starting to emerge out of the upmost continuum in second order in perturbation. We explained this effect by the antiferromagnetic coupling between neighboring spins in the Kitaev model, therefore connecting the number of antibound states to the number of nearest-neighbor configurations. By calculating the spectral weights in the $n$QP sector for $n\leq 3$, we further showed that the spectral contribution of the 2QP sector is non-negligible. In particular, for increasing Kitaev couplings, we found the dominant 2QP spectral intensity for the three antibound states being well separated in energy from the 2QP continua. Our results are therefore fully consistent with DMRG calculations by Gohlke et al.~\cite{Gohlke2018} confirming their conjecture about the existence of antibound states.

We further observed rising standard deviations when extrapolating the upper 1QP band to intermediate couplings within the polarized phase. This can be traced back physically to quasiparticle decay of the upper 1QP excitations into the lowest 2QP continuum which is beyond our perturbative pCUT ansatz. An extension of our calculation to nonperturbative CUTs treating the inherently nonperturbative quasiparticle decay would therefore be desirable. A potential route would be the use of adjusted quasiparticle generators as introduced in \cite{Fischer2010}, which does not seek for a full block diagonalization of the effective Hamiltonian.

In addition, there are several interesting aspects which would deserve further investigations in the future. Clearly, one could consider a general direction of the magnetic field, nonuniform Kitaev couplings, or larger spin values. Furthermore, the theoretical description of inelastic light scattering like Raman scattering, resonant inelastic X-ray scattering, or infrared absorption would be of interest, see, e.g., the works by Glamazda et al.~\cite{Glamazda2016} and Sandilands et al.~\cite{Sandilands2015}. 
Finally, the addition of a Heisenberg interaction, as done in \cite{Perreault2015,Knolle2014,Pan2021,Yamamoto2020,Chaloupka2013,Jiang2011}, would be highly relevant for a more realistic description of existing Kitaev materials \cite{Kitagawa2018}.

\section*{Acknowledgments}

We thank Matthias Gohlke for sharing the DMRG data and for fruitful discussions. A.S.\ thanks Matthias M{\"u}hlhauser for suggesting the idea for the cluster generation. K.P.S.\ acknowledges financial support by the German Science Foundation (DFG) through the grant SCHM 2511/11-1. 

\appendix*
\section{Cluster creation}
For calculating the effective processes using pCUT, we define clusters to calculate the processes in the thermodynamic limit without finite-size effects, while keeping the size as small as possible to reduce computation time and memory usage.

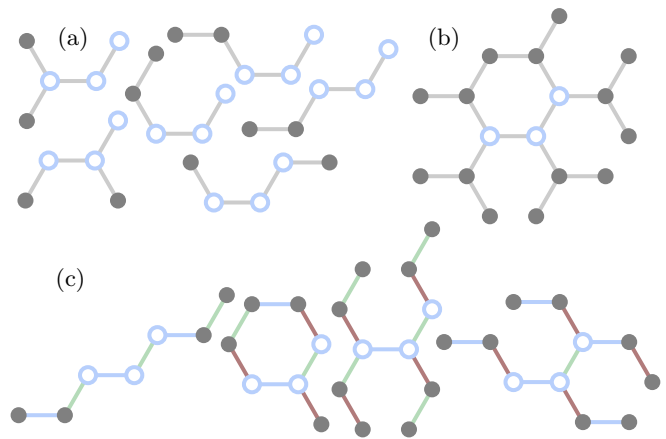
\begin{figure}
    \def\svgwidth{3.4in}
\begingroup%
  \makeatletter%
  \providecommand\color[2][]{%
    \errmessage{(Inkscape) Color is used for the text in Inkscape, but the package 'color.sty' is not loaded}%
    \renewcommand\color[2][]{}%
  }%
  \providecommand\transparent[1]{%
    \errmessage{(Inkscape) Transparency is used (non-zero) for the text in Inkscape, but the package 'transparent.sty' is not loaded}%
    \renewcommand\transparent[1]{}%
  }%
  \providecommand\rotatebox[2]{#2}%
  \newcommand*\fsize{\dimexpr\f@size pt\relax}%
  \newcommand*\lineheight[1]{\fontsize{\fsize}{#1\fsize}\selectfont}%
  \ifx\svgwidth\undefined%
    \setlength{\unitlength}{608.09037781bp}%
    \ifx\svgscale\undefined%
      \relax%
    \else%
      \setlength{\unitlength}{\unitlength * \real{\svgscale}}%
    \fi%
  \else%
    \setlength{\unitlength}{\svgwidth}%
  \fi%
  \global\let\svgwidth\undefined%
  \global\let\svgscale\undefined%
  \makeatother%
  \begin{picture}(1,0.6612795)%
    \lineheight{1}%
    \setlength\tabcolsep{0pt}%
    \put(0,0){\includegraphics[width=\unitlength,page=1]{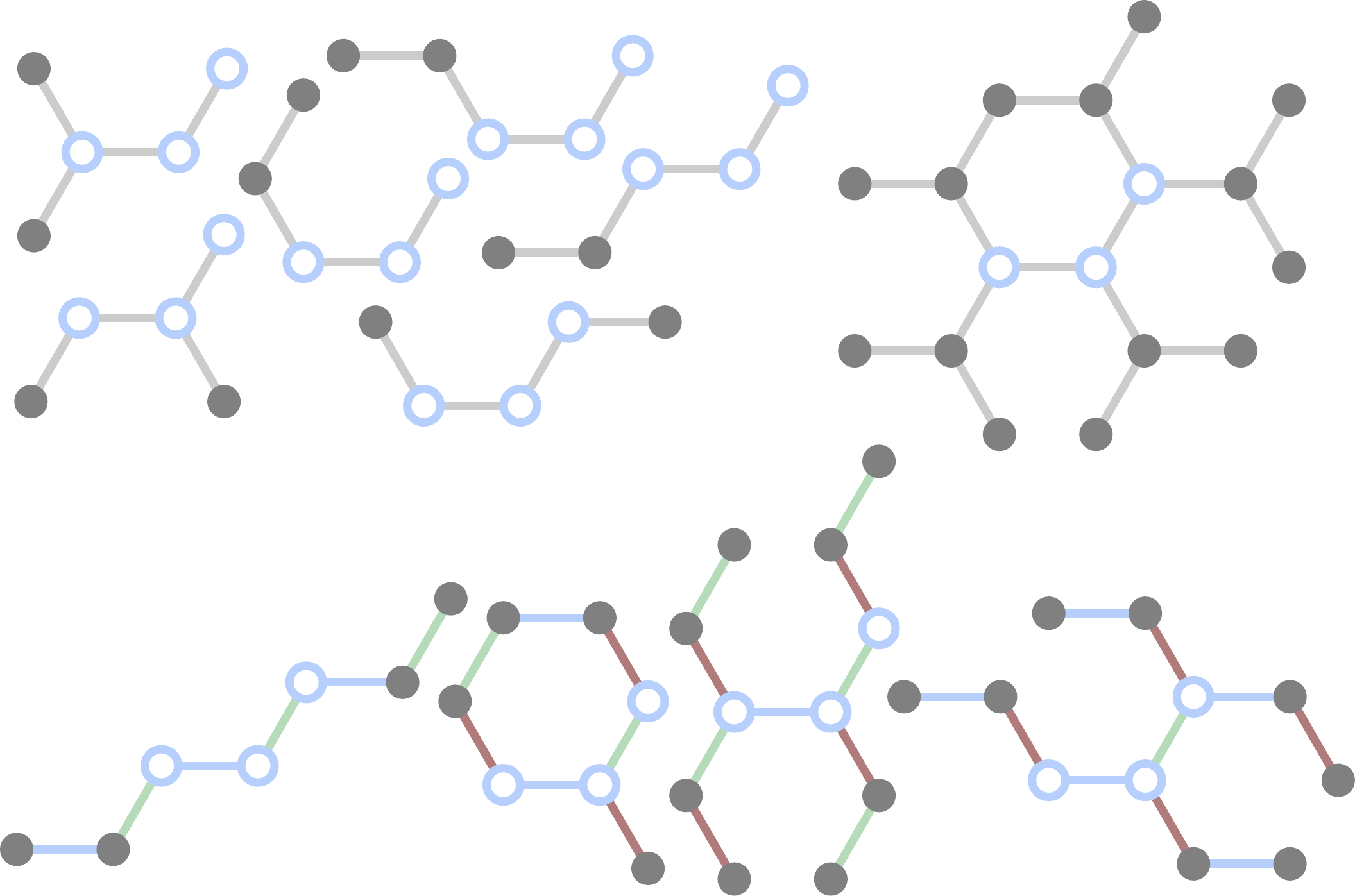}}%
    \put(0.09609708,0.60495668){\color[rgb]{0,0,0}\makebox(0,0)[t]{\lineheight{1.25}\smash{\begin{tabular}[t]{c}(a)\end{tabular}}}}%
    \put(0.09226958,0.23362957){\color[rgb]{0,0,0}\makebox(0,0)[t]{\lineheight{1.25}\smash{\begin{tabular}[t]{c}(c)\end{tabular}}}}%
    \put(0.66728679,0.60481308){\color[rgb]{0,0,0}\makebox(0,0)[t]{\lineheight{1.25}\smash{\begin{tabular}[t]{c}(b)\end{tabular}}}}%
  \end{picture}%
\endgroup%

    \labelsubfig{0}{Fig:HopPro_a}
    \labelsubfig{1}{Fig:HopPro_b}
    \labelsubfig{2}{Fig:HopPro_c}
    \caption{Cluster creation for processes of Eq.~\eqref{Eq:ProcessOnCluster} including three sites marked in blue for order four. (a): A selection of the 21 subclusters calculated by the adapted algorithm including a bond (colored in light gray) of all three blue sites. (b): The final cluster when combining the subcluster. (c): The minimized clusters of (b), when taking the bond types into account. The $k_\beta$ values from left to right are $k_x=k_y=2,k_z=0;\; k_x=k_y=1,k_z=2;\; k_x=k_z=1,k_y=2;\; k_x=2,k_y=k_z=1$. }
    \label{Fig:HoppingProcesses}
\end{figure}

\paragraph*{General approach}
To create the clusters we adapt the algorithm presented in \cite{Ruecker_2000}. Starting at an arbitrary bond, the algorithm creates a set $\mathcal S$ of all connected subclusters of the original cluster $\mathcal C$ including all bonds. Using the linked cluster theorem we know that quasiparticles can move at most $k$ sites for order $k$, as we only have nearest-neighboring processes in Eq.~\eqref{Eq:RotatedHamiltonian}. Therefore, we apply the algorithm on a sufficiently large cluster of $k$ sites spread in all directions, to be sure that all possible processes in order $k$ are included. As we have to calculate processes for $O \in \{\Hn{n}, \On{n}^\alpha(\vec r)\}$, we generally have terms of the form
\begin{align}
    \braket{i_1,\dots,i_n|O|j_1,\dots,j_m}
    \label{Eq:ProcessOnCluster}
\end{align}
with $m=0$ for calculations with $\On{n}^\alpha(\vec r)$ and $n=m$ for calculations with $\Hn{n}$. For all correlated processes in Eq.~\eqref{Eq:ProcessOnCluster}, $O$ has to act on a linked cluster including at least one of the bonds of all sites $i_1,\dots,i_n,j_1,\dots,j_m$. For $\On{n}^\alpha(\vec r)$ also a bond adjacent to site $r$ has to be included. So, we only have to take into account all subclusters $s\in \mathcal S$ that fulfill these conditions and include $k$ bonds. Each possible process of Eq.~\eqref{Eq:ProcessOnCluster} can be calculated by one of these subclusters. Combining all resulting clusters by adding up all bonds, we obtain the minimum cluster to calculate Eq.~\eqref{Eq:ProcessOnCluster} in the thermodynamic limit for order $k$. An example is illustrated in Fig.~\ref{Fig:HoppingProcesses} for three sites colored in blue with $k=4$. Combining all possible subclusters (with a selection shown in \ref{Fig:HoppingProcesses}\ref{Fig:HopPro_a}), we obtain the minimum cluster in \ref{Fig:HoppingProcesses}\ref{Fig:HopPro_b}.

\paragraph*{Minimizing clusters using bond types}
To further minimize the clusters, we take the different bond types $x,y,z$ into account. As an individual term in the resulting series is proportional to $J_x^{k_x} J_y^{k_y} J_z^{k_z}$, we know that during this process at most $k_\beta$ different bonds of type $\beta$ are involved, as a process can potentially act multiple times on the same bond. Therefore, we replace the above condition for the valid subclusters to include $k$ bonds. Instead we demand the subclusters to contain $k_\beta$ bonds of bond type $\beta \in \{x,y,z\}$, with $k_x+k_y+k_z\leq k$. We split the calculation of Eq.~\eqref{Eq:ProcessOnCluster} on a single cluster into a number of calculations on smaller clusters by varying $k_x,k_y,k_z$. 

In some cases the resulting cluster has $k_x+k_y+k_z< k$ and cannot be expanded in terms of a specific \mbox{bond type $\beta$}, e.g., for $\beta=x$ with $k_x=1,k_y=k_z=0$. For this case we enlarge $k_\beta$ to match $k_x+k_y+k_z= k$. To reduce the number of calculations, we check if any subcluster is a subset of another subcluster and use the results of the larger one. An example is given in Fig.~\ref{Fig:HoppingProcesses}\ref{Fig:HopPro_c}, where the cluster in Fig.~\ref{Fig:HoppingProcesses}\ref{Fig:HopPro_b} is split into subclusters of different $k_\beta$ values. The different bond types are drawn in different colors as done in Fig.~\ref{Fig:GridXYZ}. 

In contrast to before, we cannot use all terms of the pCUT calculation on these smaller clusters but only those terms which are proportional to $J_x^{q_x} J_y^{q_y} J_z^{q_z}$ with $q_\beta \leq k_\beta$ for all~$\beta$. The rest of the terms have to be discarded. By adding up all valid terms of all subclusters, we obtain the final expression for Eq.~\eqref{Eq:ProcessOnCluster}.

This approach is especially beneficial when calculating series for nonuniform $J_\beta$ configurations. Otherwise the effect of smaller clusters is compensated by the higher number of perturbation parameters. We also tried a further optimization by introducing more perturbation parameters to minimize the cluster size but ended with larger computation times due to the increased number of perturbation parameters.

\newpage
\bibliography{literature}

\begin{thebibliography}{91}%
\makeatletter
\providecommand \@ifxundefined [1]{%
 \@ifx{#1\undefined}
}%
\providecommand \@ifnum [1]{%
 \ifnum #1\expandafter \@firstoftwo
 \else \expandafter \@secondoftwo
 \fi
}%
\providecommand \@ifx [1]{%
 \ifx #1\expandafter \@firstoftwo
 \else \expandafter \@secondoftwo
 \fi
}%
\providecommand \natexlab [1]{#1}%
\providecommand \enquote  [1]{``#1''}%
\providecommand \bibnamefont  [1]{#1}%
\providecommand \bibfnamefont [1]{#1}%
\providecommand \citenamefont [1]{#1}%
\providecommand \href@noop [0]{\@secondoftwo}%
\providecommand \href [0]{\begingroup \@sanitize@url \@href}%
\providecommand \@href[1]{\@@startlink{#1}\@@href}%
\providecommand \@@href[1]{\endgroup#1\@@endlink}%
\providecommand \@sanitize@url [0]{\catcode `\\12\catcode `\$12\catcode
  `\&12\catcode `\#12\catcode `\^12\catcode `\_12\catcode `\%12\relax}%
\providecommand \@@startlink[1]{}%
\providecommand \@@endlink[0]{}%
\providecommand \url  [0]{\begingroup\@sanitize@url \@url }%
\providecommand \@url [1]{\endgroup\@href {#1}{\urlprefix }}%
\providecommand \urlprefix  [0]{URL }%
\providecommand \Eprint [0]{\href }%
\providecommand \doibase [0]{https://doi.org/}%
\providecommand \selectlanguage [0]{\@gobble}%
\providecommand \bibinfo  [0]{\@secondoftwo}%
\providecommand \bibfield  [0]{\@secondoftwo}%
\providecommand \translation [1]{[#1]}%
\providecommand \BibitemOpen [0]{}%
\providecommand \bibitemStop [0]{}%
\providecommand \bibitemNoStop [0]{.\EOS\space}%
\providecommand \EOS [0]{\spacefactor3000\relax}%
\providecommand \BibitemShut  [1]{\csname bibitem#1\endcsname}%
\let\auto@bib@innerbib\@empty
\bibitem [{\citenamefont {Savary}\ and\ \citenamefont
  {Balents}(2016)}]{Savary2016}%
  \BibitemOpen
  \bibfield  {author} {\bibinfo {author} {\bibfnamefont {L.}~\bibnamefont
  {Savary}}\ and\ \bibinfo {author} {\bibfnamefont {L.}~\bibnamefont
  {Balents}},\ }\bibfield  {title} {\bibinfo {title} {{Quantum spin liquids: a
  review}},\ }\href {https://doi.org/10.1088/0034-4885/80/1/016502} {\bibfield
  {journal} {\bibinfo  {journal} {Rep. Progr. Phys.}\ }\textbf {\bibinfo
  {volume} {80}},\ \bibinfo {pages} {016502} (\bibinfo {year}
  {2016})}\BibitemShut {NoStop}%
\bibitem [{\citenamefont {Balents}(2010)}]{Balents2010}%
  \BibitemOpen
  \bibfield  {author} {\bibinfo {author} {\bibfnamefont {L.}~\bibnamefont
  {Balents}},\ }\bibfield  {title} {\bibinfo {title} {{Spin liquids in
  frustrated magnets}},\ }\href {https://doi.org/10.1038/nature08917}
  {\bibfield  {journal} {\bibinfo  {journal} {Nature}\ }\textbf {\bibinfo
  {volume} {464}},\ \bibinfo {pages} {199} (\bibinfo {year}
  {2010})}\BibitemShut {NoStop}%
\bibitem [{\citenamefont {Kitaev}(2003)}]{Kitaev2003}%
  \BibitemOpen
  \bibfield  {author} {\bibinfo {author} {\bibfnamefont {A.}~\bibnamefont
  {Kitaev}},\ }\bibfield  {title} {\bibinfo {title} {{Fault-tolerant quantum
  computation by anyons}},\ }\href
  {https://doi.org/10.1016/s0003-4916(02)00018-0} {\bibfield  {journal}
  {\bibinfo  {journal} {Ann. Phys. (NY)}\ }\textbf {\bibinfo {volume} {303}},\
  \bibinfo {pages} {2} (\bibinfo {year} {2003})}\BibitemShut {NoStop}%
\bibitem [{\citenamefont {Kitaev}(2006)}]{Kitaev_2006}%
  \BibitemOpen
  \bibfield  {author} {\bibinfo {author} {\bibfnamefont {A.}~\bibnamefont
  {Kitaev}},\ }\bibfield  {title} {\bibinfo {title} {{Anyons in an exactly
  solved model and beyond}},\ }\href
  {https://doi.org/10.1016/j.aop.2005.10.005} {\bibfield  {journal} {\bibinfo
  {journal} {Ann. Phys. (NY)}\ }\textbf {\bibinfo {volume} {321}},\ \bibinfo
  {pages} {2} (\bibinfo {year} {2006})}\BibitemShut {NoStop}%
\bibitem [{\citenamefont {Freedman}\ \emph {et~al.}(2002)\citenamefont
  {Freedman}, \citenamefont {Larsen},\ and\ \citenamefont
  {Wang}}]{Freedman2002}%
  \BibitemOpen
  \bibfield  {author} {\bibinfo {author} {\bibfnamefont {M.~H.}\ \bibnamefont
  {Freedman}}, \bibinfo {author} {\bibfnamefont {M.}~\bibnamefont {Larsen}},\
  and\ \bibinfo {author} {\bibfnamefont {Z.}~\bibnamefont {Wang}},\ }\bibfield
  {title} {\bibinfo {title} {{A modular functor which is universal for quantum
  computation}},\ }\href {https://doi.org/10.1007/s002200200645} {\bibfield
  {journal} {\bibinfo  {journal} {Comm. Math. Phys.}\ }\textbf {\bibinfo
  {volume} {227}},\ \bibinfo {pages} {605} (\bibinfo {year}
  {2002})}\BibitemShut {NoStop}%
\bibitem [{\citenamefont {Preskill}(1998)}]{Preskill1998}%
  \BibitemOpen
  \bibfield  {author} {\bibinfo {author} {\bibfnamefont {J.}~\bibnamefont
  {Preskill}},\ }\bibinfo {title} {{Fault-tolerant quantum computation}},\ in\
  \href {https://doi.org/10.1142/9789812385253_0008} {\emph {\bibinfo
  {booktitle} {{Introduction to Quantum Computation and Information}}}}\
  (\bibinfo  {publisher} {World Scientific, Singapore},\ \bibinfo {year}
  {1998})\ pp.\ \bibinfo {pages} {213--269}\BibitemShut {NoStop}%
\bibitem [{\citenamefont {Castelvecchi}(2017)}]{Castelvecchi2017}%
  \BibitemOpen
  \bibfield  {author} {\bibinfo {author} {\bibfnamefont {D.}~\bibnamefont
  {Castelvecchi}},\ }\bibfield  {title} {\bibinfo {title} {{Quantum computers
  ready to leap out of the lab}},\ }\href {https://doi.org/10.1038/541009a}
  {\bibfield  {journal} {\bibinfo  {journal} {Nature}\ }\textbf {\bibinfo
  {volume} {541}},\ \bibinfo {pages} {9} (\bibinfo {year} {2017})}\BibitemShut
  {NoStop}%
\bibitem [{\citenamefont {Arute}\ \emph {et~al.}(2019)\citenamefont {Arute},
  \citenamefont {Arya}, \citenamefont {Babbush}, \citenamefont {Bacon},
  \citenamefont {Bardin}, \citenamefont {Barends}, \citenamefont {Biswas},
  \citenamefont {Boixo}, \citenamefont {Brandao}, \citenamefont {Buell} \emph
  {et~al.}}]{Arute2019}%
  \BibitemOpen
  \bibfield  {author} {\bibinfo {author} {\bibfnamefont {F.}~\bibnamefont
  {Arute}}, \bibinfo {author} {\bibfnamefont {K.}~\bibnamefont {Arya}},
  \bibinfo {author} {\bibfnamefont {R.}~\bibnamefont {Babbush}}, \bibinfo
  {author} {\bibfnamefont {D.}~\bibnamefont {Bacon}}, \bibinfo {author}
  {\bibfnamefont {J.~C.}\ \bibnamefont {Bardin}}, \bibinfo {author}
  {\bibfnamefont {R.}~\bibnamefont {Barends}}, \bibinfo {author} {\bibfnamefont
  {R.}~\bibnamefont {Biswas}}, \bibinfo {author} {\bibfnamefont
  {S.}~\bibnamefont {Boixo}}, \bibinfo {author} {\bibfnamefont {F.~G. S.~L.}\
  \bibnamefont {Brandao}}, \bibinfo {author} {\bibfnamefont {D.~A.}\
  \bibnamefont {Buell}}, \emph {et~al.},\ }\bibfield  {title} {\bibinfo {title}
  {{Quantum supremacy using a programmable superconducting processor}},\ }\href
  {https://doi.org/10.1038/s41586-019-1666-5} {\bibfield  {journal} {\bibinfo
  {journal} {Nature}\ }\textbf {\bibinfo {volume} {574}},\ \bibinfo {pages}
  {505} (\bibinfo {year} {2019})}\BibitemShut {NoStop}%
\bibitem [{\citenamefont {Friis}\ \emph {et~al.}(2018)\citenamefont {Friis},
  \citenamefont {Marty}, \citenamefont {Maier}, \citenamefont {Hempel},
  \citenamefont {Holzäpfel}, \citenamefont {Jurcevic}, \citenamefont {Plenio},
  \citenamefont {Huber}, \citenamefont {Roos}, \citenamefont {Blatt},\ and\
  \citenamefont {Lanyon}}]{Friis2018}%
  \BibitemOpen
  \bibfield  {author} {\bibinfo {author} {\bibfnamefont {N.}~\bibnamefont
  {Friis}}, \bibinfo {author} {\bibfnamefont {O.}~\bibnamefont {Marty}},
  \bibinfo {author} {\bibfnamefont {C.}~\bibnamefont {Maier}}, \bibinfo
  {author} {\bibfnamefont {C.}~\bibnamefont {Hempel}}, \bibinfo {author}
  {\bibfnamefont {M.}~\bibnamefont {Holzäpfel}}, \bibinfo {author}
  {\bibfnamefont {P.}~\bibnamefont {Jurcevic}}, \bibinfo {author}
  {\bibfnamefont {M.~B.}\ \bibnamefont {Plenio}}, \bibinfo {author}
  {\bibfnamefont {M.}~\bibnamefont {Huber}}, \bibinfo {author} {\bibfnamefont
  {C.}~\bibnamefont {Roos}}, \bibinfo {author} {\bibfnamefont {R.}~\bibnamefont
  {Blatt}},\ and\ \bibinfo {author} {\bibfnamefont {B.}~\bibnamefont
  {Lanyon}},\ }\bibfield  {title} {\bibinfo {title} {{Observation of Entangled
  States of a Fully Controlled 20-Qubit System}},\ }\href
  {https://doi.org/10.1103/physrevx.8.021012} {\bibfield  {journal} {\bibinfo
  {journal} {Phys. Rev. X}\ }\textbf {\bibinfo {volume} {8}},\ \bibinfo {pages}
  {021012} (\bibinfo {year} {2018})}\BibitemShut {NoStop}%
\bibitem [{\citenamefont {Liu}\ \emph {et~al.}(2011)\citenamefont {Liu},
  \citenamefont {Berlijn}, \citenamefont {Yin}, \citenamefont {Ku},
  \citenamefont {Tsvelik}, \citenamefont {Kim}, \citenamefont {Gretarsson},
  \citenamefont {Singh}, \citenamefont {Gegenwart},\ and\ \citenamefont
  {Hill}}]{Liu2011}%
  \BibitemOpen
  \bibfield  {author} {\bibinfo {author} {\bibfnamefont {X.}~\bibnamefont
  {Liu}}, \bibinfo {author} {\bibfnamefont {T.}~\bibnamefont {Berlijn}},
  \bibinfo {author} {\bibfnamefont {W.-G.}\ \bibnamefont {Yin}}, \bibinfo
  {author} {\bibfnamefont {W.}~\bibnamefont {Ku}}, \bibinfo {author}
  {\bibfnamefont {A.}~\bibnamefont {Tsvelik}}, \bibinfo {author} {\bibfnamefont
  {Y.-J.}\ \bibnamefont {Kim}}, \bibinfo {author} {\bibfnamefont
  {H.}~\bibnamefont {Gretarsson}}, \bibinfo {author} {\bibfnamefont
  {Y.}~\bibnamefont {Singh}}, \bibinfo {author} {\bibfnamefont
  {P.}~\bibnamefont {Gegenwart}},\ and\ \bibinfo {author} {\bibfnamefont
  {J.~P.}\ \bibnamefont {Hill}},\ }\bibfield  {title} {\bibinfo {title}
  {{Long-range magnetic ordering in {Na${}_{2}$IrO${}_{3}$}}},\ }\href
  {https://doi.org/10.1103/PhysRevB.83.220403} {\bibfield  {journal} {\bibinfo
  {journal} {Phys. Rev. B}\ }\textbf {\bibinfo {volume} {83}},\ \bibinfo
  {pages} {220403} (\bibinfo {year} {2011})}\BibitemShut {NoStop}%
\bibitem [{\citenamefont {Choi}\ \emph {et~al.}(2012)\citenamefont {Choi},
  \citenamefont {Coldea}, \citenamefont {Kolmogorov}, \citenamefont
  {Lancaster}, \citenamefont {Mazin}, \citenamefont {Blundell}, \citenamefont
  {Radaelli}, \citenamefont {Singh}, \citenamefont {Gegenwart}, \citenamefont
  {Choi}, \citenamefont {Cheong}, \citenamefont {Baker}, \citenamefont
  {Stock},\ and\ \citenamefont {Taylor}}]{Choi2012}%
  \BibitemOpen
  \bibfield  {author} {\bibinfo {author} {\bibfnamefont {S.~K.}\ \bibnamefont
  {Choi}}, \bibinfo {author} {\bibfnamefont {R.}~\bibnamefont {Coldea}},
  \bibinfo {author} {\bibfnamefont {A.~N.}\ \bibnamefont {Kolmogorov}},
  \bibinfo {author} {\bibfnamefont {T.}~\bibnamefont {Lancaster}}, \bibinfo
  {author} {\bibfnamefont {I.~I.}\ \bibnamefont {Mazin}}, \bibinfo {author}
  {\bibfnamefont {S.~J.}\ \bibnamefont {Blundell}}, \bibinfo {author}
  {\bibfnamefont {P.~G.}\ \bibnamefont {Radaelli}}, \bibinfo {author}
  {\bibfnamefont {Y.}~\bibnamefont {Singh}}, \bibinfo {author} {\bibfnamefont
  {P.}~\bibnamefont {Gegenwart}}, \bibinfo {author} {\bibfnamefont {K.~R.}\
  \bibnamefont {Choi}}, \bibinfo {author} {\bibfnamefont {S.-W.}\ \bibnamefont
  {Cheong}}, \bibinfo {author} {\bibfnamefont {P.~J.}\ \bibnamefont {Baker}},
  \bibinfo {author} {\bibfnamefont {C.}~\bibnamefont {Stock}},\ and\ \bibinfo
  {author} {\bibfnamefont {J.}~\bibnamefont {Taylor}},\ }\bibfield  {title}
  {\bibinfo {title} {{Spin Waves and Revised Crystal Structure of Honeycomb
  Iridate {${\mathrm{Na}}_{2}{\mathrm{IrO}}_{3}$}}},\ }\href
  {https://doi.org/10.1103/PhysRevLett.108.127204} {\bibfield  {journal}
  {\bibinfo  {journal} {Phys. Rev. Lett.}\ }\textbf {\bibinfo {volume} {108}},\
  \bibinfo {pages} {127204} (\bibinfo {year} {2012})}\BibitemShut {NoStop}%
\bibitem [{\citenamefont {Gardner}\ \emph {et~al.}(2010)\citenamefont
  {Gardner}, \citenamefont {Gingras},\ and\ \citenamefont
  {Greedan}}]{Gradner2010}%
  \BibitemOpen
  \bibfield  {author} {\bibinfo {author} {\bibfnamefont {J.~S.}\ \bibnamefont
  {Gardner}}, \bibinfo {author} {\bibfnamefont {M.~J.~P.}\ \bibnamefont
  {Gingras}},\ and\ \bibinfo {author} {\bibfnamefont {J.~E.}\ \bibnamefont
  {Greedan}},\ }\bibfield  {title} {\bibinfo {title} {{Magnetic pyrochlore
  oxides}},\ }\href {https://doi.org/10.1103/RevModPhys.82.53} {\bibfield
  {journal} {\bibinfo  {journal} {Rev. Mod. Phys.}\ }\textbf {\bibinfo {volume}
  {82}},\ \bibinfo {pages} {53} (\bibinfo {year} {2010})}\BibitemShut {NoStop}%
\bibitem [{\citenamefont {Ross}\ \emph {et~al.}(2009)\citenamefont {Ross},
  \citenamefont {Ruff}, \citenamefont {Adams}, \citenamefont {Gardner},
  \citenamefont {Dabkowska}, \citenamefont {Qiu}, \citenamefont {Copley},\ and\
  \citenamefont {Gaulin}}]{Ross2009}%
  \BibitemOpen
  \bibfield  {author} {\bibinfo {author} {\bibfnamefont {K.~A.}\ \bibnamefont
  {Ross}}, \bibinfo {author} {\bibfnamefont {J.~P.~C.}\ \bibnamefont {Ruff}},
  \bibinfo {author} {\bibfnamefont {C.~P.}\ \bibnamefont {Adams}}, \bibinfo
  {author} {\bibfnamefont {J.~S.}\ \bibnamefont {Gardner}}, \bibinfo {author}
  {\bibfnamefont {H.~A.}\ \bibnamefont {Dabkowska}}, \bibinfo {author}
  {\bibfnamefont {Y.}~\bibnamefont {Qiu}}, \bibinfo {author} {\bibfnamefont
  {J.~R.~D.}\ \bibnamefont {Copley}},\ and\ \bibinfo {author} {\bibfnamefont
  {B.~D.}\ \bibnamefont {Gaulin}},\ }\bibfield  {title} {\bibinfo {title}
  {{Two-Dimensional Kagome Correlations and Field Induced Order in the
  Ferromagnetic {$XY$} Pyrochlore
  {${\mathrm{Yb}}_{2}{\mathrm{Ti}}_{2}{\mathbf{O}}_{7}$}}},\ }\href
  {https://doi.org/10.1103/PhysRevLett.103.227202} {\bibfield  {journal}
  {\bibinfo  {journal} {Phys Rev Lett}\ }\textbf {\bibinfo {volume} {103}},\
  \bibinfo {pages} {227202} (\bibinfo {year} {2009})}\BibitemShut {NoStop}%
\bibitem [{\citenamefont {Ross}\ \emph {et~al.}(2011)\citenamefont {Ross},
  \citenamefont {Savary}, \citenamefont {Gaulin},\ and\ \citenamefont
  {Balents}}]{Ross2011}%
  \BibitemOpen
  \bibfield  {author} {\bibinfo {author} {\bibfnamefont {K.~A.}\ \bibnamefont
  {Ross}}, \bibinfo {author} {\bibfnamefont {L.}~\bibnamefont {Savary}},
  \bibinfo {author} {\bibfnamefont {B.~D.}\ \bibnamefont {Gaulin}},\ and\
  \bibinfo {author} {\bibfnamefont {L.}~\bibnamefont {Balents}},\ }\bibfield
  {title} {\bibinfo {title} {{Quantum Excitations in Quantum Spin Ice}},\
  }\href {https://doi.org/10.1103/PhysRevX.1.021002} {\bibfield  {journal}
  {\bibinfo  {journal} {Phys. Rev. X}\ }\textbf {\bibinfo {volume} {1}},\
  \bibinfo {pages} {021002} (\bibinfo {year} {2011})}\BibitemShut {NoStop}%
\bibitem [{\citenamefont {Tokiwa}\ \emph {et~al.}(2016)\citenamefont {Tokiwa},
  \citenamefont {Yamashita}, \citenamefont {Udagawa}, \citenamefont {Kittaka},
  \citenamefont {Sakakibara}, \citenamefont {Terazawa}, \citenamefont
  {Shimoyama}, \citenamefont {Terashima}, \citenamefont {Yasui}, \citenamefont
  {Shibauchi},\ and\ \citenamefont {Matsuda}}]{Tokiwa2016}%
  \BibitemOpen
  \bibfield  {author} {\bibinfo {author} {\bibfnamefont {Y.}~\bibnamefont
  {Tokiwa}}, \bibinfo {author} {\bibfnamefont {T.}~\bibnamefont {Yamashita}},
  \bibinfo {author} {\bibfnamefont {M.}~\bibnamefont {Udagawa}}, \bibinfo
  {author} {\bibfnamefont {S.}~\bibnamefont {Kittaka}}, \bibinfo {author}
  {\bibfnamefont {T.}~\bibnamefont {Sakakibara}}, \bibinfo {author}
  {\bibfnamefont {D.}~\bibnamefont {Terazawa}}, \bibinfo {author}
  {\bibfnamefont {Y.}~\bibnamefont {Shimoyama}}, \bibinfo {author}
  {\bibfnamefont {T.}~\bibnamefont {Terashima}}, \bibinfo {author}
  {\bibfnamefont {Y.}~\bibnamefont {Yasui}}, \bibinfo {author} {\bibfnamefont
  {T.}~\bibnamefont {Shibauchi}},\ and\ \bibinfo {author} {\bibfnamefont
  {Y.}~\bibnamefont {Matsuda}},\ }\bibfield  {title} {\bibinfo {title}
  {{Possible observation of highly itinerant quantum magnetic monopoles in the
  frustrated pyrochlore {$\mathrm{Yb}_2\mathrm{Ti}_2\mathrm{O}_7$}}},\ }\href
  {https://doi.org/10.1038/ncomms10807} {\bibfield  {journal} {\bibinfo
  {journal} {Nat. Commun.}\ }\textbf {\bibinfo {volume} {7}},\ \bibinfo {pages}
  {10807} (\bibinfo {year} {2016})}\BibitemShut {NoStop}%
\bibitem [{\citenamefont {F\aa{}k}\ \emph {et~al.}(2012)\citenamefont
  {F\aa{}k}, \citenamefont {Kermarrec}, \citenamefont {Messio}, \citenamefont
  {Bernu}, \citenamefont {Lhuillier}, \citenamefont {Bert}, \citenamefont
  {Mendels}, \citenamefont {Koteswararao}, \citenamefont {Bouquet},
  \citenamefont {Ollivier}, \citenamefont {Hillier}, \citenamefont {Amato},
  \citenamefont {Colman},\ and\ \citenamefont {Wills}}]{Fak2012}%
  \BibitemOpen
  \bibfield  {author} {\bibinfo {author} {\bibfnamefont {B.}~\bibnamefont
  {F\aa{}k}}, \bibinfo {author} {\bibfnamefont {E.}~\bibnamefont {Kermarrec}},
  \bibinfo {author} {\bibfnamefont {L.}~\bibnamefont {Messio}}, \bibinfo
  {author} {\bibfnamefont {B.}~\bibnamefont {Bernu}}, \bibinfo {author}
  {\bibfnamefont {C.}~\bibnamefont {Lhuillier}}, \bibinfo {author}
  {\bibfnamefont {F.}~\bibnamefont {Bert}}, \bibinfo {author} {\bibfnamefont
  {P.}~\bibnamefont {Mendels}}, \bibinfo {author} {\bibfnamefont
  {B.}~\bibnamefont {Koteswararao}}, \bibinfo {author} {\bibfnamefont
  {F.}~\bibnamefont {Bouquet}}, \bibinfo {author} {\bibfnamefont
  {J.}~\bibnamefont {Ollivier}}, \bibinfo {author} {\bibfnamefont {A.~D.}\
  \bibnamefont {Hillier}}, \bibinfo {author} {\bibfnamefont {A.}~\bibnamefont
  {Amato}}, \bibinfo {author} {\bibfnamefont {R.~H.}\ \bibnamefont {Colman}},\
  and\ \bibinfo {author} {\bibfnamefont {A.~S.}\ \bibnamefont {Wills}},\
  }\bibfield  {title} {\bibinfo {title} {{Kapellasite: A Kagome Quantum Spin
  Liquid with Competing Interactions}},\ }\href
  {https://doi.org/10.1103/PhysRevLett.109.037208} {\bibfield  {journal}
  {\bibinfo  {journal} {Phys. Rev. Lett.}\ }\textbf {\bibinfo {volume} {109}},\
  \bibinfo {pages} {037208} (\bibinfo {year} {2012})}\BibitemShut {NoStop}%
\bibitem [{\citenamefont {Qi}\ \emph {et~al.}(2009)\citenamefont {Qi},
  \citenamefont {Xu},\ and\ \citenamefont {Sachdev}}]{Qi2009}%
  \BibitemOpen
  \bibfield  {author} {\bibinfo {author} {\bibfnamefont {Y.}~\bibnamefont
  {Qi}}, \bibinfo {author} {\bibfnamefont {C.}~\bibnamefont {Xu}},\ and\
  \bibinfo {author} {\bibfnamefont {S.}~\bibnamefont {Sachdev}},\ }\bibfield
  {title} {\bibinfo {title} {{Dynamics and Transport of the ${Z}_{2}$ Spin
  Liquid: Application to
  $\ensuremath{\kappa}\mathrm{\text{-}}(\mathrm{ET}{)}_{2}{{\mathrm{Cu}}}_{2}(\mathrm{CN}{)}_{3}$}},\
  }\href {https://doi.org/10.1103/PhysRevLett.102.176401} {\bibfield  {journal}
  {\bibinfo  {journal} {Phys. Rev. Lett.}\ }\textbf {\bibinfo {volume} {102}},\
  \bibinfo {pages} {176401} (\bibinfo {year} {2009})}\BibitemShut {NoStop}%
\bibitem [{\citenamefont {Han}\ \emph {et~al.}(2012)\citenamefont {Han},
  \citenamefont {Helton}, \citenamefont {Chu}, \citenamefont {Nocera},
  \citenamefont {Rodriguez-Rivera}, \citenamefont {Broholm},\ and\
  \citenamefont {Lee}}]{Han2012}%
  \BibitemOpen
  \bibfield  {author} {\bibinfo {author} {\bibfnamefont {T.-H.}\ \bibnamefont
  {Han}}, \bibinfo {author} {\bibfnamefont {J.~S.}\ \bibnamefont {Helton}},
  \bibinfo {author} {\bibfnamefont {S.}~\bibnamefont {Chu}}, \bibinfo {author}
  {\bibfnamefont {D.~G.}\ \bibnamefont {Nocera}}, \bibinfo {author}
  {\bibfnamefont {J.~A.}\ \bibnamefont {Rodriguez-Rivera}}, \bibinfo {author}
  {\bibfnamefont {C.}~\bibnamefont {Broholm}},\ and\ \bibinfo {author}
  {\bibfnamefont {Y.~S.}\ \bibnamefont {Lee}},\ }\bibfield  {title} {\bibinfo
  {title} {{Fractionalized excitations in the spin-liquid state of a
  kagome-lattice antiferromagnet}},\ }\href
  {https://doi.org/10.1038/nature11659} {\bibfield  {journal} {\bibinfo
  {journal} {Nature}\ }\textbf {\bibinfo {volume} {492}},\ \bibinfo {pages}
  {406} (\bibinfo {year} {2012})}\BibitemShut {NoStop}%
\bibitem [{\citenamefont {Dodds}\ \emph {et~al.}(2013)\citenamefont {Dodds},
  \citenamefont {Bhattacharjee},\ and\ \citenamefont {Kim}}]{Dodds2013}%
  \BibitemOpen
  \bibfield  {author} {\bibinfo {author} {\bibfnamefont {T.}~\bibnamefont
  {Dodds}}, \bibinfo {author} {\bibfnamefont {S.}~\bibnamefont
  {Bhattacharjee}},\ and\ \bibinfo {author} {\bibfnamefont {Y.~B.}\
  \bibnamefont {Kim}},\ }\bibfield  {title} {\bibinfo {title} {{Quantum spin
  liquids in the absence of spin-rotation symmetry: Application to
  herbertsmithite}},\ }\href {https://doi.org/10.1103/PhysRevB.88.224413}
  {\bibfield  {journal} {\bibinfo  {journal} {Phys. Rev. B}\ }\textbf {\bibinfo
  {volume} {88}},\ \bibinfo {pages} {224413} (\bibinfo {year}
  {2013})}\BibitemShut {NoStop}%
\bibitem [{\citenamefont {Punk}\ \emph {et~al.}(2014)\citenamefont {Punk},
  \citenamefont {Chowdhury},\ and\ \citenamefont {Sachdev}}]{Punk2014}%
  \BibitemOpen
  \bibfield  {author} {\bibinfo {author} {\bibfnamefont {M.}~\bibnamefont
  {Punk}}, \bibinfo {author} {\bibfnamefont {D.}~\bibnamefont {Chowdhury}},\
  and\ \bibinfo {author} {\bibfnamefont {S.}~\bibnamefont {Sachdev}},\
  }\bibfield  {title} {\bibinfo {title} {{Topological excitations and the
  dynamic structure factor of spin liquids on the kagome lattice}},\ }\href
  {https://doi.org/10.1038/nphys2887} {\bibfield  {journal} {\bibinfo
  {journal} {Nat. Phys.}\ }\textbf {\bibinfo {volume} {10}},\ \bibinfo {pages}
  {289} (\bibinfo {year} {2014})}\BibitemShut {NoStop}%
\bibitem [{\citenamefont {Morampudi}\ \emph {et~al.}(2017)\citenamefont
  {Morampudi}, \citenamefont {Turner}, \citenamefont {Pollmann},\ and\
  \citenamefont {Wilczek}}]{Morampudi2017}%
  \BibitemOpen
  \bibfield  {author} {\bibinfo {author} {\bibfnamefont {S.~C.}\ \bibnamefont
  {Morampudi}}, \bibinfo {author} {\bibfnamefont {A.~M.}\ \bibnamefont
  {Turner}}, \bibinfo {author} {\bibfnamefont {F.}~\bibnamefont {Pollmann}},\
  and\ \bibinfo {author} {\bibfnamefont {F.}~\bibnamefont {Wilczek}},\
  }\bibfield  {title} {\bibinfo {title} {{Statistics of Fractionalized
  Excitations through Threshold Spectroscopy}},\ }\href
  {https://doi.org/10.1103/PhysRevLett.118.227201} {\bibfield  {journal}
  {\bibinfo  {journal} {Phys. Rev. Lett.}\ }\textbf {\bibinfo {volume} {118}},\
  \bibinfo {pages} {227201} (\bibinfo {year} {2017})}\BibitemShut {NoStop}%
\bibitem [{\citenamefont {Schmidt}\ \emph {et~al.}(2008)\citenamefont
  {Schmidt}, \citenamefont {Dusuel},\ and\ \citenamefont
  {Vidal}}]{Schmidt_2003}%
  \BibitemOpen
  \bibfield  {author} {\bibinfo {author} {\bibfnamefont {K.~P.}\ \bibnamefont
  {Schmidt}}, \bibinfo {author} {\bibfnamefont {S.}~\bibnamefont {Dusuel}},\
  and\ \bibinfo {author} {\bibfnamefont {J.}~\bibnamefont {Vidal}},\ }\bibfield
   {title} {\bibinfo {title} {{Emergent Fermions and Anyons in the {K}itaev
  Model}},\ }\href {https://doi.org/10.1103/PhysRevLett.100.057208} {\bibfield
  {journal} {\bibinfo  {journal} {Phys. Rev. Lett.}\ }\textbf {\bibinfo
  {volume} {100}},\ \bibinfo {pages} {057208} (\bibinfo {year}
  {2008})}\BibitemShut {NoStop}%
\bibitem [{\citenamefont {Jackeli}\ and\ \citenamefont
  {Khaliullin}(2009)}]{Jackeli2009}%
  \BibitemOpen
  \bibfield  {author} {\bibinfo {author} {\bibfnamefont {G.}~\bibnamefont
  {Jackeli}}\ and\ \bibinfo {author} {\bibfnamefont {G.}~\bibnamefont
  {Khaliullin}},\ }\bibfield  {title} {\bibinfo {title} {{Mott Insulators in
  the Strong Spin-Orbit Coupling Limit: From {H}eisenberg to a Quantum
  {C}ompass and {K}itaev Models}},\ }\href
  {https://doi.org/10.1103/PhysRevLett.102.017205} {\bibfield  {journal}
  {\bibinfo  {journal} {Phys. Rev. Lett.}\ }\textbf {\bibinfo {volume} {102}},\
  \bibinfo {pages} {017205} (\bibinfo {year} {2009})}\BibitemShut {NoStop}%
\bibitem [{\citenamefont {Chaloupka}\ \emph {et~al.}(2010)\citenamefont
  {Chaloupka}, \citenamefont {Jackeli},\ and\ \citenamefont
  {Khaliullin}}]{Chaloupka_2010}%
  \BibitemOpen
  \bibfield  {author} {\bibinfo {author} {\bibfnamefont {J.}~\bibnamefont
  {Chaloupka}}, \bibinfo {author} {\bibfnamefont {G.}~\bibnamefont {Jackeli}},\
  and\ \bibinfo {author} {\bibfnamefont {G.}~\bibnamefont {Khaliullin}},\
  }\bibfield  {title} {\bibinfo {title} {{Kitaev-{Heisenberg} Model on a
  Honeycomb Lattice: Possible Exotic Phases in {Iridium} Oxides
  {$A_2\mathrm{IrO}_3$}}},\ }\href
  {https://doi.org/10.1103/physrevlett.105.027204} {\bibfield  {journal}
  {\bibinfo  {journal} {Phys. Rev. Lett.}\ }\textbf {\bibinfo {volume} {105}},\
  \bibinfo {pages} {027204} (\bibinfo {year} {2010})}\BibitemShut {NoStop}%
\bibitem [{\citenamefont {Chaloupka}\ \emph {et~al.}(2013)\citenamefont
  {Chaloupka}, \citenamefont {Jackeli},\ and\ \citenamefont
  {Khaliullin}}]{Chaloupka2013}%
  \BibitemOpen
  \bibfield  {author} {\bibinfo {author} {\bibfnamefont {J.}~\bibnamefont
  {Chaloupka}}, \bibinfo {author} {\bibfnamefont {G.}~\bibnamefont {Jackeli}},\
  and\ \bibinfo {author} {\bibfnamefont {G.}~\bibnamefont {Khaliullin}},\
  }\bibfield  {title} {\bibinfo {title} {{Zigzag Magnetic Order in the Iridium
  Oxide {${\mathrm{Na}}_{2}{\mathrm{IrO}}_{3}$}}},\ }\href
  {https://doi.org/10.1103/PhysRevLett.110.097204} {\bibfield  {journal}
  {\bibinfo  {journal} {Phys. Rev. Lett.}\ }\textbf {\bibinfo {volume} {110}},\
  \bibinfo {pages} {097204} (\bibinfo {year} {2013})}\BibitemShut {NoStop}%
\bibitem [{\citenamefont {Plumb}\ \emph {et~al.}(2014)\citenamefont {Plumb},
  \citenamefont {Clancy}, \citenamefont {Sandilands}, \citenamefont {Shankar},
  \citenamefont {Hu}, \citenamefont {Burch}, \citenamefont {Kee},\ and\
  \citenamefont {Kim}}]{Plumb2014}%
  \BibitemOpen
  \bibfield  {author} {\bibinfo {author} {\bibfnamefont {K.~W.}\ \bibnamefont
  {Plumb}}, \bibinfo {author} {\bibfnamefont {J.~P.}\ \bibnamefont {Clancy}},
  \bibinfo {author} {\bibfnamefont {L.~J.}\ \bibnamefont {Sandilands}},
  \bibinfo {author} {\bibfnamefont {V.~V.}\ \bibnamefont {Shankar}}, \bibinfo
  {author} {\bibfnamefont {Y.~F.}\ \bibnamefont {Hu}}, \bibinfo {author}
  {\bibfnamefont {K.~S.}\ \bibnamefont {Burch}}, \bibinfo {author}
  {\bibfnamefont {H.-Y.}\ \bibnamefont {Kee}},\ and\ \bibinfo {author}
  {\bibfnamefont {Y.-J.}\ \bibnamefont {Kim}},\ }\bibfield  {title} {\bibinfo
  {title} {{$\upalpha$-{$\mathrm{RuCl}_3$}: A spin-orbit assisted {Mott}
  insulator on a honeycomb lattice}},\ }\href
  {https://doi.org/10.1103/PhysRevB.90.041112} {\bibfield  {journal} {\bibinfo
  {journal} {Phys. Rev. B}\ }\textbf {\bibinfo {volume} {90}},\ \bibinfo
  {pages} {041112} (\bibinfo {year} {2014})}\BibitemShut {NoStop}%
\bibitem [{\citenamefont {Sears}\ \emph {et~al.}(2015)\citenamefont {Sears},
  \citenamefont {Songvilay}, \citenamefont {Plumb}, \citenamefont {Clancy},
  \citenamefont {Qiu}, \citenamefont {Zhao}, \citenamefont {Parshall},\ and\
  \citenamefont {Kim}}]{Sears2015}%
  \BibitemOpen
  \bibfield  {author} {\bibinfo {author} {\bibfnamefont {J.~A.}\ \bibnamefont
  {Sears}}, \bibinfo {author} {\bibfnamefont {M.}~\bibnamefont {Songvilay}},
  \bibinfo {author} {\bibfnamefont {K.~W.}\ \bibnamefont {Plumb}}, \bibinfo
  {author} {\bibfnamefont {J.~P.}\ \bibnamefont {Clancy}}, \bibinfo {author}
  {\bibfnamefont {Y.}~\bibnamefont {Qiu}}, \bibinfo {author} {\bibfnamefont
  {Y.}~\bibnamefont {Zhao}}, \bibinfo {author} {\bibfnamefont {D.}~\bibnamefont
  {Parshall}},\ and\ \bibinfo {author} {\bibfnamefont {Y.-J.}\ \bibnamefont
  {Kim}},\ }\bibfield  {title} {\bibinfo {title} {{Magnetic order in
  $\upalpha$-{$\mathrm{RuCl}_3$}: A honeycomb-lattice quantum magnet with
  strong spin-orbit coupling}},\ }\href
  {https://doi.org/10.1103/PhysRevB.91.144420} {\bibfield  {journal} {\bibinfo
  {journal} {Phys. Rev. B}\ }\textbf {\bibinfo {volume} {91}},\ \bibinfo
  {pages} {144420} (\bibinfo {year} {2015})}\BibitemShut {NoStop}%
\bibitem [{\citenamefont {Winter}\ \emph {et~al.}(2017)\citenamefont {Winter},
  \citenamefont {Tsirlin}, \citenamefont {Daghofer}, \citenamefont {van~den
  Brink}, \citenamefont {Singh}, \citenamefont {Gegenwart},\ and\ \citenamefont
  {Valent{\'{\i}}}}]{Winter2017}%
  \BibitemOpen
  \bibfield  {author} {\bibinfo {author} {\bibfnamefont {S.~M.}\ \bibnamefont
  {Winter}}, \bibinfo {author} {\bibfnamefont {A.~A.}\ \bibnamefont {Tsirlin}},
  \bibinfo {author} {\bibfnamefont {M.}~\bibnamefont {Daghofer}}, \bibinfo
  {author} {\bibfnamefont {J.}~\bibnamefont {van~den Brink}}, \bibinfo {author}
  {\bibfnamefont {Y.}~\bibnamefont {Singh}}, \bibinfo {author} {\bibfnamefont
  {P.}~\bibnamefont {Gegenwart}},\ and\ \bibinfo {author} {\bibfnamefont
  {R.}~\bibnamefont {Valent{\'{\i}}}},\ }\bibfield  {title} {\bibinfo {title}
  {{Models and materials for generalized {K}itaev magnetism}},\ }\href
  {https://doi.org/10.1088/1361-648x/aa8cf5} {\bibfield  {journal} {\bibinfo
  {journal} {J. Phys.: Condens. Matter}\ }\textbf {\bibinfo {volume} {29}},\
  \bibinfo {pages} {493002} (\bibinfo {year} {2017})}\BibitemShut {NoStop}%
\bibitem [{\citenamefont {Kasahara}\ \emph {et~al.}(2018)\citenamefont
  {Kasahara}, \citenamefont {Ohnishi}, \citenamefont {Mizukami}, \citenamefont
  {Tanaka}, \citenamefont {Ma}, \citenamefont {Sugii}, \citenamefont {Kurita},
  \citenamefont {Tanaka}, \citenamefont {Nasu}, \citenamefont {Motome},
  \citenamefont {Shibauchi},\ and\ \citenamefont {Matsuda}}]{Kasahara2018}%
  \BibitemOpen
  \bibfield  {author} {\bibinfo {author} {\bibfnamefont {Y.}~\bibnamefont
  {Kasahara}}, \bibinfo {author} {\bibfnamefont {T.}~\bibnamefont {Ohnishi}},
  \bibinfo {author} {\bibfnamefont {Y.}~\bibnamefont {Mizukami}}, \bibinfo
  {author} {\bibfnamefont {O.}~\bibnamefont {Tanaka}}, \bibinfo {author}
  {\bibfnamefont {S.}~\bibnamefont {Ma}}, \bibinfo {author} {\bibfnamefont
  {K.}~\bibnamefont {Sugii}}, \bibinfo {author} {\bibfnamefont
  {N.}~\bibnamefont {Kurita}}, \bibinfo {author} {\bibfnamefont
  {H.}~\bibnamefont {Tanaka}}, \bibinfo {author} {\bibfnamefont
  {J.}~\bibnamefont {Nasu}}, \bibinfo {author} {\bibfnamefont {Y.}~\bibnamefont
  {Motome}}, \bibinfo {author} {\bibfnamefont {T.}~\bibnamefont {Shibauchi}},\
  and\ \bibinfo {author} {\bibfnamefont {Y.}~\bibnamefont {Matsuda}},\
  }\bibfield  {title} {\bibinfo {title} {{Majorana quantization and
  half-integer thermal quantum {H}all effect in a {K}itaev spin liquid}},\
  }\href {https://doi.org/10.1038/s41586-018-0274-0} {\bibfield  {journal}
  {\bibinfo  {journal} {Nature}\ }\textbf {\bibinfo {volume} {559}},\ \bibinfo
  {pages} {227} (\bibinfo {year} {2018})}\BibitemShut {NoStop}%
\bibitem [{\citenamefont {Sanders}\ \emph {et~al.}(2021)\citenamefont
  {Sanders}, \citenamefont {Mole}, \citenamefont {Liu}, \citenamefont {Brown},
  \citenamefont {Yu}, \citenamefont {Ling},\ and\ \citenamefont
  {Rachel}}]{Sanders2021}%
  \BibitemOpen
  \bibfield  {author} {\bibinfo {author} {\bibfnamefont {A.~L.}\ \bibnamefont
  {Sanders}}, \bibinfo {author} {\bibfnamefont {R.~A.}\ \bibnamefont {Mole}},
  \bibinfo {author} {\bibfnamefont {J.}~\bibnamefont {Liu}}, \bibinfo {author}
  {\bibfnamefont {A.~J.}\ \bibnamefont {Brown}}, \bibinfo {author}
  {\bibfnamefont {D.}~\bibnamefont {Yu}}, \bibinfo {author} {\bibfnamefont
  {C.~D.}\ \bibnamefont {Ling}},\ and\ \bibinfo {author} {\bibfnamefont
  {S.}~\bibnamefont {Rachel}},\ }\href@noop {} {\bibinfo {title} {{Dominant
  {K}itaev interactions in the honeycomb materials {Na$_3$Co$_2$SbO$_6$ and
  Na$_2$Co$_2$TeO$_6$}}}} (\bibinfo {year} {2021}),\ \Eprint
  {https://arxiv.org/abs/2112.12254} {arXiv:2112.12254 [cond-mat.str-el]}
  \BibitemShut {NoStop}%
\bibitem [{\citenamefont {Wolter}\ \emph {et~al.}(2017)\citenamefont {Wolter},
  \citenamefont {Corredor}, \citenamefont {Janssen}, \citenamefont {Nenkov},
  \citenamefont {Schönecker}, \citenamefont {Do}, \citenamefont {Choi},
  \citenamefont {Albrecht}, \citenamefont {Hunger}, \citenamefont {Doert},
  \citenamefont {Vojta},\ and\ \citenamefont {Büchner}}]{Wolter_2017}%
  \BibitemOpen
  \bibfield  {author} {\bibinfo {author} {\bibfnamefont {A.~U.~B.}\
  \bibnamefont {Wolter}}, \bibinfo {author} {\bibfnamefont {L.~T.}\
  \bibnamefont {Corredor}}, \bibinfo {author} {\bibfnamefont {L.}~\bibnamefont
  {Janssen}}, \bibinfo {author} {\bibfnamefont {K.}~\bibnamefont {Nenkov}},
  \bibinfo {author} {\bibfnamefont {S.}~\bibnamefont {Schönecker}}, \bibinfo
  {author} {\bibfnamefont {S.-H.}\ \bibnamefont {Do}}, \bibinfo {author}
  {\bibfnamefont {K.-Y.}\ \bibnamefont {Choi}}, \bibinfo {author}
  {\bibfnamefont {R.}~\bibnamefont {Albrecht}}, \bibinfo {author}
  {\bibfnamefont {J.}~\bibnamefont {Hunger}}, \bibinfo {author} {\bibfnamefont
  {T.}~\bibnamefont {Doert}}, \bibinfo {author} {\bibfnamefont
  {M.}~\bibnamefont {Vojta}},\ and\ \bibinfo {author} {\bibfnamefont
  {B.}~\bibnamefont {Büchner}},\ }\bibfield  {title} {\bibinfo {title}
  {{Field-induced quantum criticality in the {Kitaev} system
  $\upalpha$-$\mathrm{RuCl}_3$}},\ }\href
  {https://doi.org/10.1103/physrevb.96.041405} {\bibfield  {journal} {\bibinfo
  {journal} {Phys. Rev. B}\ }\textbf {\bibinfo {volume} {96}},\ \bibinfo
  {pages} {041405} (\bibinfo {year} {2017})}\BibitemShut {NoStop}%
\bibitem [{\citenamefont {Trebst}\ and\ \citenamefont
  {Hickey}(2022)}]{Trebst2022}%
  \BibitemOpen
  \bibfield  {author} {\bibinfo {author} {\bibfnamefont {S.}~\bibnamefont
  {Trebst}}\ and\ \bibinfo {author} {\bibfnamefont {C.}~\bibnamefont
  {Hickey}},\ }\bibfield  {title} {\bibinfo {title} {{Kitaev materials}},\
  }\href {https://doi.org/10.1016/j.physrep.2021.11.003} {\bibfield  {journal}
  {\bibinfo  {journal} {Phys. Rep.}\ }\textbf {\bibinfo {volume} {950}},\
  \bibinfo {pages} {1} (\bibinfo {year} {2022})}\BibitemShut {NoStop}%
\bibitem [{\citenamefont {Singh}\ \emph {et~al.}(2012)\citenamefont {Singh},
  \citenamefont {Manni}, \citenamefont {Reuther}, \citenamefont {Berlijn},
  \citenamefont {Thomale}, \citenamefont {Ku}, \citenamefont {Trebst},\ and\
  \citenamefont {Gegenwart}}]{Singh_2012}%
  \BibitemOpen
  \bibfield  {author} {\bibinfo {author} {\bibfnamefont {Y.}~\bibnamefont
  {Singh}}, \bibinfo {author} {\bibfnamefont {S.}~\bibnamefont {Manni}},
  \bibinfo {author} {\bibfnamefont {J.}~\bibnamefont {Reuther}}, \bibinfo
  {author} {\bibfnamefont {T.}~\bibnamefont {Berlijn}}, \bibinfo {author}
  {\bibfnamefont {R.}~\bibnamefont {Thomale}}, \bibinfo {author} {\bibfnamefont
  {W.}~\bibnamefont {Ku}}, \bibinfo {author} {\bibfnamefont {S.}~\bibnamefont
  {Trebst}},\ and\ \bibinfo {author} {\bibfnamefont {P.}~\bibnamefont
  {Gegenwart}},\ }\bibfield  {title} {\bibinfo {title} {{Relevance of the
  {Heisenberg-Kitaev} Model for the Honeycomb Lattice Iridates
  {$A_2\mathrm{IrO}_3$}}},\ }\href
  {https://doi.org/10.1103/physrevlett.108.127203} {\bibfield  {journal}
  {\bibinfo  {journal} {Phys. Rev. Lett.}\ }\textbf {\bibinfo {volume} {108}},\
  \bibinfo {pages} {127203} (\bibinfo {year} {2012})}\BibitemShut {NoStop}%
\bibitem [{\citenamefont {Ye}\ \emph {et~al.}(2012)\citenamefont {Ye},
  \citenamefont {Chi}, \citenamefont {Cao}, \citenamefont {Chakoumakos},
  \citenamefont {Fernandez-Baca}, \citenamefont {Custelcean}, \citenamefont
  {Qi}, \citenamefont {Korneta},\ and\ \citenamefont {Cao}}]{Ye_2012}%
  \BibitemOpen
  \bibfield  {author} {\bibinfo {author} {\bibfnamefont {F.}~\bibnamefont
  {Ye}}, \bibinfo {author} {\bibfnamefont {S.}~\bibnamefont {Chi}}, \bibinfo
  {author} {\bibfnamefont {H.}~\bibnamefont {Cao}}, \bibinfo {author}
  {\bibfnamefont {B.~C.}\ \bibnamefont {Chakoumakos}}, \bibinfo {author}
  {\bibfnamefont {J.~A.}\ \bibnamefont {Fernandez-Baca}}, \bibinfo {author}
  {\bibfnamefont {R.}~\bibnamefont {Custelcean}}, \bibinfo {author}
  {\bibfnamefont {T.~F.}\ \bibnamefont {Qi}}, \bibinfo {author} {\bibfnamefont
  {O.~B.}\ \bibnamefont {Korneta}},\ and\ \bibinfo {author} {\bibfnamefont
  {G.}~\bibnamefont {Cao}},\ }\bibfield  {title} {\bibinfo {title} {{Direct
  evidence of a zigzag spin-chain structure in the honeycomb lattice: A neutron
  and x-ray diffraction investigation of single-crystal
  {$\mathrm{Na}_2\mathrm{IrO}_3$}}},\ }\href
  {https://doi.org/10.1103/physrevb.85.180403} {\bibfield  {journal} {\bibinfo
  {journal} {Phys. Rev. B}\ }\textbf {\bibinfo {volume} {85}},\ \bibinfo
  {pages} {180403} (\bibinfo {year} {2012})}\BibitemShut {NoStop}%
\bibitem [{\citenamefont {Yadav}\ \emph {et~al.}(2016)\citenamefont {Yadav},
  \citenamefont {Bogdanov}, \citenamefont {Katukuri}, \citenamefont
  {Nishimoto}, \citenamefont {van~den Brink},\ and\ \citenamefont
  {Hozoi}}]{Yadav2016}%
  \BibitemOpen
  \bibfield  {author} {\bibinfo {author} {\bibfnamefont {R.}~\bibnamefont
  {Yadav}}, \bibinfo {author} {\bibfnamefont {N.~A.}\ \bibnamefont {Bogdanov}},
  \bibinfo {author} {\bibfnamefont {V.~M.}\ \bibnamefont {Katukuri}}, \bibinfo
  {author} {\bibfnamefont {S.}~\bibnamefont {Nishimoto}}, \bibinfo {author}
  {\bibfnamefont {J.}~\bibnamefont {van~den Brink}},\ and\ \bibinfo {author}
  {\bibfnamefont {L.}~\bibnamefont {Hozoi}},\ }\bibfield  {title} {\bibinfo
  {title} {{Kitaev exchange and field-induced quantum spin-liquid states in
  honeycomb $\upalpha$-{$\mathrm{RuCl}_3$}}},\ }\href
  {https://doi.org/10.1038/srep37925} {\bibfield  {journal} {\bibinfo
  {journal} {Sci. Rep.}\ }\textbf {\bibinfo {volume} {6}},\ \bibinfo {pages}
  {37925} (\bibinfo {year} {2016})}\BibitemShut {NoStop}%
\bibitem [{\citenamefont {Janssen}\ \emph {et~al.}(2016)\citenamefont
  {Janssen}, \citenamefont {Andrade},\ and\ \citenamefont
  {Vojta}}]{Janssen2016}%
  \BibitemOpen
  \bibfield  {author} {\bibinfo {author} {\bibfnamefont {L.}~\bibnamefont
  {Janssen}}, \bibinfo {author} {\bibfnamefont {E.~C.}\ \bibnamefont
  {Andrade}},\ and\ \bibinfo {author} {\bibfnamefont {M.}~\bibnamefont
  {Vojta}},\ }\bibfield  {title} {\bibinfo {title} {{Honeycomb-Lattice
  {H}eisenberg-{K}}itaev model in a magnetic field: Spin canting,
  metamagnetism, and vortex crystals},\ }\href
  {https://doi.org/10.1103/PhysRevLett.117.277202} {\bibfield  {journal}
  {\bibinfo  {journal} {Phys. Rev. Lett.}\ }\textbf {\bibinfo {volume} {117}},\
  \bibinfo {pages} {277202} (\bibinfo {year} {2016})}\BibitemShut {NoStop}%
\bibitem [{\citenamefont {Zheng}\ \emph {et~al.}(2017)\citenamefont {Zheng},
  \citenamefont {Ran}, \citenamefont {Li}, \citenamefont {Wang}, \citenamefont
  {Wang}, \citenamefont {Liu}, \citenamefont {Liu}, \citenamefont {Normand},
  \citenamefont {Wen},\ and\ \citenamefont {Yu}}]{Zheng2017}%
  \BibitemOpen
  \bibfield  {author} {\bibinfo {author} {\bibfnamefont {J.}~\bibnamefont
  {Zheng}}, \bibinfo {author} {\bibfnamefont {K.}~\bibnamefont {Ran}}, \bibinfo
  {author} {\bibfnamefont {T.}~\bibnamefont {Li}}, \bibinfo {author}
  {\bibfnamefont {J.}~\bibnamefont {Wang}}, \bibinfo {author} {\bibfnamefont
  {P.}~\bibnamefont {Wang}}, \bibinfo {author} {\bibfnamefont {B.}~\bibnamefont
  {Liu}}, \bibinfo {author} {\bibfnamefont {Z.-X.}\ \bibnamefont {Liu}},
  \bibinfo {author} {\bibfnamefont {B.}~\bibnamefont {Normand}}, \bibinfo
  {author} {\bibfnamefont {J.}~\bibnamefont {Wen}},\ and\ \bibinfo {author}
  {\bibfnamefont {W.}~\bibnamefont {Yu}},\ }\bibfield  {title} {\bibinfo
  {title} {{Gapless Spin Excitations in the Field-Induced Quantum Spin Liquid
  Phase of {$\upalpha\text{-}{\mathrm{RuCl}}_{3}$}}},\ }\href
  {https://doi.org/10.1103/PhysRevLett.119.227208} {\bibfield  {journal}
  {\bibinfo  {journal} {Phys. Rev. Lett.}\ }\textbf {\bibinfo {volume} {119}},\
  \bibinfo {pages} {227208} (\bibinfo {year} {2017})}\BibitemShut {NoStop}%
\bibitem [{\citenamefont {Baek}\ \emph {et~al.}(2017)\citenamefont {Baek},
  \citenamefont {Do}, \citenamefont {Choi}, \citenamefont {Kwon}, \citenamefont
  {Wolter}, \citenamefont {Nishimoto}, \citenamefont {van~den Brink},\ and\
  \citenamefont {Büchner}}]{Baek_2017}%
  \BibitemOpen
  \bibfield  {author} {\bibinfo {author} {\bibfnamefont {S.-H.}\ \bibnamefont
  {Baek}}, \bibinfo {author} {\bibfnamefont {S.-H.}\ \bibnamefont {Do}},
  \bibinfo {author} {\bibfnamefont {K.-Y.}\ \bibnamefont {Choi}}, \bibinfo
  {author} {\bibfnamefont {Y.}~\bibnamefont {Kwon}}, \bibinfo {author}
  {\bibfnamefont {A.}~\bibnamefont {Wolter}}, \bibinfo {author} {\bibfnamefont
  {S.}~\bibnamefont {Nishimoto}}, \bibinfo {author} {\bibfnamefont
  {J.}~\bibnamefont {van~den Brink}},\ and\ \bibinfo {author} {\bibfnamefont
  {B.}~\bibnamefont {Büchner}},\ }\bibfield  {title} {\bibinfo {title}
  {{Evidence for a Field-Induced Quantum Spin Liquid in
  {$\upalpha$-$\mathrm{RuCl}_3$}}},\ }\href
  {https://doi.org/10.1103/physrevlett.119.037201} {\bibfield  {journal}
  {\bibinfo  {journal} {Phys. Rev. Lett.}\ }\textbf {\bibinfo {volume} {119}},\
  \bibinfo {pages} {037201} (\bibinfo {year} {2017})}\BibitemShut {NoStop}%
\bibitem [{\citenamefont {Sears}\ \emph {et~al.}(2017)\citenamefont {Sears},
  \citenamefont {Zhao}, \citenamefont {Xu}, \citenamefont {Lynn},\ and\
  \citenamefont {Kim}}]{Sears_2017}%
  \BibitemOpen
  \bibfield  {author} {\bibinfo {author} {\bibfnamefont {J.~A.}\ \bibnamefont
  {Sears}}, \bibinfo {author} {\bibfnamefont {Y.}~\bibnamefont {Zhao}},
  \bibinfo {author} {\bibfnamefont {Z.}~\bibnamefont {Xu}}, \bibinfo {author}
  {\bibfnamefont {J.~W.}\ \bibnamefont {Lynn}},\ and\ \bibinfo {author}
  {\bibfnamefont {Y.-J.}\ \bibnamefont {Kim}},\ }\bibfield  {title} {\bibinfo
  {title} {{Phase diagram of {$\upalpha$-$\mathrm{RuCl}_3$} in an in-plane
  magnetic field}},\ }\href {https://doi.org/10.1103/physrevb.95.180411}
  {\bibfield  {journal} {\bibinfo  {journal} {Phys. Rev. B}\ }\textbf {\bibinfo
  {volume} {95}},\ \bibinfo {pages} {180411} (\bibinfo {year}
  {2017})}\BibitemShut {NoStop}%
\bibitem [{\citenamefont {Leahy}\ \emph {et~al.}(2017)\citenamefont {Leahy},
  \citenamefont {Pocs}, \citenamefont {Siegfried}, \citenamefont {Graf},
  \citenamefont {Do}, \citenamefont {Choi}, \citenamefont {Normand},\ and\
  \citenamefont {Lee}}]{Leahy_2017}%
  \BibitemOpen
  \bibfield  {author} {\bibinfo {author} {\bibfnamefont {I.~A.}\ \bibnamefont
  {Leahy}}, \bibinfo {author} {\bibfnamefont {C.~A.}\ \bibnamefont {Pocs}},
  \bibinfo {author} {\bibfnamefont {P.~E.}\ \bibnamefont {Siegfried}}, \bibinfo
  {author} {\bibfnamefont {D.}~\bibnamefont {Graf}}, \bibinfo {author}
  {\bibfnamefont {S.-H.}\ \bibnamefont {Do}}, \bibinfo {author} {\bibfnamefont
  {K.-Y.}\ \bibnamefont {Choi}}, \bibinfo {author} {\bibfnamefont
  {B.}~\bibnamefont {Normand}},\ and\ \bibinfo {author} {\bibfnamefont
  {M.}~\bibnamefont {Lee}},\ }\bibfield  {title} {\bibinfo {title} {{Anomalous
  Thermal Conductivity and Magnetic Torque Response in the Honeycomb Magnet
  $\upalpha$-$\mathrm{RuCl}_3$}},\ }\href
  {https://doi.org/10.1103/physrevlett.118.187203} {\bibfield  {journal}
  {\bibinfo  {journal} {Phys. Rev. Lett.}\ }\textbf {\bibinfo {volume} {118}},\
  \bibinfo {pages} {187203} (\bibinfo {year} {2017})}\BibitemShut {NoStop}%
\bibitem [{\citenamefont {Yokoi}\ \emph {et~al.}(2021)\citenamefont {Yokoi},
  \citenamefont {Ma}, \citenamefont {Kasahara}, \citenamefont {Kasahara},
  \citenamefont {Shibauchi}, \citenamefont {Kurita}, \citenamefont {Tanaka},
  \citenamefont {Nasu}, \citenamefont {Motome}, \citenamefont {Hickey},
  \citenamefont {Trebst},\ and\ \citenamefont {Matsuda}}]{Yokoi2021}%
  \BibitemOpen
  \bibfield  {author} {\bibinfo {author} {\bibfnamefont {T.}~\bibnamefont
  {Yokoi}}, \bibinfo {author} {\bibfnamefont {S.}~\bibnamefont {Ma}}, \bibinfo
  {author} {\bibfnamefont {Y.}~\bibnamefont {Kasahara}}, \bibinfo {author}
  {\bibfnamefont {S.}~\bibnamefont {Kasahara}}, \bibinfo {author}
  {\bibfnamefont {T.}~\bibnamefont {Shibauchi}}, \bibinfo {author}
  {\bibfnamefont {N.}~\bibnamefont {Kurita}}, \bibinfo {author} {\bibfnamefont
  {H.}~\bibnamefont {Tanaka}}, \bibinfo {author} {\bibfnamefont
  {J.}~\bibnamefont {Nasu}}, \bibinfo {author} {\bibfnamefont {Y.}~\bibnamefont
  {Motome}}, \bibinfo {author} {\bibfnamefont {C.}~\bibnamefont {Hickey}},
  \bibinfo {author} {\bibfnamefont {S.}~\bibnamefont {Trebst}},\ and\ \bibinfo
  {author} {\bibfnamefont {Y.}~\bibnamefont {Matsuda}},\ }\bibfield  {title}
  {\bibinfo {title} {{Half-integer quantized anomalous thermal {H}all effect in
  the {K}itaev material candidate $\upalpha$-{RuCl}${}_3$}},\ }\href
  {https://doi.org/10.1126/science.aay5551} {\bibfield  {journal} {\bibinfo
  {journal} {Science}\ }\textbf {\bibinfo {volume} {373}},\ \bibinfo {pages}
  {568} (\bibinfo {year} {2021})}\BibitemShut {NoStop}%
\bibitem [{\citenamefont {Lin}\ \emph {et~al.}(2021)\citenamefont {Lin},
  \citenamefont {Jeong}, \citenamefont {Kim}, \citenamefont {Wang},
  \citenamefont {Huang}, \citenamefont {Masuda}, \citenamefont {Asai},
  \citenamefont {Itoh}, \citenamefont {Günther}, \citenamefont {Russina},
  \citenamefont {Lu}, \citenamefont {Sheng}, \citenamefont {Wang},
  \citenamefont {Wang}, \citenamefont {Wang}, \citenamefont {Ren},
  \citenamefont {Xi}, \citenamefont {Tong}, \citenamefont {Ling}, \citenamefont
  {Liu}, \citenamefont {Wu}, \citenamefont {Mei}, \citenamefont {Qu},
  \citenamefont {Zhou}, \citenamefont {Wang}, \citenamefont {Park},
  \citenamefont {Wan},\ and\ \citenamefont {Ma}}]{Lin_2021}%
  \BibitemOpen
  \bibfield  {author} {\bibinfo {author} {\bibfnamefont {G.}~\bibnamefont
  {Lin}}, \bibinfo {author} {\bibfnamefont {J.}~\bibnamefont {Jeong}}, \bibinfo
  {author} {\bibfnamefont {C.}~\bibnamefont {Kim}}, \bibinfo {author}
  {\bibfnamefont {Y.}~\bibnamefont {Wang}}, \bibinfo {author} {\bibfnamefont
  {Q.}~\bibnamefont {Huang}}, \bibinfo {author} {\bibfnamefont
  {T.}~\bibnamefont {Masuda}}, \bibinfo {author} {\bibfnamefont
  {S.}~\bibnamefont {Asai}}, \bibinfo {author} {\bibfnamefont {S.}~\bibnamefont
  {Itoh}}, \bibinfo {author} {\bibfnamefont {G.}~\bibnamefont {Günther}},
  \bibinfo {author} {\bibfnamefont {M.}~\bibnamefont {Russina}}, \bibinfo
  {author} {\bibfnamefont {Z.}~\bibnamefont {Lu}}, \bibinfo {author}
  {\bibfnamefont {J.}~\bibnamefont {Sheng}}, \bibinfo {author} {\bibfnamefont
  {L.}~\bibnamefont {Wang}}, \bibinfo {author} {\bibfnamefont {J.}~\bibnamefont
  {Wang}}, \bibinfo {author} {\bibfnamefont {G.}~\bibnamefont {Wang}}, \bibinfo
  {author} {\bibfnamefont {Q.}~\bibnamefont {Ren}}, \bibinfo {author}
  {\bibfnamefont {C.}~\bibnamefont {Xi}}, \bibinfo {author} {\bibfnamefont
  {W.}~\bibnamefont {Tong}}, \bibinfo {author} {\bibfnamefont {L.}~\bibnamefont
  {Ling}}, \bibinfo {author} {\bibfnamefont {Z.}~\bibnamefont {Liu}}, \bibinfo
  {author} {\bibfnamefont {L.}~\bibnamefont {Wu}}, \bibinfo {author}
  {\bibfnamefont {J.}~\bibnamefont {Mei}}, \bibinfo {author} {\bibfnamefont
  {Z.}~\bibnamefont {Qu}}, \bibinfo {author} {\bibfnamefont {H.}~\bibnamefont
  {Zhou}}, \bibinfo {author} {\bibfnamefont {X.}~\bibnamefont {Wang}}, \bibinfo
  {author} {\bibfnamefont {J.-G.}\ \bibnamefont {Park}}, \bibinfo {author}
  {\bibfnamefont {Y.}~\bibnamefont {Wan}},\ and\ \bibinfo {author}
  {\bibfnamefont {J.}~\bibnamefont {Ma}},\ }\bibfield  {title} {\bibinfo
  {title} {{Field-induced quantum spin disordered state in spin-1/2 honeycomb
  magnet $\mathrm{Na}_2\mathrm{Co}_2\mathrm{TeO}_6$}},\ }\href
  {https://doi.org/10.1038/s41467-021-25567-7} {\bibfield  {journal} {\bibinfo
  {journal} {Nat. Commun.}\ }\textbf {\bibinfo {volume} {12}},\ \bibinfo
  {pages} {5559} (\bibinfo {year} {2021})}\BibitemShut {NoStop}%
\bibitem [{\citenamefont {Hentrich}\ \emph {et~al.}(2018)\citenamefont
  {Hentrich}, \citenamefont {Wolter}, \citenamefont {Zotos}, \citenamefont
  {Brenig}, \citenamefont {Nowak}, \citenamefont {Isaeva}, \citenamefont
  {Doert}, \citenamefont {Banerjee}, \citenamefont {Lampen-Kelley},
  \citenamefont {Mandrus}, \citenamefont {Nagler}, \citenamefont {Sears},
  \citenamefont {Kim}, \citenamefont {Büchner},\ and\ \citenamefont
  {Hess}}]{Hentrich_2018}%
  \BibitemOpen
  \bibfield  {author} {\bibinfo {author} {\bibfnamefont {R.}~\bibnamefont
  {Hentrich}}, \bibinfo {author} {\bibfnamefont {A.~U.}\ \bibnamefont
  {Wolter}}, \bibinfo {author} {\bibfnamefont {X.}~\bibnamefont {Zotos}},
  \bibinfo {author} {\bibfnamefont {W.}~\bibnamefont {Brenig}}, \bibinfo
  {author} {\bibfnamefont {D.}~\bibnamefont {Nowak}}, \bibinfo {author}
  {\bibfnamefont {A.}~\bibnamefont {Isaeva}}, \bibinfo {author} {\bibfnamefont
  {T.}~\bibnamefont {Doert}}, \bibinfo {author} {\bibfnamefont
  {A.}~\bibnamefont {Banerjee}}, \bibinfo {author} {\bibfnamefont
  {P.}~\bibnamefont {Lampen-Kelley}}, \bibinfo {author} {\bibfnamefont {D.~G.}\
  \bibnamefont {Mandrus}}, \bibinfo {author} {\bibfnamefont {S.~E.}\
  \bibnamefont {Nagler}}, \bibinfo {author} {\bibfnamefont {J.}~\bibnamefont
  {Sears}}, \bibinfo {author} {\bibfnamefont {Y.-J.}\ \bibnamefont {Kim}},
  \bibinfo {author} {\bibfnamefont {B.}~\bibnamefont {Büchner}},\ and\
  \bibinfo {author} {\bibfnamefont {C.}~\bibnamefont {Hess}},\ }\bibfield
  {title} {\bibinfo {title} {{Unusual Phonon Heat Transport in
  {$\upalpha$-$\mathrm{RuCl}_3$} : Strong Spin-Phonon Scattering and
  Field-Induced Spin Gap}},\ }\href
  {https://doi.org/10.1103/physrevlett.120.117204} {\bibfield  {journal}
  {\bibinfo  {journal} {Phys. Rev. Lett.}\ }\textbf {\bibinfo {volume} {120}},\
  \bibinfo {pages} {117204} (\bibinfo {year} {2018})}\BibitemShut {NoStop}%
\bibitem [{\citenamefont {Yu}\ \emph {et~al.}(2018)\citenamefont {Yu},
  \citenamefont {Xu}, \citenamefont {Ran}, \citenamefont {Ni}, \citenamefont
  {Huang}, \citenamefont {Wang}, \citenamefont {Wen},\ and\ \citenamefont
  {Li}}]{Yu_2018}%
  \BibitemOpen
  \bibfield  {author} {\bibinfo {author} {\bibfnamefont {Y.}~\bibnamefont
  {Yu}}, \bibinfo {author} {\bibfnamefont {Y.}~\bibnamefont {Xu}}, \bibinfo
  {author} {\bibfnamefont {K.}~\bibnamefont {Ran}}, \bibinfo {author}
  {\bibfnamefont {J.}~\bibnamefont {Ni}}, \bibinfo {author} {\bibfnamefont
  {Y.}~\bibnamefont {Huang}}, \bibinfo {author} {\bibfnamefont
  {J.}~\bibnamefont {Wang}}, \bibinfo {author} {\bibfnamefont {J.}~\bibnamefont
  {Wen}},\ and\ \bibinfo {author} {\bibfnamefont {S.}~\bibnamefont {Li}},\
  }\bibfield  {title} {\bibinfo {title} {{Ultralow-Temperature Thermal
  Conductivity of the Kitaev Honeycomb Magnet {$\upalpha$-$\mathrm{RuCl}_3$}
  across the Field-Induced Phase Transition}},\ }\href
  {https://doi.org/10.1103/physrevlett.120.067202} {\bibfield  {journal}
  {\bibinfo  {journal} {Phys. Rev. Lett.}\ }\textbf {\bibinfo {volume} {120}},\
  \bibinfo {pages} {067202} (\bibinfo {year} {2018})}\BibitemShut {NoStop}%
\bibitem [{\citenamefont {Modic}\ \emph {et~al.}(2020)\citenamefont {Modic},
  \citenamefont {McDonald}, \citenamefont {Ruff}, \citenamefont {Bachmann},
  \citenamefont {Lai}, \citenamefont {Palmstrom}, \citenamefont {Graf},
  \citenamefont {Chan}, \citenamefont {Balakirev}, \citenamefont {Betts},
  \citenamefont {Boebinger}, \citenamefont {Schmidt}, \citenamefont {Lawler},
  \citenamefont {Sokolov}, \citenamefont {Moll}, \citenamefont {Ramshaw},\ and\
  \citenamefont {Shekhter}}]{Modic_2020}%
  \BibitemOpen
  \bibfield  {author} {\bibinfo {author} {\bibfnamefont {K.~A.}\ \bibnamefont
  {Modic}}, \bibinfo {author} {\bibfnamefont {R.~D.}\ \bibnamefont {McDonald}},
  \bibinfo {author} {\bibfnamefont {J.~P.~C.}\ \bibnamefont {Ruff}}, \bibinfo
  {author} {\bibfnamefont {M.~D.}\ \bibnamefont {Bachmann}}, \bibinfo {author}
  {\bibfnamefont {Y.}~\bibnamefont {Lai}}, \bibinfo {author} {\bibfnamefont
  {J.~C.}\ \bibnamefont {Palmstrom}}, \bibinfo {author} {\bibfnamefont
  {D.}~\bibnamefont {Graf}}, \bibinfo {author} {\bibfnamefont {M.~K.}\
  \bibnamefont {Chan}}, \bibinfo {author} {\bibfnamefont {F.~F.}\ \bibnamefont
  {Balakirev}}, \bibinfo {author} {\bibfnamefont {J.~B.}\ \bibnamefont
  {Betts}}, \bibinfo {author} {\bibfnamefont {G.~S.}\ \bibnamefont
  {Boebinger}}, \bibinfo {author} {\bibfnamefont {M.}~\bibnamefont {Schmidt}},
  \bibinfo {author} {\bibfnamefont {M.~J.}\ \bibnamefont {Lawler}}, \bibinfo
  {author} {\bibfnamefont {D.~A.}\ \bibnamefont {Sokolov}}, \bibinfo {author}
  {\bibfnamefont {P.~J.~W.}\ \bibnamefont {Moll}}, \bibinfo {author}
  {\bibfnamefont {B.~J.}\ \bibnamefont {Ramshaw}},\ and\ \bibinfo {author}
  {\bibfnamefont {A.}~\bibnamefont {Shekhter}},\ }\bibfield  {title} {\bibinfo
  {title} {{Scale-invariant magnetic anisotropy in {$\mathrm{RuCl}_3$} at high
  magnetic fields}},\ }\href {https://doi.org/10.1038/s41567-020-1028-0}
  {\bibfield  {journal} {\bibinfo  {journal} {Nat. Phys.}\ }\textbf {\bibinfo
  {volume} {17}},\ \bibinfo {pages} {240} (\bibinfo {year} {2020})}\BibitemShut
  {NoStop}%
\bibitem [{\citenamefont {Gohlke}\ \emph {et~al.}(2018)\citenamefont {Gohlke},
  \citenamefont {Moessner},\ and\ \citenamefont {Pollmann}}]{Gohlke2018}%
  \BibitemOpen
  \bibfield  {author} {\bibinfo {author} {\bibfnamefont {M.}~\bibnamefont
  {Gohlke}}, \bibinfo {author} {\bibfnamefont {R.}~\bibnamefont {Moessner}},\
  and\ \bibinfo {author} {\bibfnamefont {F.}~\bibnamefont {Pollmann}},\
  }\bibfield  {title} {\bibinfo {title} {{Dynamical and topological properties
  of the {Kitaev} model in a [111] magnetic field}},\ }\href
  {https://doi.org/10.1103/PhysRevB.98.014418} {\bibfield  {journal} {\bibinfo
  {journal} {Phys. Rev. B}\ }\textbf {\bibinfo {volume} {98}},\ \bibinfo
  {pages} {014418} (\bibinfo {year} {2018})}\BibitemShut {NoStop}%
\bibitem [{\citenamefont {Zhu}\ \emph {et~al.}(2018)\citenamefont {Zhu},
  \citenamefont {Kimchi}, \citenamefont {Sheng},\ and\ \citenamefont
  {Fu}}]{Zhu2018}%
  \BibitemOpen
  \bibfield  {author} {\bibinfo {author} {\bibfnamefont {Z.}~\bibnamefont
  {Zhu}}, \bibinfo {author} {\bibfnamefont {I.}~\bibnamefont {Kimchi}},
  \bibinfo {author} {\bibfnamefont {D.~N.}\ \bibnamefont {Sheng}},\ and\
  \bibinfo {author} {\bibfnamefont {L.}~\bibnamefont {Fu}},\ }\bibfield
  {title} {\bibinfo {title} {{Robust non-{Abelian} spin liquid and a possible
  intermediate phase in the antiferromagnetic {Kitaev} model with magnetic
  field}},\ }\href {https://doi.org/10.1103/PhysRevB.97.241110} {\bibfield
  {journal} {\bibinfo  {journal} {Phys. Rev. B}\ }\textbf {\bibinfo {volume}
  {97}},\ \bibinfo {pages} {241110} (\bibinfo {year} {2018})}\BibitemShut
  {NoStop}%
\bibitem [{\citenamefont {Hickey}\ and\ \citenamefont
  {Trebst}(2019)}]{TrebstPhases2019}%
  \BibitemOpen
  \bibfield  {author} {\bibinfo {author} {\bibfnamefont {C.}~\bibnamefont
  {Hickey}}\ and\ \bibinfo {author} {\bibfnamefont {S.}~\bibnamefont
  {Trebst}},\ }\bibfield  {title} {\bibinfo {title} {{{Emergence of a
  field-driven {$U(1)$} spin liquid in the {K}itaev honeycomb model}}},\ }\href
  {https://doi.org/10.1038/s41467-019-08459-9} {\bibfield  {journal} {\bibinfo
  {journal} {Nat. Commun.}\ }\textbf {\bibinfo {volume} {10}},\ \bibinfo
  {pages} {530} (\bibinfo {year} {2019})}\BibitemShut {NoStop}%
\bibitem [{\citenamefont {Jiang}\ \emph {et~al.}(2011)\citenamefont {Jiang},
  \citenamefont {Gu}, \citenamefont {Qi},\ and\ \citenamefont
  {Trebst}}]{Jiang2011}%
  \BibitemOpen
  \bibfield  {author} {\bibinfo {author} {\bibfnamefont {H.-C.}\ \bibnamefont
  {Jiang}}, \bibinfo {author} {\bibfnamefont {Z.-C.}\ \bibnamefont {Gu}},
  \bibinfo {author} {\bibfnamefont {X.-L.}\ \bibnamefont {Qi}},\ and\ \bibinfo
  {author} {\bibfnamefont {S.}~\bibnamefont {Trebst}},\ }\bibfield  {title}
  {\bibinfo {title} {{Possible proximity of the {M}ott insulating iridate
  {Na${}_{2}$IrO${}_{3}$} to a topological phase: Phase diagram of the
  {H}eisenberg-{K}itaev model in a magnetic field}},\ }\href
  {https://doi.org/10.1103/PhysRevB.83.245104} {\bibfield  {journal} {\bibinfo
  {journal} {Phys. Rev. B}\ }\textbf {\bibinfo {volume} {83}},\ \bibinfo
  {pages} {245104} (\bibinfo {year} {2011})}\BibitemShut {NoStop}%
\bibitem [{\citenamefont {Nasu}\ \emph {et~al.}(2018)\citenamefont {Nasu},
  \citenamefont {Kato}, \citenamefont {Kamiya},\ and\ \citenamefont
  {Motome}}]{Nasu2018}%
  \BibitemOpen
  \bibfield  {author} {\bibinfo {author} {\bibfnamefont {J.}~\bibnamefont
  {Nasu}}, \bibinfo {author} {\bibfnamefont {Y.}~\bibnamefont {Kato}}, \bibinfo
  {author} {\bibfnamefont {Y.}~\bibnamefont {Kamiya}},\ and\ \bibinfo {author}
  {\bibfnamefont {Y.}~\bibnamefont {Motome}},\ }\bibfield  {title} {\bibinfo
  {title} {{Successive {Majorana} topological transitions driven by a magnetic
  field in the {Kitaev} model}},\ }\href
  {https://doi.org/10.1103/physrevb.98.060416} {\bibfield  {journal} {\bibinfo
  {journal} {Phys. Rev. B}\ }\textbf {\bibinfo {volume} {98}},\ \bibinfo
  {pages} {060416} (\bibinfo {year} {2018})}\BibitemShut {NoStop}%
\bibitem [{\citenamefont {Liang}\ \emph {et~al.}(2018)\citenamefont {Liang},
  \citenamefont {Jiang}, \citenamefont {Chen}, \citenamefont {Li},\ and\
  \citenamefont {Wang}}]{Liang2018}%
  \BibitemOpen
  \bibfield  {author} {\bibinfo {author} {\bibfnamefont {S.}~\bibnamefont
  {Liang}}, \bibinfo {author} {\bibfnamefont {M.-H.}\ \bibnamefont {Jiang}},
  \bibinfo {author} {\bibfnamefont {W.}~\bibnamefont {Chen}}, \bibinfo {author}
  {\bibfnamefont {J.-X.}\ \bibnamefont {Li}},\ and\ \bibinfo {author}
  {\bibfnamefont {Q.-H.}\ \bibnamefont {Wang}},\ }\bibfield  {title} {\bibinfo
  {title} {{Intermediate gapless phase and topological phase transition of the
  {K}itaev model in a uniform magnetic field}},\ }\href
  {https://doi.org/10.1103/PhysRevB.98.054433} {\bibfield  {journal} {\bibinfo
  {journal} {Phys. Rev. B}\ }\textbf {\bibinfo {volume} {98}},\ \bibinfo
  {pages} {054433} (\bibinfo {year} {2018})}\BibitemShut {NoStop}%
\bibitem [{\citenamefont {Jahromi}\ \emph {et~al.}(2021)\citenamefont
  {Jahromi}, \citenamefont {H{\"{o}}rmann}, \citenamefont {Adelhardt},
  \citenamefont {Fey}, \citenamefont {Orus},\ and\ \citenamefont
  {Schmidt}}]{Jahromi2021}%
  \BibitemOpen
  \bibfield  {author} {\bibinfo {author} {\bibfnamefont {S.~S.}\ \bibnamefont
  {Jahromi}}, \bibinfo {author} {\bibfnamefont {M.}~\bibnamefont
  {H{\"{o}}rmann}}, \bibinfo {author} {\bibfnamefont {P.}~\bibnamefont
  {Adelhardt}}, \bibinfo {author} {\bibfnamefont {S.}~\bibnamefont {Fey}},
  \bibinfo {author} {\bibfnamefont {R.}~\bibnamefont {Orus}},\ and\ \bibinfo
  {author} {\bibfnamefont {K.~P.}\ \bibnamefont {Schmidt}},\ }\href@noop {}
  {\bibinfo {title} {{Kitaev honeycomb antiferromagnet in a field: quantum
  phase diagram for general spin}}} (\bibinfo {year} {2021}),\ \Eprint
  {https://arxiv.org/abs/2111.06132} {arXiv:2111.06132 [cond-mat.str-el]}
  \BibitemShut {NoStop}%
\bibitem [{\citenamefont {Jiang}\ \emph {et~al.}(2018)\citenamefont {Jiang},
  \citenamefont {Wang}, \citenamefont {Huang},\ and\ \citenamefont
  {Lu}}]{Jiang_2018}%
  \BibitemOpen
  \bibfield  {author} {\bibinfo {author} {\bibfnamefont {H.-C.}\ \bibnamefont
  {Jiang}}, \bibinfo {author} {\bibfnamefont {C.-Y.}\ \bibnamefont {Wang}},
  \bibinfo {author} {\bibfnamefont {B.}~\bibnamefont {Huang}},\ and\ \bibinfo
  {author} {\bibfnamefont {Y.-M.}\ \bibnamefont {Lu}},\ }\href@noop {}
  {\bibinfo {title} {{Field induced quantum spin liquid with spinon Fermi
  surfaces in the {Kitaev} model}}} (\bibinfo {year} {2018}),\ \Eprint
  {https://arxiv.org/abs/1809.08247} {arXiv:1809.08247 [cond-mat.str-el]}
  \BibitemShut {NoStop}%
\bibitem [{\citenamefont {Jiang}\ \emph {et~al.}(2020)\citenamefont {Jiang},
  \citenamefont {Liang}, \citenamefont {Chen}, \citenamefont {Qi},
  \citenamefont {Li},\ and\ \citenamefont {Wang}}]{Jiang2020}%
  \BibitemOpen
  \bibfield  {author} {\bibinfo {author} {\bibfnamefont {M.-H.}\ \bibnamefont
  {Jiang}}, \bibinfo {author} {\bibfnamefont {S.}~\bibnamefont {Liang}},
  \bibinfo {author} {\bibfnamefont {W.}~\bibnamefont {Chen}}, \bibinfo {author}
  {\bibfnamefont {Y.}~\bibnamefont {Qi}}, \bibinfo {author} {\bibfnamefont
  {J.-X.}\ \bibnamefont {Li}},\ and\ \bibinfo {author} {\bibfnamefont {Q.-H.}\
  \bibnamefont {Wang}},\ }\bibfield  {title} {\bibinfo {title} {{Tuning
  Topological Orders by a Conical Magnetic Field in the {Kitaev} Model}},\
  }\href {https://doi.org/10.1103/physrevlett.125.177203} {\bibfield  {journal}
  {\bibinfo  {journal} {Phys. Rev. Lett.}\ }\textbf {\bibinfo {volume} {125}},\
  \bibinfo {pages} {177203} (\bibinfo {year} {2020})}\BibitemShut {NoStop}%
\bibitem [{\citenamefont {Yao}\ and\ \citenamefont
  {Kivelson}(2007)}]{Yao_2007}%
  \BibitemOpen
  \bibfield  {author} {\bibinfo {author} {\bibfnamefont {H.}~\bibnamefont
  {Yao}}\ and\ \bibinfo {author} {\bibfnamefont {S.~A.}\ \bibnamefont
  {Kivelson}},\ }\bibfield  {title} {\bibinfo {title} {{Exact Chiral Spin
  Liquid with Non-{Abelian} Anyons}},\ }\href
  {https://doi.org/10.1103/physrevlett.99.247203} {\bibfield  {journal}
  {\bibinfo  {journal} {Phys. Rev. Lett.}\ }\textbf {\bibinfo {volume} {99}},\
  \bibinfo {pages} {247203} (\bibinfo {year} {2007})}\BibitemShut {NoStop}%
\bibitem [{\citenamefont {Yang}\ \emph {et~al.}(2007)\citenamefont {Yang},
  \citenamefont {Zhou},\ and\ \citenamefont {Sun}}]{Yang_2007}%
  \BibitemOpen
  \bibfield  {author} {\bibinfo {author} {\bibfnamefont {S.}~\bibnamefont
  {Yang}}, \bibinfo {author} {\bibfnamefont {D.~L.}\ \bibnamefont {Zhou}},\
  and\ \bibinfo {author} {\bibfnamefont {C.~P.}\ \bibnamefont {Sun}},\
  }\bibfield  {title} {\bibinfo {title} {{Mosaic spin models with topological
  order}},\ }\href {https://doi.org/10.1103/physrevb.76.180404} {\bibfield
  {journal} {\bibinfo  {journal} {Phys. Rev. B}\ }\textbf {\bibinfo {volume}
  {76}},\ \bibinfo {pages} {180404} (\bibinfo {year} {2007})}\BibitemShut
  {NoStop}%
\bibitem [{\citenamefont {Becker}\ \emph {et~al.}(2015)\citenamefont {Becker},
  \citenamefont {Hermanns}, \citenamefont {Bauer}, \citenamefont {Garst},\ and\
  \citenamefont {Trebst}}]{Becker_2015}%
  \BibitemOpen
  \bibfield  {author} {\bibinfo {author} {\bibfnamefont {M.}~\bibnamefont
  {Becker}}, \bibinfo {author} {\bibfnamefont {M.}~\bibnamefont {Hermanns}},
  \bibinfo {author} {\bibfnamefont {B.}~\bibnamefont {Bauer}}, \bibinfo
  {author} {\bibfnamefont {M.}~\bibnamefont {Garst}},\ and\ \bibinfo {author}
  {\bibfnamefont {S.}~\bibnamefont {Trebst}},\ }\bibfield  {title} {\bibinfo
  {title} {{Spin-orbit physics of $j=1/2$ {Mott} insulators on the triangular
  lattice}},\ }\href {https://doi.org/10.1103/physrevb.91.155135} {\bibfield
  {journal} {\bibinfo  {journal} {Phys. Rev. B}\ }\textbf {\bibinfo {volume}
  {91}},\ \bibinfo {pages} {155135} (\bibinfo {year} {2015})}\BibitemShut
  {NoStop}%
\bibitem [{\citenamefont {Hickey}\ \emph {et~al.}(2021)\citenamefont {Hickey},
  \citenamefont {Gohlke}, \citenamefont {Berke},\ and\ \citenamefont
  {Trebst}}]{Hickey_2021}%
  \BibitemOpen
  \bibfield  {author} {\bibinfo {author} {\bibfnamefont {C.}~\bibnamefont
  {Hickey}}, \bibinfo {author} {\bibfnamefont {M.}~\bibnamefont {Gohlke}},
  \bibinfo {author} {\bibfnamefont {C.}~\bibnamefont {Berke}},\ and\ \bibinfo
  {author} {\bibfnamefont {S.}~\bibnamefont {Trebst}},\ }\bibfield  {title}
  {\bibinfo {title} {{Generic field-driven phenomena in {Kitaev} spin liquids:
  Canted magnetism and proximate spin liquid physics}},\ }\href
  {https://doi.org/10.1103/physrevb.103.064417} {\bibfield  {journal} {\bibinfo
   {journal} {Phys. Rev. B}\ }\textbf {\bibinfo {volume} {103}},\ \bibinfo
  {pages} {064417} (\bibinfo {year} {2021})}\BibitemShut {NoStop}%
\bibitem [{\citenamefont {Zhu}\ \emph {et~al.}(2020)\citenamefont {Zhu},
  \citenamefont {Weng},\ and\ \citenamefont {Sheng}}]{Zhu_2020}%
  \BibitemOpen
  \bibfield  {author} {\bibinfo {author} {\bibfnamefont {Z.}~\bibnamefont
  {Zhu}}, \bibinfo {author} {\bibfnamefont {Z.-Y.}\ \bibnamefont {Weng}},\ and\
  \bibinfo {author} {\bibfnamefont {D.~N.}\ \bibnamefont {Sheng}},\ }\bibfield
  {title} {\bibinfo {title} {{Magnetic field induced spin liquids in {$S=1$}
  {Kitaev} honeycomb model}},\ }\href
  {https://doi.org/10.1103/PhysRevResearch.2.022047} {\bibfield  {journal}
  {\bibinfo  {journal} {Phys. Rev. Research}\ }\textbf {\bibinfo {volume}
  {2}},\ \bibinfo {pages} {022047} (\bibinfo {year} {2020})}\BibitemShut
  {NoStop}%
\bibitem [{\citenamefont {Hickey}\ \emph {et~al.}(2020)\citenamefont {Hickey},
  \citenamefont {Berke}, \citenamefont {Stavropoulos}, \citenamefont {Kee},\
  and\ \citenamefont {Trebst}}]{Trebst_2020_S1}%
  \BibitemOpen
  \bibfield  {author} {\bibinfo {author} {\bibfnamefont {C.}~\bibnamefont
  {Hickey}}, \bibinfo {author} {\bibfnamefont {C.}~\bibnamefont {Berke}},
  \bibinfo {author} {\bibfnamefont {P.~P.}\ \bibnamefont {Stavropoulos}},
  \bibinfo {author} {\bibfnamefont {H.-Y.}\ \bibnamefont {Kee}},\ and\ \bibinfo
  {author} {\bibfnamefont {S.}~\bibnamefont {Trebst}},\ }\bibfield  {title}
  {\bibinfo {title} {{Field-driven gapless spin liquid in the spin-1 {Kitaev}
  honeycomb model}},\ }\href {https://doi.org/10.1103/PhysRevResearch.2.023361}
  {\bibfield  {journal} {\bibinfo  {journal} {Phys. Rev. Research}\ }\textbf
  {\bibinfo {volume} {2}},\ \bibinfo {pages} {023361} (\bibinfo {year}
  {2020})}\BibitemShut {NoStop}%
\bibitem [{\citenamefont {Lee}\ \emph {et~al.}(2020)\citenamefont {Lee},
  \citenamefont {Kawashima},\ and\ \citenamefont {Kim}}]{Lee2020}%
  \BibitemOpen
  \bibfield  {author} {\bibinfo {author} {\bibfnamefont {H.-Y.}\ \bibnamefont
  {Lee}}, \bibinfo {author} {\bibfnamefont {N.}~\bibnamefont {Kawashima}},\
  and\ \bibinfo {author} {\bibfnamefont {Y.~B.}\ \bibnamefont {Kim}},\
  }\bibfield  {title} {\bibinfo {title} {{Tensor network wavefunction of
  {$S=1$} {Kitaev} spin liquids}},\ }\href
  {https://doi.org/10.1103/PhysRevResearch.2.033318} {\bibfield  {journal}
  {\bibinfo  {journal} {Phys. Rev. Research}\ }\textbf {\bibinfo {volume}
  {2}},\ \bibinfo {pages} {033318} (\bibinfo {year} {2020})}\BibitemShut
  {NoStop}%
\bibitem [{\citenamefont {Knetter}\ and\ \citenamefont
  {Uhrig}(2000)}]{Knetter2000}%
  \BibitemOpen
  \bibfield  {author} {\bibinfo {author} {\bibfnamefont {C.}~\bibnamefont
  {Knetter}}\ and\ \bibinfo {author} {\bibfnamefont {G.}~\bibnamefont
  {Uhrig}},\ }\bibfield  {title} {\bibinfo {title} {{Perturbation theory by
  flow equations: Dimerized and frustrated {$S = 1/2$} chain}},\ }\href
  {https://doi.org/10.1007/s100510050026} {\bibfield  {journal} {\bibinfo
  {journal} {Eur. Phys. J. B}\ }\textbf {\bibinfo {volume} {13}},\ \bibinfo
  {pages} {209} (\bibinfo {year} {2000})}\BibitemShut {NoStop}%
\bibitem [{\citenamefont {Knetter}\ \emph
  {et~al.}(2003{\natexlab{a}})\citenamefont {Knetter}, \citenamefont
  {Schmidt},\ and\ \citenamefont {Uhrig}}]{Knetter2003}%
  \BibitemOpen
  \bibfield  {author} {\bibinfo {author} {\bibfnamefont {C.}~\bibnamefont
  {Knetter}}, \bibinfo {author} {\bibfnamefont {K.~P.}\ \bibnamefont
  {Schmidt}},\ and\ \bibinfo {author} {\bibfnamefont {G.~S.}\ \bibnamefont
  {Uhrig}},\ }\bibfield  {title} {\bibinfo {title} {{High order perturbation
  theory for spectral densities of multi-particle excitations: {$S = 1/2$}
  two-leg {Heisenberg} ladder}},\ }\href
  {https://doi.org/10.1140/epjb/e2004-00008-2} {\bibfield  {journal} {\bibinfo
  {journal} {Eur. Phys. J. B}\ }\textbf {\bibinfo {volume} {36}},\ \bibinfo
  {pages} {525} (\bibinfo {year} {2003}{\natexlab{a}})}\BibitemShut {NoStop}%
\bibitem [{\citenamefont {Jin}\ \emph {et~al.}(2021)\citenamefont {Jin},
  \citenamefont {Tu},\ and\ \citenamefont {Zhou}}]{Jin2021}%
  \BibitemOpen
  \bibfield  {author} {\bibinfo {author} {\bibfnamefont {H.-K.}\ \bibnamefont
  {Jin}}, \bibinfo {author} {\bibfnamefont {H.-H.}\ \bibnamefont {Tu}},\ and\
  \bibinfo {author} {\bibfnamefont {Y.}~\bibnamefont {Zhou}},\ }\bibfield
  {title} {\bibinfo {title} {{Density matrix renormalization group boosted by
  {Gutzwiller} projected wave functions}},\ }\href
  {https://doi.org/10.1103/physrevb.104.l020409} {\bibfield  {journal}
  {\bibinfo  {journal} {Phys. Rev. B}\ }\textbf {\bibinfo {volume} {104}},\
  \bibinfo {pages} {l020409} (\bibinfo {year} {2021})}\BibitemShut {NoStop}%
\bibitem [{\citenamefont {Zhang}\ \emph {et~al.}(2022)\citenamefont {Zhang},
  \citenamefont {Hal{\'{a}}sz},\ and\ \citenamefont {Batista}}]{Zhang2022}%
  \BibitemOpen
  \bibfield  {author} {\bibinfo {author} {\bibfnamefont {S.-S.}\ \bibnamefont
  {Zhang}}, \bibinfo {author} {\bibfnamefont {G.~B.}\ \bibnamefont
  {Hal{\'{a}}sz}},\ and\ \bibinfo {author} {\bibfnamefont {C.~D.}\ \bibnamefont
  {Batista}},\ }\bibfield  {title} {\bibinfo {title} {{Theory of the {Kitaev}
  model in a [111] magnetic field}},\ }\href
  {https://doi.org/10.1038/s41467-022-28014-3} {\bibfield  {journal} {\bibinfo
  {journal} {Nat. Commun.}\ }\textbf {\bibinfo {volume} {13}},\ \bibinfo
  {pages} {399} (\bibinfo {year} {2022})}\BibitemShut {NoStop}%
\bibitem [{\citenamefont {Matsubara}\ and\ \citenamefont
  {Matsuda}(1956)}]{Matsubara1956}%
  \BibitemOpen
  \bibfield  {author} {\bibinfo {author} {\bibfnamefont {T.}~\bibnamefont
  {Matsubara}}\ and\ \bibinfo {author} {\bibfnamefont {H.}~\bibnamefont
  {Matsuda}},\ }\bibfield  {title} {\bibinfo {title} {{A Lattice Model of
  Liquid Helium}},\ }\href {https://doi.org/10.1143/ptp.16.416} {\bibfield
  {journal} {\bibinfo  {journal} {Prog. Theor. Phys.}\ }\textbf {\bibinfo
  {volume} {16}},\ \bibinfo {pages} {416} (\bibinfo {year} {1956})}\BibitemShut
  {NoStop}%
\bibitem [{\citenamefont {Guo}(2012)}]{Guo2012}%
  \BibitemOpen
  \bibfield  {author} {\bibinfo {author} {\bibfnamefont {H.}~\bibnamefont
  {Guo}},\ }\bibfield  {title} {\bibinfo {title} {{Hard-core bosons in
  one-dimensional interacting topological bands}},\ }\href
  {https://doi.org/10.1103/PhysRevA.86.055604} {\bibfield  {journal} {\bibinfo
  {journal} {Phys. Rev. A}\ }\textbf {\bibinfo {volume} {86}},\ \bibinfo
  {pages} {055604} (\bibinfo {year} {2012})}\BibitemShut {NoStop}%
\bibitem [{\citenamefont {Baker}\ and\ \citenamefont
  {Graves-Morris}(1996)}]{PadeApproximants1996}%
  \BibitemOpen
  \bibfield  {author} {\bibinfo {author} {\bibfnamefont {G.~A.}\ \bibnamefont
  {Baker}}\ and\ \bibinfo {author} {\bibfnamefont {P.}~\bibnamefont
  {Graves-Morris}},\ }\href {https://doi.org/10.1017/CBO9780511530074} {\emph
  {\bibinfo {title} {{Pad{\'{e}} Approximants}}}},\ \bibinfo {edition} {2nd}\
  ed.,\ Encyclopedia of Mathematics and its Applications\ (\bibinfo
  {publisher} {Cambridge University Press, Cambridge, UK},\ \bibinfo {year}
  {1996})\BibitemShut {NoStop}%
\bibitem [{\citenamefont {Press}\ \emph {et~al.}(2007)\citenamefont {Press},
  \citenamefont {Teukolsky}, \citenamefont {Vetterling},\ and\ \citenamefont
  {Flannery}}]{NumericalRecipies2007}%
  \BibitemOpen
  \bibfield  {author} {\bibinfo {author} {\bibfnamefont {W.}~\bibnamefont
  {Press}}, \bibinfo {author} {\bibfnamefont {S.}~\bibnamefont {Teukolsky}},
  \bibinfo {author} {\bibfnamefont {W.}~\bibnamefont {Vetterling}},\ and\
  \bibinfo {author} {\bibfnamefont {B.}~\bibnamefont {Flannery}},\ }\href
  {https://www.cambridge.org/de/academic/subjects/mathematics/numerical-recipes/numerical-recipes-art-scientific-computing-3rd-edition?format=HB}
  {\emph {\bibinfo {title} {{Numerical Recipes: The Art of Scientific
  Computing}}}},\ \bibinfo {edition} {3rd}\ ed.\ (\bibinfo  {publisher}
  {Cambridge University Press, Cambridge, UK},\ \bibinfo {year}
  {2007})\BibitemShut {NoStop}%
\bibitem [{\citenamefont {Knetter}\ \emph
  {et~al.}(2003{\natexlab{b}})\citenamefont {Knetter}, \citenamefont
  {Schmidt},\ and\ \citenamefont {Uhrig}}]{Knetter2003_JoPA}%
  \BibitemOpen
  \bibfield  {author} {\bibinfo {author} {\bibfnamefont {C.}~\bibnamefont
  {Knetter}}, \bibinfo {author} {\bibfnamefont {K.~P.}\ \bibnamefont
  {Schmidt}},\ and\ \bibinfo {author} {\bibfnamefont {G.}~\bibnamefont
  {Uhrig}},\ }\bibfield  {title} {\bibinfo {title} {{The structure of operators
  in effective particle-conserving models}},\ }\href
  {https://doi.org/10.1088/0305-4470/36/29/302} {\bibfield  {journal} {\bibinfo
   {journal} {J. Phys. A}\ }\textbf {\bibinfo {volume} {36}},\ \bibinfo {pages}
  {7889} (\bibinfo {year} {2003}{\natexlab{b}})}\BibitemShut {NoStop}%
\bibitem [{\citenamefont {Knetter}(2003)}]{PhdKnetter2003}%
  \BibitemOpen
  \bibfield  {author} {\bibinfo {author} {\bibfnamefont {C.}~\bibnamefont
  {Knetter}},\ }\emph {\bibinfo {title} {{Perturbative Continuous Unitary
  Transformations: Spectral Properties of Low Dimensional Spin Systems}}},\
  \href {https://kups.ub.uni-koeln.de/942/} {Ph.D. thesis},\ \bibinfo  {school}
  {Universit{\"a}t zu K{\"o}ln} (\bibinfo {year} {2003})\BibitemShut {NoStop}%
\bibitem [{\citenamefont {Knetter}\ \emph {et~al.}(2001)\citenamefont
  {Knetter}, \citenamefont {Schmidt}, \citenamefont {Gr\"uninger},\ and\
  \citenamefont {Uhrig}}]{Knetter2001}%
  \BibitemOpen
  \bibfield  {author} {\bibinfo {author} {\bibfnamefont {C.}~\bibnamefont
  {Knetter}}, \bibinfo {author} {\bibfnamefont {K.~P.}\ \bibnamefont
  {Schmidt}}, \bibinfo {author} {\bibfnamefont {M.}~\bibnamefont
  {Gr\"uninger}},\ and\ \bibinfo {author} {\bibfnamefont {G.~S.}\ \bibnamefont
  {Uhrig}},\ }\bibfield  {title} {\bibinfo {title} {{Fractional and Integer
  Excitations in Quantum Antiferromagnetic Spin $1/2$ Ladders}},\ }\href
  {https://doi.org/10.1103/PhysRevLett.87.167204} {\bibfield  {journal}
  {\bibinfo  {journal} {Phys. Rev. Lett.}\ }\textbf {\bibinfo {volume} {87}},\
  \bibinfo {pages} {167204} (\bibinfo {year} {2001})}\BibitemShut {NoStop}%
\bibitem [{\citenamefont {Gelfand}\ \emph {et~al.}(1990)\citenamefont
  {Gelfand}, \citenamefont {Singh},\ and\ \citenamefont {Huse}}]{Gelfand1990}%
  \BibitemOpen
  \bibfield  {author} {\bibinfo {author} {\bibfnamefont {M.}~\bibnamefont
  {Gelfand}}, \bibinfo {author} {\bibfnamefont {R.}~\bibnamefont {Singh}},\
  and\ \bibinfo {author} {\bibfnamefont {D.}~\bibnamefont {Huse}},\ }\bibfield
  {title} {\bibinfo {title} {{Perturbation expansions for quantum many-body
  systems}},\ }\href {https://doi.org/10.1007/BF01334744} {\bibfield  {journal}
  {\bibinfo  {journal} {J. Stat. Phys.}\ }\textbf {\bibinfo {volume} {59}},\
  \bibinfo {pages} {1093} (\bibinfo {year} {1990})}\BibitemShut {NoStop}%
\bibitem [{\citenamefont {C{\"o}ster}(2015)}]{PhdCoester2015}%
  \BibitemOpen
  \bibfield  {author} {\bibinfo {author} {\bibfnamefont {K.}~\bibnamefont
  {C{\"o}ster}},\ }\emph {\bibinfo {title} {{Quasiparticle pictures and graphs
  - from perturbative to non-perturbative linked-cluster expansions}}},\ \href
  {https://doi.org/10.17877/DE290R-16955} {Ph.D. thesis},\ \bibinfo  {school}
  {TU Dortmund} (\bibinfo {year} {2015})\BibitemShut {NoStop}%
\bibitem [{\citenamefont {Schmidt}(2004)}]{SchmidtPhD}%
  \BibitemOpen
  \bibfield  {author} {\bibinfo {author} {\bibfnamefont {K.~P.}\ \bibnamefont
  {Schmidt}},\ }\emph {\bibinfo {title} {{Spectral properties of quasi
  one-dimensional quantum antiferromagnets perturbative continuous unitary
  transformations}}},\ \href
  {http://kups.ub.uni-koeln.de/volltexte/2004/1316/pdf/Diss2.pdf} {Ph.D.
  thesis},\ \bibinfo  {school} {Universit{\"a}t zu K{\"{o}}ln} (\bibinfo {year}
  {2004})\BibitemShut {NoStop}%
\bibitem [{\citenamefont {Feynman}(1965)}]{Feynman_lectures}%
  \BibitemOpen
  \bibfield  {author} {\bibinfo {author} {\bibfnamefont {R.~P.}\ \bibnamefont
  {Feynman}},\ }\href {https://www.feynmanlectures.caltech.edu/info/} {\emph
  {\bibinfo {title} {{The {F}eynman lectures on physics}}}}\ (\bibinfo
  {publisher} {Addison-Wesley Boston, MA},\ \bibinfo {year}
  {1963--1965})\BibitemShut {NoStop}%
\bibitem [{\citenamefont {Adelhardt}\ \emph {et~al.}(2020)\citenamefont
  {Adelhardt}, \citenamefont {Koziol}, \citenamefont {Schellenberger},\ and\
  \citenamefont {Schmidt}}]{Adelhardt2020}%
  \BibitemOpen
  \bibfield  {author} {\bibinfo {author} {\bibfnamefont {P.}~\bibnamefont
  {Adelhardt}}, \bibinfo {author} {\bibfnamefont {J.~A.}\ \bibnamefont
  {Koziol}}, \bibinfo {author} {\bibfnamefont {A.}~\bibnamefont
  {Schellenberger}},\ and\ \bibinfo {author} {\bibfnamefont {K.~P.}\
  \bibnamefont {Schmidt}},\ }\bibfield  {title} {\bibinfo {title} {{Quantum
  criticality and excitations of a long-range anisotropic {XY} chain in a
  transverse field}},\ }\href {https://doi.org/10.1103/physrevb.102.174424}
  {\bibfield  {journal} {\bibinfo  {journal} {Phys. Rev. B}\ }\textbf {\bibinfo
  {volume} {102}},\ \bibinfo {pages} {174424} (\bibinfo {year}
  {2020})}\BibitemShut {NoStop}%
\bibitem [{\citenamefont {Lenke}\ \emph {et~al.}(2021)\citenamefont {Lenke},
  \citenamefont {M\"uhlhauser},\ and\ \citenamefont
  {Schmidt}}]{lenke2021highorder}%
  \BibitemOpen
  \bibfield  {author} {\bibinfo {author} {\bibfnamefont {L.}~\bibnamefont
  {Lenke}}, \bibinfo {author} {\bibfnamefont {M.}~\bibnamefont
  {M\"uhlhauser}},\ and\ \bibinfo {author} {\bibfnamefont {K.~P.}\ \bibnamefont
  {Schmidt}},\ }\bibfield  {title} {\bibinfo {title} {{High-order series
  expansion of non-{Hermitian} quantum spin models}},\ }\href
  {https://doi.org/10.1103/PhysRevB.104.195137} {\bibfield  {journal} {\bibinfo
   {journal} {Phys. Rev. B}\ }\textbf {\bibinfo {volume} {104}},\ \bibinfo
  {pages} {195137} (\bibinfo {year} {2021})}\BibitemShut {NoStop}%
\bibitem [{\citenamefont {Hafez}\ and\ \citenamefont
  {Jafari}(2010)}]{Hafez2010}%
  \BibitemOpen
  \bibfield  {author} {\bibinfo {author} {\bibfnamefont {M.}~\bibnamefont
  {Hafez}}\ and\ \bibinfo {author} {\bibfnamefont {S.~A.}\ \bibnamefont
  {Jafari}},\ }\bibfield  {title} {\bibinfo {title} {{Excitation Spectrum of
  One-dimensional Extended Ionic {Hubbard} Model}},\ }\href
  {https://doi.org/10.1140/epjb/e2010-10509-x} {\bibfield  {journal} {\bibinfo
  {journal} {Eur. Phys. J. B}\ }\textbf {\bibinfo {volume} {78}},\ \bibinfo
  {pages} {323} (\bibinfo {year} {2010})}\BibitemShut {NoStop}%
\bibitem [{\citenamefont {Verresen}\ \emph {et~al.}(2019)\citenamefont
  {Verresen}, \citenamefont {Moessner},\ and\ \citenamefont
  {Pollmann}}]{Verresen_2019}%
  \BibitemOpen
  \bibfield  {author} {\bibinfo {author} {\bibfnamefont {R.}~\bibnamefont
  {Verresen}}, \bibinfo {author} {\bibfnamefont {R.}~\bibnamefont {Moessner}},\
  and\ \bibinfo {author} {\bibfnamefont {F.}~\bibnamefont {Pollmann}},\
  }\bibfield  {title} {\bibinfo {title} {{Avoided quasiparticle decay from
  strong quantum interactions}},\ }\href
  {https://doi.org/10.1038/s41567-019-0535-3} {\bibfield  {journal} {\bibinfo
  {journal} {Nat. Phys.}\ }\textbf {\bibinfo {volume} {15}},\ \bibinfo {pages}
  {750} (\bibinfo {year} {2019})}\BibitemShut {NoStop}%
\bibitem [{\citenamefont {Fischer}\ \emph {et~al.}(2010)\citenamefont
  {Fischer}, \citenamefont {Duffe},\ and\ \citenamefont {Uhrig}}]{Fischer2010}%
  \BibitemOpen
  \bibfield  {author} {\bibinfo {author} {\bibfnamefont {T.}~\bibnamefont
  {Fischer}}, \bibinfo {author} {\bibfnamefont {S.}~\bibnamefont {Duffe}},\
  and\ \bibinfo {author} {\bibfnamefont {G.~S.}\ \bibnamefont {Uhrig}},\
  }\bibfield  {title} {\bibinfo {title} {{Adapted continuous unitary
  transformation to treat systems with quasiparticles of finite lifetime}},\
  }\href {https://doi.org/10.1088/1367-2630/12/3/033048} {\bibfield  {journal}
  {\bibinfo  {journal} {New J. Phys.}\ }\textbf {\bibinfo {volume} {12}},\
  \bibinfo {pages} {033048} (\bibinfo {year} {2010})}\BibitemShut {NoStop}%
\bibitem [{Note1()}]{Note1}%
  \BibitemOpen
  \bibinfo {note} {We sort out defective poles using the approach of Adelhardt
  et al.~\cite {Adelhardt2020}. Therefore we vary $n,m$ for $n+m\leq k-1$ and
  group the resulting extrapolants in families of constant $n-m$. If a family
  inhibits less than two members after removing the extrapolants with
  nonphysical poles, we discard the whole family. For approximating $f$, we
  take the mean of the highest order extrapolant of all families \cite
  {Adelhardt2020}}\BibitemShut {NoStop}%
\bibitem [{Note2()}]{Note2}%
  \BibitemOpen
  \bibinfo {note} {In discussion with Gohlke an error of $1/(4\pi )$ in the
  normalization of Fig.~7(a) in \cite {Gohlke2018} was found. Adding this
  factor to our data, we obtain reasonable matching results.}\BibitemShut
  {Stop}%
\bibitem [{\citenamefont {Glamazda}\ \emph {et~al.}(2016)\citenamefont
  {Glamazda}, \citenamefont {Lemmens}, \citenamefont {Do}, \citenamefont
  {Choi},\ and\ \citenamefont {Choi}}]{Glamazda2016}%
  \BibitemOpen
  \bibfield  {author} {\bibinfo {author} {\bibfnamefont {A.}~\bibnamefont
  {Glamazda}}, \bibinfo {author} {\bibfnamefont {P.}~\bibnamefont {Lemmens}},
  \bibinfo {author} {\bibfnamefont {S.~H.}\ \bibnamefont {Do}}, \bibinfo
  {author} {\bibfnamefont {Y.~S.}\ \bibnamefont {Choi}},\ and\ \bibinfo
  {author} {\bibfnamefont {K.~Y.}\ \bibnamefont {Choi}},\ }\bibfield  {title}
  {\bibinfo {title} {{Raman spectroscopic signature of fractionalized
  excitations in the harmonic-honeycomb iridates $\upbeta$- and
  {$\upgamma$-$\mathrm{Li}_2\mathrm{IrO}_3$}}},\ }\href
  {https://doi.org/10.1038/ncomms12286} {\bibfield  {journal} {\bibinfo
  {journal} {Nat. Commun.}\ }\textbf {\bibinfo {volume} {7}},\ \bibinfo {pages}
  {12286} (\bibinfo {year} {2016})}\BibitemShut {NoStop}%
\bibitem [{\citenamefont {Sandilands}\ \emph {et~al.}(2015)\citenamefont
  {Sandilands}, \citenamefont {Tian}, \citenamefont {Plumb}, \citenamefont
  {Kim},\ and\ \citenamefont {Burch}}]{Sandilands2015}%
  \BibitemOpen
  \bibfield  {author} {\bibinfo {author} {\bibfnamefont {L.~J.}\ \bibnamefont
  {Sandilands}}, \bibinfo {author} {\bibfnamefont {Y.}~\bibnamefont {Tian}},
  \bibinfo {author} {\bibfnamefont {K.~W.}\ \bibnamefont {Plumb}}, \bibinfo
  {author} {\bibfnamefont {Y.-J.}\ \bibnamefont {Kim}},\ and\ \bibinfo {author}
  {\bibfnamefont {K.~S.}\ \bibnamefont {Burch}},\ }\bibfield  {title} {\bibinfo
  {title} {{Scattering Continuum and Possible Fractionalized Excitations in
  $\upalpha$-{$\mathrm{RuCl}_3$}}},\ }\href
  {https://doi.org/10.1103/physrevlett.114.147201} {\bibfield  {journal}
  {\bibinfo  {journal} {Phys. Rev. Lett.}\ }\textbf {\bibinfo {volume} {114}},\
  \bibinfo {pages} {147201} (\bibinfo {year} {2015})}\BibitemShut {NoStop}%
\bibitem [{\citenamefont {Perreault}\ \emph {et~al.}(2015)\citenamefont
  {Perreault}, \citenamefont {Knolle}, \citenamefont {Perkins},\ and\
  \citenamefont {Burnell}}]{Perreault2015}%
  \BibitemOpen
  \bibfield  {author} {\bibinfo {author} {\bibfnamefont {B.}~\bibnamefont
  {Perreault}}, \bibinfo {author} {\bibfnamefont {J.}~\bibnamefont {Knolle}},
  \bibinfo {author} {\bibfnamefont {N.~B.}\ \bibnamefont {Perkins}},\ and\
  \bibinfo {author} {\bibfnamefont {F.~J.}\ \bibnamefont {Burnell}},\
  }\bibfield  {title} {\bibinfo {title} {{Theory of {Raman} response in
  three-dimensional {K}itaev spin liquids: Application to $\ensuremath{\beta}$-
  and {$\ensuremath{\gamma}$-${\mathrm{Li}}_{2}{\mathrm{IrO}}_{3}$}
  compounds}},\ }\href {https://doi.org/10.1103/PhysRevB.92.094439} {\bibfield
  {journal} {\bibinfo  {journal} {Phys. Rev. B}\ }\textbf {\bibinfo {volume}
  {92}},\ \bibinfo {pages} {094439} (\bibinfo {year} {2015})}\BibitemShut
  {NoStop}%
\bibitem [{\citenamefont {Knolle}\ \emph {et~al.}(2014)\citenamefont {Knolle},
  \citenamefont {Chern}, \citenamefont {Kovrizhin}, \citenamefont {Moessner},\
  and\ \citenamefont {Perkins}}]{Knolle2014}%
  \BibitemOpen
  \bibfield  {author} {\bibinfo {author} {\bibfnamefont {J.}~\bibnamefont
  {Knolle}}, \bibinfo {author} {\bibfnamefont {G.-W.}\ \bibnamefont {Chern}},
  \bibinfo {author} {\bibfnamefont {D.~L.}\ \bibnamefont {Kovrizhin}}, \bibinfo
  {author} {\bibfnamefont {R.}~\bibnamefont {Moessner}},\ and\ \bibinfo
  {author} {\bibfnamefont {N.~B.}\ \bibnamefont {Perkins}},\ }\bibfield
  {title} {\bibinfo {title} {{Raman Scattering Signatures of {K}itaev Spin
  Liquids in {$\mathrm {A}_2 \mathrm{IrO}_3$} Iridates with {A=Na} or {Li}}},\
  }\href {https://doi.org/10.1103/physrevlett.113.187201} {\bibfield  {journal}
  {\bibinfo  {journal} {Phys. Rev. Lett.}\ }\textbf {\bibinfo {volume} {113}},\
  \bibinfo {pages} {187201} (\bibinfo {year} {2014})}\BibitemShut {NoStop}%
\bibitem [{\citenamefont {Pan}\ \emph {et~al.}(2021)\citenamefont {Pan},
  \citenamefont {Jin}, \citenamefont {Ji}, \citenamefont {Zhang},\ and\
  \citenamefont {Yu}}]{Pan2021}%
  \BibitemOpen
  \bibfield  {author} {\bibinfo {author} {\bibfnamefont {J.}~\bibnamefont
  {Pan}}, \bibinfo {author} {\bibfnamefont {F.}~\bibnamefont {Jin}}, \bibinfo
  {author} {\bibfnamefont {J.}~\bibnamefont {Ji}}, \bibinfo {author}
  {\bibfnamefont {Q.}~\bibnamefont {Zhang}},\ and\ \bibinfo {author}
  {\bibfnamefont {R.}~\bibnamefont {Yu}},\ }\href@noop {} {\bibinfo {title}
  {{Two-magnon {R}aman scattering in antiferromagnetic phases of frustrated
  spin models on the honeycomb lattice}}} (\bibinfo {year} {2021}),\ \Eprint
  {https://arxiv.org/abs/2104.01903} {arXiv:2104.01903 [cond-mat.str-el]}
  \BibitemShut {NoStop}%
\bibitem [{\citenamefont {Yamamoto}\ and\ \citenamefont
  {Kimura}(2020)}]{Yamamoto2020}%
  \BibitemOpen
  \bibfield  {author} {\bibinfo {author} {\bibfnamefont {S.}~\bibnamefont
  {Yamamoto}}\ and\ \bibinfo {author} {\bibfnamefont {T.}~\bibnamefont
  {Kimura}},\ }\bibfield  {title} {\bibinfo {title} {{Raman Scattering
  Polarization and Single Spinon Identification in Two-Dimensional {Kitaev}
  Quantum Spin Liquids}},\ }\href {https://doi.org/10.7566/jpsj.89.063701}
  {\bibfield  {journal} {\bibinfo  {journal} {J. Phys. Soc. Japan}\ }\textbf
  {\bibinfo {volume} {89}},\ \bibinfo {pages} {063701} (\bibinfo {year}
  {2020})}\BibitemShut {NoStop}%
\bibitem [{\citenamefont {Kitagawa}\ \emph {et~al.}(2018)\citenamefont
  {Kitagawa}, \citenamefont {Takayama}, \citenamefont {Matsumoto},
  \citenamefont {Kato}, \citenamefont {Takano}, \citenamefont {Kishimoto},
  \citenamefont {Bette}, \citenamefont {Dinnebier}, \citenamefont {Jackeli},\
  and\ \citenamefont {Takagi}}]{Kitagawa2018}%
  \BibitemOpen
  \bibfield  {author} {\bibinfo {author} {\bibfnamefont {K.}~\bibnamefont
  {Kitagawa}}, \bibinfo {author} {\bibfnamefont {T.}~\bibnamefont {Takayama}},
  \bibinfo {author} {\bibfnamefont {Y.}~\bibnamefont {Matsumoto}}, \bibinfo
  {author} {\bibfnamefont {A.}~\bibnamefont {Kato}}, \bibinfo {author}
  {\bibfnamefont {R.}~\bibnamefont {Takano}}, \bibinfo {author} {\bibfnamefont
  {Y.}~\bibnamefont {Kishimoto}}, \bibinfo {author} {\bibfnamefont
  {S.}~\bibnamefont {Bette}}, \bibinfo {author} {\bibfnamefont
  {R.}~\bibnamefont {Dinnebier}}, \bibinfo {author} {\bibfnamefont
  {G.}~\bibnamefont {Jackeli}},\ and\ \bibinfo {author} {\bibfnamefont
  {H.}~\bibnamefont {Takagi}},\ }\bibfield  {title} {\bibinfo {title} {{A
  spin--orbital-entangled quantum liquid on a honeycomb lattice}},\ }\href
  {https://doi.org/10.1038/nature25482} {\bibfield  {journal} {\bibinfo
  {journal} {Nature}\ }\textbf {\bibinfo {volume} {554}},\ \bibinfo {pages}
  {341} (\bibinfo {year} {2018})}\BibitemShut {NoStop}%
\bibitem [{\citenamefont {R{\"u}cker}\ and\ \citenamefont
  {R{\"u}cker}(2000)}]{Ruecker_2000}%
  \BibitemOpen
  \bibfield  {author} {\bibinfo {author} {\bibfnamefont {G.}~\bibnamefont
  {R{\"u}cker}}\ and\ \bibinfo {author} {\bibfnamefont {C.}~\bibnamefont
  {R{\"u}cker}},\ }\bibfield  {title} {\bibinfo {title} {{Automatic enumeration
  of all connected subgraphs}},\ }\href
  {https://match.pmf.kg.ac.rs/electronic_versions/Match41/match41_145-149.pdf}
  {\bibfield  {journal} {\bibinfo  {journal} {MATCH Commun. Math. Comput.
  Chem.}\ }\textbf {\bibinfo {volume} {41}},\ \bibinfo {pages} {145} (\bibinfo
  {year} {2000})}\BibitemShut {NoStop}%
\end{thebibliography}%

\end{document}